\documentclass[preprint,12pt]{elsarticle}

\graphicspath{{./}{Figures/}}

\usepackage{amssymb}
\usepackage{amsmath}
\usepackage{bm}
\usepackage{subcaption}
\usepackage{float}
\usepackage{geometry}
 \usepackage{setspace}
\usepackage{multirow}
\usepackage{siunitx}
\usepackage{nicefrac}
\usepackage[dvipsnames]{xcolor}
\usepackage{hyperref}
\usepackage[normalem]{ulem}
\usepackage[english]{babel}

\biboptions{sort&compress}

\newcommand{\suggestion}[1]{{\color{black} #1}}

\makeatletter
\def\ps@pprintTitle{%
  \let\@oddhead\@empty
  \let\@evenhead\@empty
  \let\@oddfoot\@empty
  \let\@evenfoot\@oddfoot
}
\makeatother

\begin{document}

\begin{frontmatter}

\title{\suggestion{Self-sorting of bidisperse particles in evaporating sessile droplets}}

\author[affil1]{Aman Kumar Jain}
\author[affil2] {Fabian Denner}
\author[affil1] {Berend van Wachem\corref{cor1}}
\ead{berend.van.wachem@multiflow.org}

\address[affil1]{Chair of Mechanical Process Engineering, Otto-von-Guericke-Universit\"{a}t Magdeburg, Universit\"atsplatz 2, 39106 Magdeburg, Germany}
\address[affil2]{Department of Mechanical Engineering, Polytechnique Montréal, Montréal, H3T 1J4, Québec, Canada}
\cortext[cor1]{Corresponding author: }

\begin{abstract}
This study investigates the dispersion and self-sorting dynamics of bidisperse particles, i.e., a mixture of two distinct particle sizes, during the evaporation of ethanol droplets on a heated substrate, focusing on the influence of surface wettability, Marangoni stresses, and relative particle density.
To this end, numerical simulations are carried out using a two-stage numerical approach: the first stage simulates the gas-liquid flow along with the heat and vapor distribution, while the second stage models the particle behavior using Lagrangian particle tracking.
The results reveal that for an ethanol droplet evaporating with a constant contact angle in the absence of thermocapillary Marangoni stresses, the flow induced by the receding motion of the contact line supersedes the capillary flow, moving the fluid from the contact line to the apex of the droplet. This flow moves the particles from the bulk of the droplet to the apex of the droplet and suppresses size-based self-sorting of the particles. 
However, in the presence of Marangoni stresses, a flow along the interface near the apex of the droplet promotes the self-sorting of particles based on their size, whereby smaller particles concentrate near the droplet apex and larger particles form an outer shell around them.
\end{abstract}

% %%Graphical abstract
% \begin{graphicalabstract}
% \includegraphics[width=\textwidth]{Figures/GraphicalAbstractPaper2.pdf}
% \end{graphicalabstract}

% %%Research highlights
% \begin{highlights}
% \item Improved two-stage VOF-DEM model for the behavior of bidisperse particles in evaporation sessile droplets.
% \item Without Marangoni flow, the flow induced by the receding motion of the contact line supersedes the capillary flow.
% \item The rate of evaporation is reduced in the presence of Marangoni flow for droplets \suggestion{with a larger contact angle.}
% % \correction{for hydrophobic droplets}.
% % owing to the heat removal from the droplet through the interface for hydrophobic droplets.
% \item Marangoni flow induces self-sorting of particles, with smaller particles concentrated at the droplet apex and larger particles at the periphery.
% \item The flow near the apex of the droplet affects the self-sorting of the particles.
% % \end{highlights}


\end{frontmatter}

%% \linenumbers

%% main text
\section{Introduction}
\label{Introduction}

Sessile droplets are encountered in many industrial and technological applications~\citep{Zang2019}, such as inkjet printing~\citep{Yoo2015}, spray cooling, and medical diagnostics~\citep{Chen2016a}.
Even though sessile droplets frequently occur, describing them accurately is a complex process, as they involve mass and heat transfer, as well as the motion of the three-phase contact line.
In addition, Marangoni convection resulting from temperature and solutal gradients as well as the wettability of the substrate affects the motion of the contact line, which in turn affects the flow within the droplet~\citep{Snoeijer2013}. 
In the literature, two main evaporation modes of sessile droplets are often distinguished~\citep{Gelderblom2022}: droplets with a constant contact radius and varying contact angle, often referred to as constant contact radius (CCR) mode, and droplets with a varying contact radius and constant contact angle, often referred to as constant contact angle (CCA) mode. In reality, depending on the receding contact angle for a given solid-liquid combination and on self-pinning due to deposits, a droplet either evaporates in the CCR mode or CCA mode~\citep{Larson2014}.
The substrate properties can, as is done in many engineering applications, be tailored to achieve a desired motion of the contact line, thereby steering the flow inside the droplet and, consequently, the particle deposition on the substrate~\citep{Iqbal2022}.

To understand the nature of the deposition, it is important to understand the heat, mass, and momentum transport within the droplet during evaporation, the behavior of the three-phase contact line, and the interaction of particles with the interface of the droplet and substrate~\citep{Larson2014, Erdem2024}.
The deposition pattern can be of several forms, with the most common being a ``coffee ring'' deposit~\citep{Deegan2000} along with the coffee-eye, concentric rings, cracks, or wrinkling~\citep{Mondal2023}. Usually, these non-uniform deposits are regarded as defects in industrial processes and efforts are made to obtain a uniform deposition, which requires an in-depth understanding of the phenomenon related to sessile droplet evaporation and particle dispersion.

In the case of the evaporation of a sessile droplet on a heated substrate, the flow inside the droplet is primarily affected by thermal Marangoni stresses and the contact line motion. In our previous work~\citep{Jain2024}, we show that for a pinned ethanol droplet evaporating on a heated substrate, the flow inside the droplet, the temperature distribution, and the particle dispersion are significantly affected by the presence of thermal Marangoni stresses. Generally, when a droplet evaporates, the CCR mode dominates the evaporation process until the receding contact angle is reached~\citep{Hu2002}. As a result, the CCR evaporation mode has been extensively studied. However, it is possible to ``choose'' or engineer the solid surface properties such that droplets evaporate in CCA mode to obtain a deposit pattern other than a coffee ring~\citep{Wang2013b}.
\citet{Dash2013} experimentally show that droplets evaporating in CCA mode on a super-hydrophobic substrate require a longer time to evaporate as predicted by a classical diffusion model. The reason for this observation is a reduced evaporation rate caused by evaporative cooling at the droplet interface, leading to a longer thermal resistance path, and the low effective conductivity of the substrate owing to the relatively small solid-liquid contact area.
For a droplet evaporating in CCA mode or with a moving contact line, along with the capillary and Marangoni flows, an additional inward flow away from the contact line is introduced  because of the receding motion of the contact line~\citep{Berteloot2008}.
These three types of flow inside an evaporating droplet with moving contact line play a dominant role in the dispersion and deposition of particles inside the droplet.

% \remove{With the complex interplay of different flows inside the depinned evaporating droplet, the possibility of a Marangoni instability exists, which is common in volatile sessile droplets~\citep{Zhu2019,Zhu2019a}. \citet{Zhu2019} experimentally showed that for a highly volatile n-hexane sessile droplet evaporating in the CCA mode, there exists a quasi-steady axisymmetric spoke-like pattern with several rolls that appear and merge in random positions as the contact line recedes. In another study, \citet{Zhu2021a} numerically predicted the occurrence of hydrothermal waves in sessile droplets evaporating in CCA mode. The results show that fan-like hydrothermal waves occur at the center of the droplet with a moderate contact angle, and a higher substrate temperature or larger initial radius promote the occurrence of irregular hydrothermal waves. In contrast, a sufficiently small initial contact radius inhibits the appearance of instabilities.}
% \aman{This part has been added to already account for Marangoni instability questions that would arise during the review. It can be removed and later added if and when asked by the reviewers.}

\citet{Nguyen2013} describe the experimental observation that the evaporation of a water droplet containing silica particles and with a moving contact line on a smooth hydrophobic surface results in an inner coffee ring deposit, containing different dendrite deposit patterns with a radius smaller than the initial contact radius.
Similarly, \citet{Das2017a} experimentally show that coating a glass substrate with silicone causes water droplets to evaporate in the CCA mode 75\% of the time, which suppresses the coffee-ring deposition of silica particles and promotes ordered crystal growth at the center of the droplet. They also suggest that a low evaporation rate of the droplet favors the formation of more ordered structures in the final aggregate.
\citet{Yang2020d} show that the hydrophobicity of the substrate can be altered by coating silicon wafers with alkysilane to experimentally study the colloidal deposition of silica particles from a water droplet evaporating in the depinned mode. They show that the particle concentration near the contact line required for the pinning of the contact line depends linearly on the receding contact angle of the surface, irrespective of the substrate material and initial particle concentration. Furthermore, they report 
that the required particle concentration decreases with increasing particle size and that the large particles of bidisperse silica suspensions dominate the self-pinning process. Another experimental study by \citet{Gupta2023} of water droplets with polystyrene particles on a hydrophobic PDMS surface, show that for a monodisperse particle population, the initial pinning stage is higher for smaller particles than for larger particles, and that the depinning of droplets containing bidisperse particles is delayed compared to monodisperse particles. They also investigated the effect of substrate heating and showed that, as the substrate temperature increases, the initial pinning stage is extended due to the combined action of thermal Marangoni and outward-driven radial evaporative flows at elevated temperatures, and the number of particles near the contact line increases, thereby promoting self-pinning.

Often, physical processes involving particle deposition from sessile droplets contain particles that vary in size~\citep{Zolotarev2022}, and the particles may be segregated or classified during the dispersion and deposition inside the droplet. \citet{Chhasatia2011} experimentally describe the deposition behavior of a mixture of micro- and nano-particles in a water droplet onto a glass substrate with different wettabilities. Their experiments show that the CCA mode dominates the evaporation process on a hydrophobic substrate, such that the particles do not deposit until the later stages of the droplet evaporation, at which point only little carrier liquid is left to facilitate particle classification.
However, increasing the surface wettability improves particle sorting, particularly near the contact line when the droplet evaporates in the CCR mode.

Several numerical studies have been conducted with the aim of understanding the behavior of evaporating sessile droplets with a moving contact line~\citep{Paul2023, Shang2024, Zhu2021, Bhardwaj2018, Erdem2024}.
% \citet{Shang2024} developed a finite-volume based method incorporating the volume-of-fluid method to perform direct numerical simulations of droplet evaporation, demonstrating that for elevated substrate temperatures, \unclear{the contribution of the Stefan flow and natural convection to the evaporation increases} \aman{Correction: Stefan flow or natural convection} while remaining almost independent of the environmental humidity.
\citet{Paul2023} applied the finite-element method to study the transients of the Stefan and Marangoni advection during the evaporation of both pinned and depinned sessile droplets.
They report that for droplets evaporating in depinned mode, the temperature and velocity fields at the intermediate stages remained qualitatively similar to those of the initial conditions, and in the later stages, evaporative cooling dominates, due to the lower conductivity from the smaller solid-liquid contact area and enhanced evaporative mass flux from liquid-vapor interface, which is inversely proportional to the droplet contact radius.
The prevailing challenges in modelling the flow inside evaporating sessile droplets are highlighted in the studies of \citet{Petsi2008} and \citet{Bhardwaj2018}. Assuming the flow inside an evaporating sessile droplet is accurately described as a two-dimensional Stokes flow, the results reported in \citet{Petsi2008} suggest that for contact angles 
less than 90$^\circ$ the flow is directed towards the center of the droplet, whereas for contact angles larger than 90$^\circ$ the flow is directed towards the contact line of the droplet.
% \citet{Petsi2008} applied Stokes flow inside a 2D droplet evaporating in a depinned state without Marangoni stress, \unclear{suggesting that for contact angles less than 90$^\circ$ the flow is directed towards the center of the droplet, whereas for contact angles greater than 90$^\circ$ the flow is directed towards the contact line of the droplet.
In \citet{Bhardwaj2018}, an expression for the evaporation mass rate for a hydrophobic surface using a scaling analysis to estimate the direction and magnitude of the characteristic evaporation-driven flow velocity inside the droplet is proposed, suggesting that for a contact angle less than 90$^\circ$ the flow is directed towards the contact line of the droplet, whereas for contact angles greater than 90$^\circ$ the flow is directed towards the center of the droplet.
These studies mainly focus on the evaporation flux and the flow inside the droplet, which eventually affects the dispersion of the particles in the droplet.

In this study, we investigate the behavior of bidisperse particles, i.e., a particle mixture with two distinct particle sizes, in sessile droplets with a moving contact line as they evaporate. To this end, we use a finite-volume method to model the gas-liquid fluid flow, in conjunction with Lagrangian particle tracking to model the particle behavior, based on our previous work~\citep{Jain2024}. 
Considering three different contact angles of the droplet and two different particle sizes, our study focuses primarily on the influence of the thermocapillary Marangoni stresses and the contact angle on the fluid flow inside an ethanol droplet evaporating on a heated substrate, as well as the resulting dispersion and size-based sorting of particles inside the droplet and at the gas-liquid interface.

The remainder of this paper is organized as follows. In Section \ref{Method}, we discuss the volume of fluid (VOF) and the discrete element model (DEM) frameworks which are used in this study. In Section \ref{CaseSetup}, we detail the configuration of various simulation cases and the different physical parameters governing the problem are given. In Section \ref{validation}, we validate the employed numerical framework and chosen mesh resolution for the evaporating sessile droplets with a moving contact line by comparing the numerical results to analytical solutions and experimental measurements. Subsequently, we investigate the flow and temperature distribution inside the droplet (Section \ref{Flow}) as well as the particle dispersion and self-sorting (Section \ref{Particles}), considering or neglecting the Marangoni stresses, and varying the relative density of the particles compared to the density of the fluid. Section \ref{conclusions} presents the conclusions of this study.

\section{Methodology}
\label{Method}
In this study, we consider bidisperse particle populations varying in size that are simulated in sessile droplets evaporating with a moving contact line on a heated substrate. To facilitate this, our existing model for pinned sessile droplets with monodisperse particles~\citep{Jain2024} is modified to account for a moving contact line and bidisperse particles.

For an evaporating ethanol droplet with a constant contact angle of $\theta_c \in \{60^{\circ}, 90^{\circ}, 120^{\circ}\}$ and an initial radius of $R_c = 0.5$~mm, the Bond number varies from $\text{Bo} = 0.049 - 0.1475 \ll 1$, suggesting that the droplet maintains a spherical shape throughout the simulation.
Similar to our previous work~\citep{Jain2024}, it can be shown that the Péclet number for vapor transport in the gaseous phase is less than 1; thus, we can assume that vapor transport in the gas phase is dominated by diffusion. However, the Péclet number for heat transport in the liquid phase due to Marangoni convection is larger than 10, suggesting that heat transport due to convection inside the liquid cannot be neglected.
The total volume fraction of the particles inside the droplet is less than 1 \% and we, thus, assume that the particles have a negligible influence on the fluid flow. This allows us to reduce the computational cost by applying one-way coupling, where the fluid influences the motion of the particles, but the particles do not influence the fluid flow.

\subsection{Gas-liquid flow model}
\label{FVM}

To simulate the gas-liquid system of an evaporating sessile droplet and its surrounding fluid, a second-order finite-volume method is employed ~\citep{Denner2014a}. This model comprises three key elements: the equations that govern the two-phase flow (Section \ref{flow_equations}), the volume of fluid (VOF) method~\citep{Denner2014a} to capture the behavior of the gas-liquid interface (Section \ref{VOF}), and a model for the heat and mass transfer (Section \ref{EvapModel}).

\subsubsection{Fluid flow equations}
\label{flow_equations}
The gas-liquid flow in and around the droplet is modelled using the one-fluid formulation of the incompressible continuity and momentum equations, given as

\begin{equation}
    \nabla \cdot \bm{u} = -\dot{m} \left( \frac{1}{\rho_\mathrm{l}} - \frac{1}{\rho_\mathrm{v}}\right),
\label{eq:continuityeqn}
\end{equation}

 \begin{equation}
     \rho \left( \frac{\partial \bm{u}}{\partial t} +
     \nabla \cdot (\bm{u} \otimes \bm{u}) \right)
     = -\nabla{p} + \nabla\cdot(\mu(\nabla{\bm{u}} + \nabla{\bm{u}^T})) + \bm{f}_{\mathrm{s}},
     \label{eq:momentum}
 \end{equation}
where $\boldsymbol{u}$ is the fluid velocity vector, $\dot{m}$ is the fluid mass flux due to evaporation, $\rho$ is the fluid density, 
% with $\rho_\text{l}$ and $\rho_\text{v}$ indicating specifically the density of the liquid and the vapor, respectively,
$p$ is the pressure, $\mu$ is the fluid viscosity, and $\bm{f}_\mathrm{s}$ is the volumetric  force representing the surface tension.
Due to the low Bond number of the considered droplet, gravity is neglected when solving the fluid flow.
The volumetric force representing surface tension in the presence of temperature variations is given as~\citep{Brackbill1992,Kothe1996}
\begin{equation}
     \bm{f}_{\mathrm{s}}  = \sigma \kappa \nabla{\alpha} + \nabla_\mathrm{s}{\sigma}|{\nabla{\alpha}}|,
\end{equation}
where $\alpha$ is the liquid volume fraction (see Section \ref{VOF}), $\kappa$ is the interface curvature, and $\nabla_\mathrm{s}{\sigma}$ is the gradient of the surface tension coefficient tangential to the interface.
The surface tension coefficient $\sigma$ is assumed to vary linearly with the temperature and is defined as $\sigma = \sigma_0 + \sigma_{\mathrm{T}}(T-T_0)$, such that
 \begin{equation}
     \nabla_\mathrm{s}{\sigma} = \sigma_{\mathrm{T}}[\nabla{T} - (\nabla{T}\cdot\bm{n})\bm{n}],
 \end{equation}
 where $T$ is the temperature, $\sigma_0$ is the surface tension coefficient at the reference temperature $T_0$, $\sigma_{\mathrm{T}}$ is the temperature coefficient of the surface tension, and $\bm{n}$ is the normal vector of the interface.

\subsubsection{Gas-liquid interface}
\label{VOF}

The gas-liquid interface is described using the VOF method, where the liquid volume fraction $\alpha$ is defined as
\begin{equation}
    \alpha =  \begin{cases}
        1  & \text{in the liquid phase,}\\
        0  & \text{in the gas phase},
    \end{cases}
\end{equation}
with $ 0<\alpha<1$ signifying the presence of the gas-liquid interface.
The value of $\alpha$ is used to define the local fluid properties, such as the density, viscosity, specific heat capacity, thermal diffusivity, and the mass diffusion coefficient.
In our work, the volume fraction evolves only as a result of the evaporative mass flux, such that
\begin{equation}
    \frac{\partial \alpha}{\partial t} = - \frac{\dot m}{\rho_\mathrm{l}}.
    \label{eq:interfaceAdvection}
\end{equation}
It should be noted that in this equation the advection term is neglected, since considering the evaporative flux for the interface advection directly eliminates the need for any special treatment of the Stefan flux~\citep{Jain2024}.
To maintain the spherical shape of the droplet, regular re-initialization is carried out.

\subsubsection{Heat and mass transfer}
\label{EvapModel}
The evaporation model combines solving for the vapor concentration and the thermal energy of the fluid.
Based on the compressive continuous species transfer model (C-CST)~\citep{Haroun2010a,Marschall2012,Deising2016, Zanutto2022}, the temperature $T$  and vapor mass concentration $c$ are calculated using a one-field energy equation,
\begin{align}
    \rho C_{\mathrm{p}} \left( \frac{\partial T}{\partial t} +\bm{u}\cdot\nabla{T}\right) &= \nabla \cdot \left({k\nabla{T}}\right) - \dot{m} L,\\
    \frac{\partial c}{\partial t} & = \nabla \cdot (D \nabla c + \bm{\phi}),
    \label{eq:SpeciesEq}
\end{align}
where $k$ represents the thermal conductivity, \(C_{\mathrm{p}}\) the specific heat capacity, \(L\) is the latent heat of vaporization, and $D$ represents the effective diffusion coefficient of the vapor.
% and $\bm{\phi}$ represents the \unclear{solvent transfer across phases owing to concentration gradient.}\aman{The term $\bm{\phi}$ is an additional flux, called CST flux, which results from the concentration jump at the fluid/fluid interface.}
The term $\bm{\phi}$ represents the flux that results from the concentration jump at the fluid/fluid interface.
Neglecting the advection term in Eq.~\eqref{eq:SpeciesEq} is justified because the Péclet number of the vapor transport is less than 1, suggesting that diffusion is the dominant transport mechanism of the vapor~\citep{Jain2024}.
The calculation of the effective diffusion coefficients $D$ and $\bm{\phi}$ has been detailed in our previous work~\citep{Zanutto2022, Jain2024}.
The calculation of the mass flux due to evaporation, \(\dot{m}\), is approximated by Fick's law, as explained by~\citet{Zanutto2022} and~\citet{Maes2020}, where \(\dot m\) is defined as
\begin{align}
    \dot m & = \frac{(D\nabla c - \bm{\phi})}{1-\alpha} \cdot \nabla \alpha ,
    \label{eq:mdot}
\end{align}

\subsection{Discrete element method}
\label{DEM}

Newton's second law of motion is applied to describe the particle motion and the discrete element method (DEM) is employed to simulate interactions between particles and between particles and the substrate~\citep{Cundall1979}. The translational and rotational motion of each particle is determined by
\begin{align}
    m_\mathrm{p}\frac{\mathrm{d}\bm{v_\mathrm{p}}}{\mathrm{d}t} & = m_\mathrm{p}\left(1-\frac{\rho_\mathrm{f}}{\rho_\mathrm{p}}\right)\bm{g} + \bm{f}_\mathrm{c} + \bm{f}_\mathrm{f} + \bm{f}_\mathrm{{adh}} +\bm{f}_\mathrm{{cap}}+\bm{f}_\mathrm{{pp-cap}} ,\label{eq:particleAdvection}
\end{align}
and
\begin{align}
    \bar{\bar{\bm{I}}} \bm{\omega}_\mathrm{p} & = \bm{T}_\mathrm{c} ,\label{eq:particleRotation}
\end{align}
 where \(m_\mathrm{p}\) is the mass of the particle, \(\bm{v}_\mathrm{p}\) is the velocity of the particle, and \(\bm{\omega}_\mathrm{p}\) is the rotational velocity of the particle. The first term on the right-hand side of Eq.~\eqref{eq:particleAdvection} represents the forces associated with gravitational acceleration \(\bm{g}\), where \(\rho_\mathrm{p}\) and \(\rho_\mathrm{f}\) are the densities of the particles and fluid, respectively. \(\bm{f}_\mathrm{c}\) represents the forces acting on the particle arising from particle-particle and particle-substrate interactions, \(\bm{f}_\mathrm{f}\) is the drag force resulting from particle-fluid interactions, \(\bm{f}_\mathrm{{adh}}\) accounts for the \suggestion{adhesive} van der Waals forces between multiple particles, \(\bm{f}_\mathrm{{cap}}\) represents the capillary force on a particle at the gas-liquid interface, and \(\bm{f}_\mathrm{{pp-cap}}\) describes the force between two particles at the gas-liquid interface. Equation \eqref{eq:particleRotation} defines the rotational acceleration of the particle, where \(\bar{\bar{\bm{I}}}\) is the moment of inertia and \(\bm{T}_\mathrm{c}\) is the sum of external torques acting on the particle due to particle-particle and particle-substrate interactions.

\subsubsection{Forces acting on the particles}
\label{sec:particleforces}

The interactions between particles and between particles and the substrate generate forces that can be modelled using a nonlinear spring-dashpot-slider system. This system is based on the physical characteristics of the particles and the degree of deformation that occurs during their interactions. The total force exerted during a collision is calculated by combining the normal and tangential components, 
 \begin{equation}
     \bm{f}_\mathrm{c} = \bm{f}_\mathrm{{c,n}}+\bm{f}_\mathrm{{c,t}},
 \end{equation}
 which are determined using the physical properties of the particles and the extent of their deformation.
 For a collision with an overlap of \(\delta_\mathrm{n}\), a relative particle velocity \(\bm{v}_\mathrm{pp}\), and a normal contact vector \(\bm{n}\), the normal force
 is given by~\citep{Tsuji1992}
\begin{align}
\ \bm{f}_\mathrm{c,n} & = \left(K_\mathrm{n}\delta_\mathrm{n}^{\frac{3}{2}} - \eta_\mathrm{n} \bm{v}_\mathrm{pp}\cdot \bm{n} \right) \bm{n},\\
\ K_\mathrm{n} & = \frac{4R^{*}}{3E^{*}},\\
\ \eta_\mathrm{n} & = \alpha_\mathrm{d}(mK_\mathrm{n})^{\frac{1}{2}}\delta_\mathrm{n}^{\frac{1}{4}},
\label{eq:Normal force}
\end{align}
where \(R^{*}\),  \(E^{*}\) and $m$ are the reduced radius, the Young's modulus, and the mass of the particles, respectively.
Following \citet{Tsuji1992}, the damping coefficient \(\alpha_\mathrm{d}\) is empirically related to the coefficient of restitution of the particle material.

The tangential contact forces resulting from particle-particle and particle-substrate interactions are determined as~\citep{Mindlin1953}
\begin{equation}
\ \bm{f}_\mathrm{c,t} = \begin{cases}
- K_\mathrm{t}\bm{\delta_\mathrm{t}} - \eta\cdot \bm{v_\mathrm{s}} &  \text{ if $|\bm{f_\mathrm{c,t}}| < \mu_\mathrm{f}|\bm{f_\mathrm{n}}|$,}\\
\mu_\mathrm{f} |\bm{f_\mathrm{n}}|\bm{t}   &  \text{otherwise, }
\end{cases}
\end{equation}
with the tangential spring constant
\begin{equation}
\ K_\mathrm{t}  = 8 \sqrt{R^{*}\delta_\mathrm{n}}G^{*},
\label{eq:Tangential force}
\end{equation}
where \(\bm{\delta_\mathrm{t}}\) denotes the cumulative tangential displacement projected onto the plane perpendicular to the collision normal, \(\bm{v}_\mathrm{s}\) denotes the slip velocity between the two particle surfaces at the point of contact, \(\mu_\mathrm{f}\) denotes the friction coefficient, \(\bm{t}\) denotes the tangential contact vector (normal to the normal contact vector), and \(G^{*}\) denotes the reduced shear modulus. The tangential spring constant can be directly related to the material properties~\citep{Mindlin1953}, where the shear modulus of a particle is calculated from the Young's modulus and the Poisson ratio.

The dominant force acting on a particle from the fluid phase is the drag force, \({\bm{f}_\mathrm{d}}\), which depends on the relative velocity between the particle and the surrounding fluid. In this work, the drag force is modelled using the expression by~\citet{Wen1966}, where the drag force is expressed as
\begin{equation}
    {\bm{f}_\mathrm{d}} = \beta \frac{V_\mathrm{p}}{(1-\epsilon_\mathrm{f})}(\bm{v}_\mathrm{f}-\bm{v}_\mathrm{p}),
\end{equation}
 where \(V_\mathrm{p}\) is the volume of particle with diameter \(d_\mathrm{p}\), \(\epsilon_\mathrm{f}\) is the local fluid volume fraction, \(\bm{v}_\mathrm{f}\) is the velocity of the fluid at the particle, and \(\bm{v}_\mathrm{p}\) is the velocity of the particle.
 \(\beta\) is the momentum transfer coefficient, given as~\citep{Wen1966}
 \begin{equation}
     \beta = C_\mathrm{d} \frac{3\epsilon_\mathrm{f}(1-\epsilon_\mathrm{f})\rho_\mathrm{f}|\bm{v}_\mathrm{f}-\bm{v}_\mathrm{p}|}{4d_\mathrm{p}}\epsilon_\mathrm{f}^{-2.65}.
 \end{equation}
The drag coefficient is defined as
\begin{equation}
C_\mathrm{d} = \begin{cases}
 \dfrac{24}{ \epsilon_\mathrm{f}\mathrm{Re}}\left(1-0.15(\epsilon_\mathrm{f}\mathrm{Re})^{0.687}\right) &  \text{if $\epsilon_\mathrm{f}\mathrm{Re} < 1000$,}\\
0.44  &  \text{if $\epsilon_\mathrm{f}\mathrm{Re} > 1000$},
\end{cases}
\end{equation}
based on the particle Reynolds number
\begin{equation}
    \mathrm{Re} = \frac{\rho_\mathrm{f} d_\mathrm{p} |\bm{v}_\mathrm{f} - \bm{v}_\mathrm{p}|}{\mu_\mathrm{f}}.
\end{equation}

The adhesion force due to van der Waals effects between particles as well as between the particles and the substrate is described using the Dejarguin-Muller-Toporov (DMT) model~\citep{Derjaguin1975}, with the adhesion force given as
\begin{equation}
    {\bm{f}_\mathrm{adh}} = -2\pi R^{*} \Delta\gamma \bm{{n}},
\end{equation}
where \(\Delta\gamma\) is the work of adhesion or the energy per unit area required to pull two infinite planar surfaces apart, and \(\bm{{n}}\) is the normal contact vector.
The DMT model offers a straightforward framework to be applied in DEM simulations~\citep{Wilson2016}.

The capillary interaction between particles and the interface exerts a force on the particles, as a result of which the particles adsorb at the interface. The capillary force on the particle due to the interface is given as~\citep{Lebedev-Stepanov2013,Jain2024} 
\begin{equation}
    \bm{f}_\mathrm{{cap}} =\begin{cases}
        2\pi r_\mathrm{l} \,  f \, {\bm{n}} & \text{if  $r_\mathrm{l}<R$}, \\
        \bm{0} & \text{otherwise},
    \end{cases}
\end{equation}
where $r_\mathrm{l}$ is the distance between the particle center and the interface, $R$ is the radius of the particle, and $f$ is the surface free energy at the liquid-particle interface.
The deformation of the surrounding interface causes particles situated on it to generate an inter-particle force, a phenomenon referred to as the ``Cheerios effect''~\citep{Vella2005}.
In this study, the force is modelled as a simplified version of the expression proposed by~\citet{Vassileva2005}, as~\citep{Li2011}
\begin{equation}
    {\bm{f}_\mathrm{{pp-cap}}} = - \pi \sigma \left( \frac{R^{*2}}{2} \, \sin^2(\beta_0) \, \tan^2(\alpha_\mathrm{c} - \beta_0) \, \frac{\bm{n}}{d}\right),
\end{equation}
where $d$ is the distance between the two particle centers, and $\beta_0$ and $\alpha_c$ are the central cone angle of the contact point and contact angle between the fluid and the particle, respectively~\citep{Vassileva2005}.

\subsubsection{Integration of the particle trajectories}

Once the forces on each particle are determined, the position vector for every particle is updated at each time interval using the Verlet algorithm~\citep{Allen1989}, 
\begin{align}
 \bm{x}_{\mathrm{p}}(t + \Delta t ) & = 2 \bm{x}_{\mathrm{p}}(t) -\bm{x}_{\mathrm{p}}(t - \Delta t) + \Delta t^2 \frac{\bm{f}_{\mathrm{p}}(t)}{m_{\mathrm{p}}} + \mathcal{O}(\Delta t^4),
\label{eq:final x vector}
\end{align}
where \(\bm{f}_{\mathrm{p}}(t)\) is the sum of all forces acting on the particle and \(\Delta t\) is the applied numerical time step.
The velocity of the particle is computed as
\begin{align}
 \bm{v}_{\mathrm{p}}(t) & = \frac{ \bm{x}_{\mathrm{p}}(t + \Delta t) -  \bm{x}_{\mathrm{p}}(t - \Delta t)}{2 \Delta t} + \mathcal{O}(\Delta t^2).
\label{eq:final v vector}
\end{align}

\subsection{Numerical procedure}
\label{algorithm}
In this study, a two-stage simulation procedure, described in detail in our previous work~\citep{Jain2024}, is used. 
In the first stage, the gas-liquid flow is simulated without accounting for the particles, while in the second stage, the behavior of the bidisperse particles is simulated and analyzed using the results from the gas-liquid flow simulations of the first stage.

In stage 1, the governing equations of the gas-liquid flow are solved using a finite-volume algorithm~\citep{Denner2014a, Denner2020}, combined with an algebraic VOF method~\citep{Denner2014e} to capture the gas-liquid interface. In stage 2, the flow data from stage 1 is used to simulate the particle behavior using a Lagrangian particle tracking method. The forces acting on the particles are computed based on the fluid flow and interface properties.

The fluid flow simulations in stage 1 are performed with a time step of $\Delta t = 5 \times 10^{-6}$~s, with the results stored at intervals of $t_\mathrm{{out}} = 5 \times 10^{-4}$~s. In stage 2, the Lagrangian particle-tracking algorithm operates with a smaller time step of $\Delta t = 5 \times 10^{-8}$~s. The storage interval in stage 1 is chosen such that the interface movement due to evaporation during each time step is significantly smaller than the smallest diameter of the particles.

\section{Case setup}
\label{CaseSetup}
This study investigates the dispersion of bidisperse particles in evaporating sessile droplets with a moving contact line, focusing on the roles of surface wettability, Marangoni stresses, and the relative density of the particles compared to the fluid.
Ethanol is selected as the solvent for the sessile droplet in all the simulations due to its higher volatility compared to water, which results in a higher evaporation rate~\citep{Gurrala2019}.
In the first simulation stage, the contact angle is varied and the effect of considering the Marangoni stresses is investigated. In the second stage, also the relative density of the particles compared with the density of the fluid is varied.

\subsection{Considered cases}

The first stage involves six simulations of depinned ethanol sessile droplet evaporation, with initial contact angles of \(\theta = \{60\,^\circ, 90\,^\circ, 120\,^\circ \}\).
For each contact angle, two simulations are performed: in one simulation the Marangoni stresses are considered, while in the other simulation the Marangoni stresses are neglected.  
These six cases are labeled as C1, C2, C3, C4, C5 and C6. Cases C1 and C2 involve a droplet with a contact angle \(\theta = 60\,^\circ\), C3 and C4 involve a droplet with a contact angle \(\theta = 90\, ^\circ\), and C5 and C6 involve a droplet with a contact angle \(\theta = 120\, ^\circ\).
Marangoni stresses are considered in the even-numbered cases and are neglected in the odd-numbered cases.
The droplet is placed in air at an ambient pressure of $101325 \, \text{Pa}$ and an ambient temperature of $25 \, ^\circ$C, with the fluid properties detailed in Table \ref{tab:fluidProperties}, and the substrate temperature is kept fixed at \(T_\mathrm{s} = 50\, ^\circ\)C.

In the second stage, bidisperse particles with radii \(R_\mathrm{p} = \{1.5~\mu \mathrm{m}, 2.5~\mu \mathrm{m}\}\) are randomly distributed within the droplet.
Particles of both sizes are added until the volume fraction associated with each particle size is 0.1 \%. The properties of the silica particles simulated in this study are given in Table \ref{tab:particleproperties}.
Using the data from a number of cases from stage 1, simulations with particles are carried out to examine
the influence of the contact angle and the Marangoni stresses on the particle dispersion inside the droplet.
The two-stage model also enables the examination of the impact of the particle density using the same fluid flow results. In this context, we also evaluate neutrally buoyant particles ($\rho_\text{p}=\rho_\text{f}$), maintaining all properties listed in Table \ref{tab:particleproperties}, except for the particle density.
For the second-stage simulations, case names from the corresponding first-stage simulations are extended with a suffix: standard silica particles are identified by the suffix S, while neutrally buoyant particles are designated with the suffix N.
The case names for both the stages are summarized in Table \ref{tab:caseNames}.

% \unclear{\bf Why limit the second stage to these particular cases?}
% \aman{Due to the constraint of computational expense, the second stage cases were limited to the one presented.}

\begin{table}[t]
	\caption{Properties of the liquid-gas system (ethanol-air) used for all the simulations in stage 1.}
    \centering
		\begin{tabular}{|l|l|l|}
		\hline
		Parameters & Ethanol & Air \\ \hline \hline
        Density (kg/\si{\metre\cubed}) & 750 & 1.23 \\
        Viscosity (Pa s) &  $0.65\times 10^{-3}$ & $1.78\times 10^{-5}$ \\
        Thermal conductivity (W/\si{\metre} K) & 0.165 &  0.046  \\
         Specific heat capacity (J/kg K) & 2750 & 1000 \\
          Latent heat of vaporisation (J/kg) &  $0.9 \times 10^6$ & - \\
          Molar mass (kg/mol) & 0.029 & 0.046 \\
          Boiling temperature (K) & 351 & - \\
           Surface tension coefficient (N/m) & 0.022 & - \\
           Temperature coefficient of surface tension (N/m K) & $-1.2\times10^{-4}$ & - \\
		\hline
	\end{tabular}
		\label{tab:fluidProperties}
\end{table}

\begin{table}[t]
	\centering
	\caption{Properties of the particles used for all the simulations in stage 2.}
	\begin{tabular}{|*{2}{l|}}%{\textwidth}{|l|c|}
		\hline
		Parameters &Value\\ \hline \hline
        Density (kg/\si{\metre\cubed})         &2650  \\ \hline
		Young's modulus (MPa)& 1.0 \\ \hline
        Poisson's ratio & 0.15 \\ \hline
        Friction coefficient & 0.3 \\ \hline
        Coefficient of restitution & 0.8	\\ \hline
        Volume fraction of particles & 0.5 \\ \hline
        Number of particles & 607 \\ \hline
	\end{tabular}
		\label{tab:particleproperties}
\end{table}

\begin{table}[t]
	\centering
	\caption{The simulation cases carried out in this work, where {C1} - {C6} have variation in substrate temperature and in taking into account the Marangoni stresses, or neglecting them. The heavier silica particles are denoted with suffix S and the neutrally buoyant particles are denoted with suffix N.}
\begin{tabular}{ |c|c||c|c| }
\hline
\multicolumn{2}{|c||} {Stage 1} & \multicolumn{2}{c|} {Stage 2}  \\
\hline
Cases & Properties & Cases & Properties \\
\hline
\hline
\multirow{2}{4em}{C1} & $\theta = 60\, ^\circ$ & C1S & Silica particles \\
& No Marangoni stresses & C1N & Neutrally buoyant particles \\
\hline

\multirow{2}{4em}{C2} & $\theta = 60\, ^\circ$ & C2S & Silica particles \\
& Marangoni stresses & C2N & Neutrally buoyant particles \\
\hline

\multirow{2}{4em}{C3} & $\theta = 90\, ^\circ$ & C3S & Silica particles \\
& No Marangoni stresses &  &  \\
\hline

\multirow{2}{4em}{C4} & $\theta = 90\, ^\circ$ & C4S & Silica particles \\
& Marangoni stresses &  &  \\
\hline

\multirow{2}{4em}{C5} & $\theta = 120\, ^\circ$ &  &  \\
& No Marangoni stresses &  &  \\
\hline

\multirow{2}{4em}{C6} & $\theta = 120\, ^\circ$ &  &  \\
& Marangoni stresses & &  \\
\hline

\end{tabular}
\label{tab:caseNames}
\end{table}

\subsection{Boundary and initial conditions}
\label{BCandIC}
Given the axisymmetric nature of the evaporation dynamics of a sessile droplet on a heated substrate, only a quarter of the droplet is simulated to create a three-dimensional domain for the particle simulations in stage 2.
In stage 1, the lower boundary in the $z$-direction, i.e., \(z=0\), represents the substrate at a fixed temperature, \(T_\mathrm{s}\), a no-slip condition (\(\bm{u} = 0\)), and no penetration of the vapor mass, that is, \(\bm{n}_\mathrm{z}\cdot\nabla{c} = 0\). A Neumann-type boundary condition is applied to the liquid volume fraction to facilitate a moving contact line.
Applying the no-slip boundary condition for a moving contact line may cause a singularity in the stress and a logarithmic singularity in the energy dissipation rate at the contact line. This was first pointed out by \citet{Huh1971}, and is usually referred to as the Huh-Scriven paradox. However, most interface-capturing techniques, such as the VOF method used in this work, utilize the velocity normal to the mesh faces to advect the volume fraction field, which implies that the methodology includes an ``implicit'' slip length at the no-slip boundary~\citep{Afkhami2009a}.
To further analyze the effect of different slip lengths, a study of fluid flow and temperature distribution evolution is presented in Appendix B, which suggests that with the given model, the no-slip boundary condition is an adequate choice. This choice of boundary condition is also consistent with other studies simulating sessile droplets with moving contact lines~\citep{Paul2023, Shang2024, Zhu2021a, Zhu2021}.
A symmetry boundary condition is used in the lower $x$- and $y$-directions, i.e., \(x=0 \text{ and } y=0\).

Earlier studies opted for a large gas-phase domain~\citep{Hu2002, Chen2017c}, typically ranging from 20 to 50 times the droplet radius, using Dirichlet boundary conditions for both temperature and vapor concentration to ensure that the vapor distribution surrounding the droplet was not influenced by the far-field boundary conditions.
To reduce the size of the computational domain without adversely impacting the simulation results, the assumption of a diffusion-dominated distribution of the vapor concentration and temperature in the gas away from the interface (\(\Delta{c} = 0\) and \(\Delta{T} = 0\)) is employed~\citep{Diddens2017b}. For the vapor concentration and the temperature, boundary conditions at the upper limits of the domain ($x,y,z = L$, where $L$ represents the domain size) are established by assuming a constant gradient of values perpendicular to the boundary. The vapor concentration field is obtained by solving the Laplace equation. This approach enables the reduction of the domain size to four times the droplet contact radius ($L = 4 R_c$), as confirmed in Section \ref{validation}.

\begin{figure}
      \centering
     \begin{subfigure}[b]{0.48\textwidth}
               \includegraphics[clip,trim={7cm 2cm 7cm 2cm}, width=\textwidth]{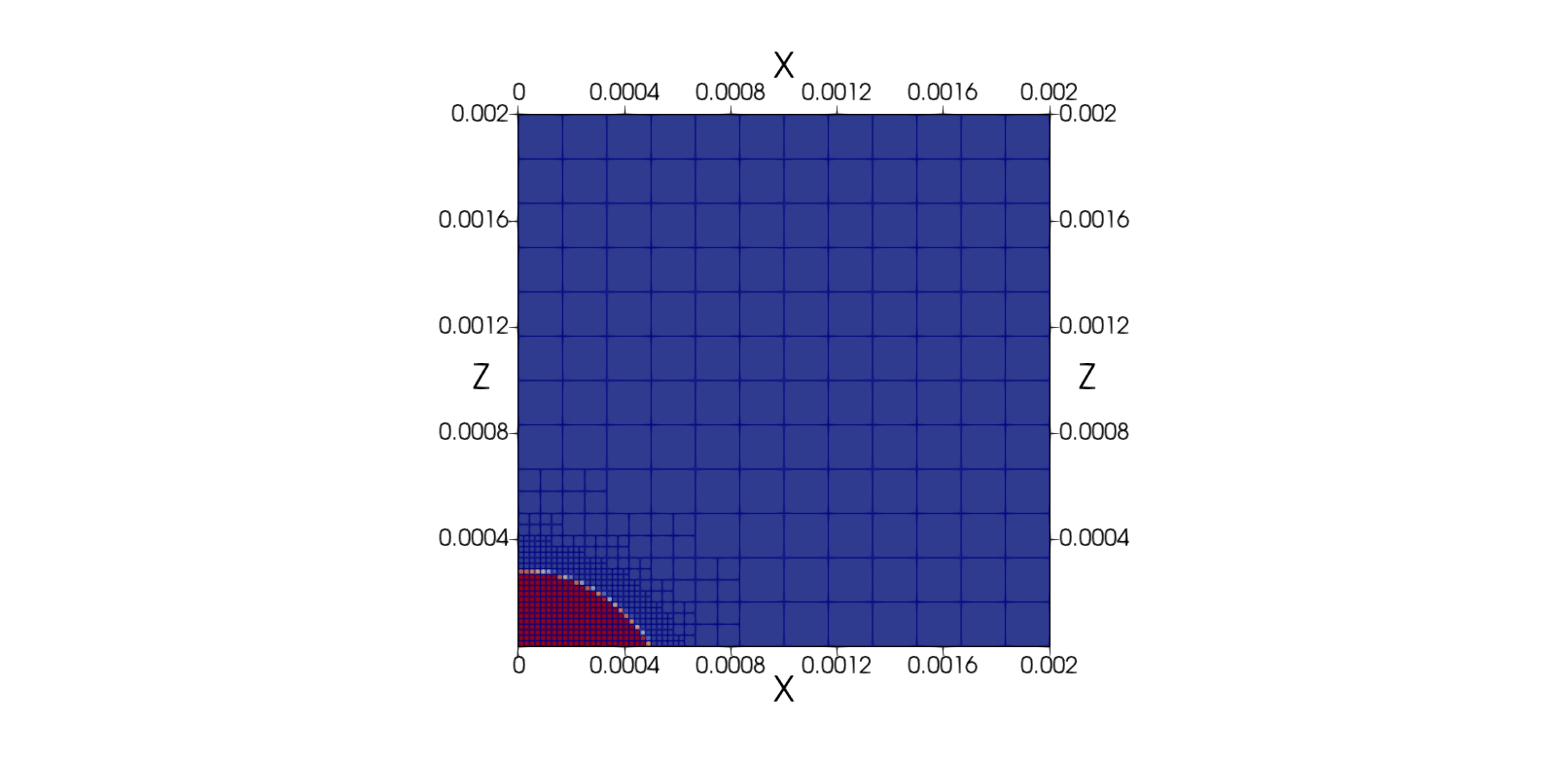}
                           \caption{}
       \label{fig:Mesh1}
       \end{subfigure}
            \begin{subfigure}[b]{0.48\textwidth}
               \includegraphics[clip,trim={7cm 2cm 7cm 2cm}, width=\textwidth]{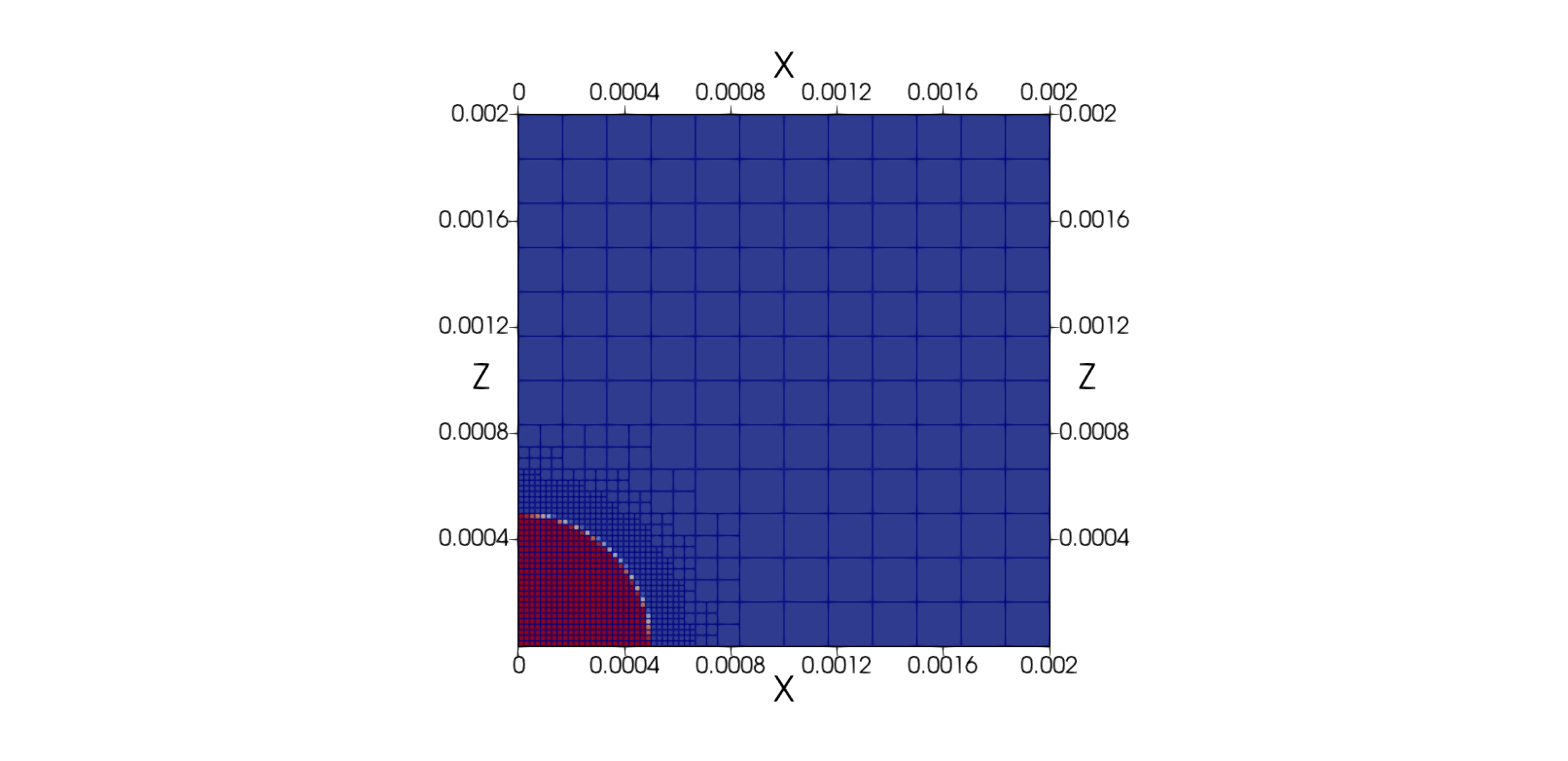}
                           \caption{}
       \label{fig:Mesh2}
       \end{subfigure}
            \begin{subfigure}[b]{0.48\textwidth}
               \includegraphics[clip,trim={7cm 2cm 7cm 2cm}, width=\textwidth]{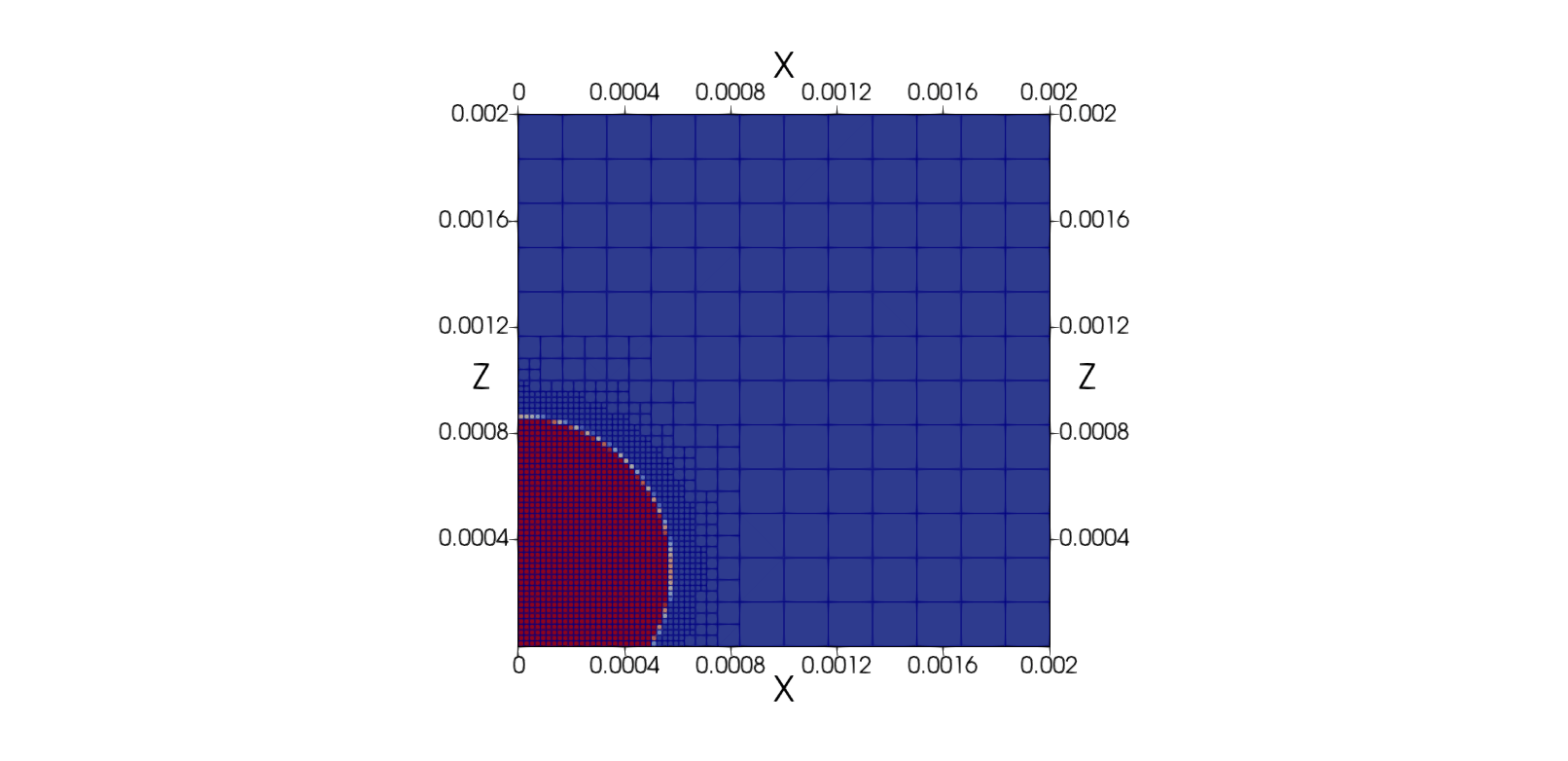}
                           \caption{}
       \label{fig:Mesh3}
       \end{subfigure}
            \begin{subfigure}[b]{0.48\textwidth}
               \includegraphics[clip,trim={0cm 0 0cm 0}, width=0.95\textwidth]{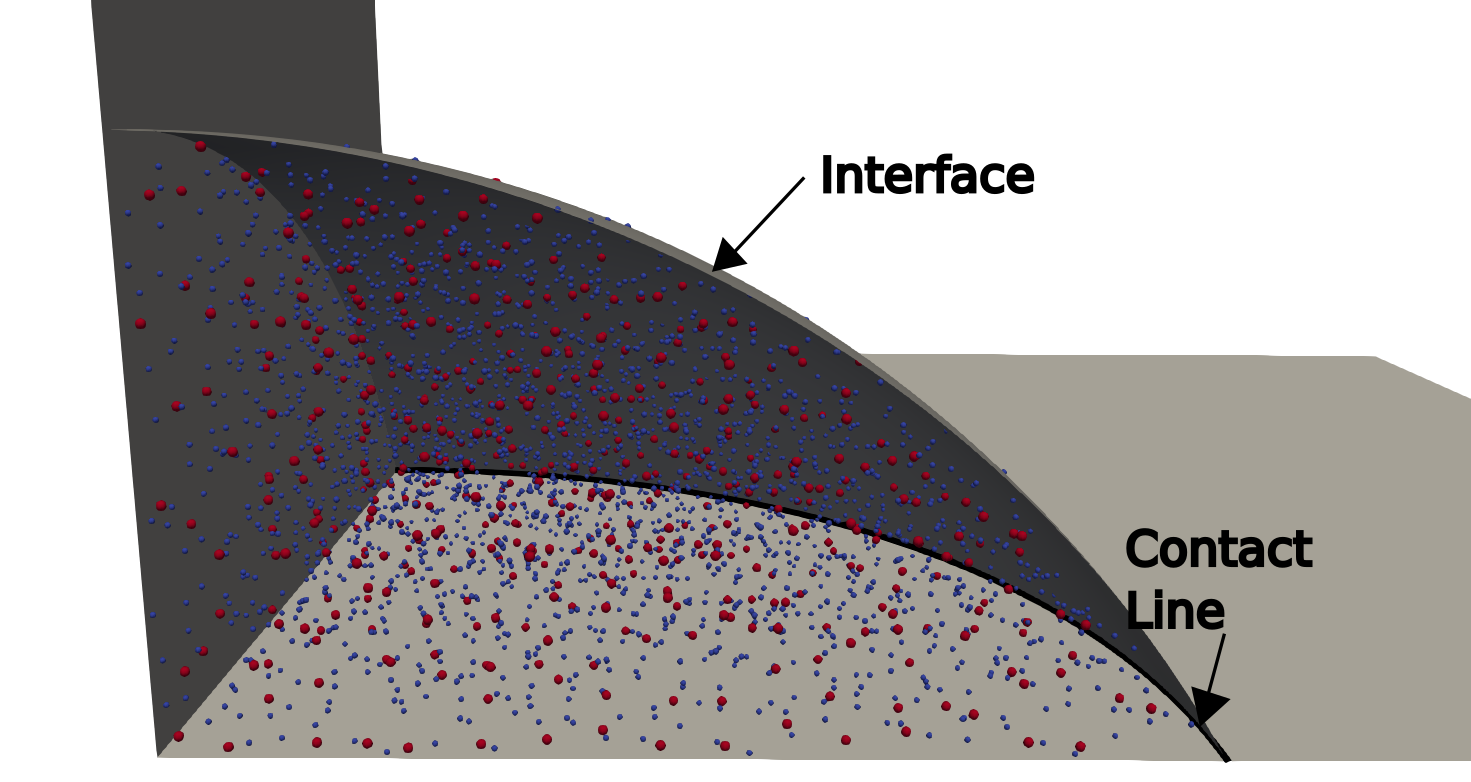}
                           \caption{}
       \label{fig:PartInitialization}
       \end{subfigure}
       \caption{A cross-section of the simulation geometry, showing the mesh refinement near the droplet and the initialisation of the color function for stage 1, \textbf{(a)} \(\theta_c = 60\,^\circ\)  \textbf{(b)} \(\theta_c = 90\,^\circ\)  \textbf{(c)} \(\theta_c = 120\,^\circ\). \textbf{(d)} The initialization of bidisperse particles inside the droplet in stage 2. The particle size depicted in the figure is significantly enlarged for visualization purposes.}
\end{figure}

The computational cost is further reduced by using an adaptively refined mesh in and around the droplet to simulate the flow in stage 1, as shown in Figure \ref{fig:Mesh1}-\ref{fig:Mesh3}.
For the initialization of the droplet, the liquid volume fraction and vapor concentration are initialized as a spherical cap. Inside the droplet, the liquid volume fraction is 1 and the vapor mass concentration is the liquid density~\citep{Zanutto2022}. The initial temperature and velocity are set to \(25\, ^\circ\mathrm{C} \text{ and } 0\), respectively, throughout the domain and the pressure is set to atmospheric pressure.
In stage 2, bidisperse particles are randomly placed inside the droplet as shown in Figure \ref{fig:PartInitialization}.

\section{Validation}
\label{validation}

\begin{figure}
      \centering
		\includegraphics[width=0.45\textwidth]{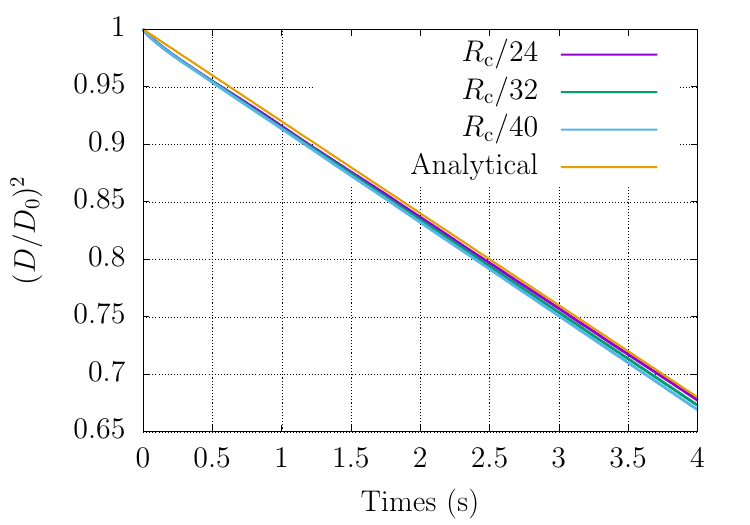}
			    \caption{Validation and mesh sensitivity study for an evaporating droplet with moving contact line,  comparing the numerical results of the evaporation of an isothermal sessile droplet with three mesh resolutions against the analytical solution given in Eq.~\eqref{eq:isothermal-analytical}.}
       \label{fig:CCAIsothermal}
\end{figure}

Various aspects of the employed model have been extensively validated in our previous study~\citep{Jain2024}. We validate the model for the evaporation of a sessile droplet with a moving contact line using the analytical solution of the volume evolution during the evaporation of an isothermal sessile droplet with a contact angle of 90$\,^\circ$~\citep{Erbil2012}.
Figure \ref{fig:CCAIsothermal} shows a comparison between the numerical results of the volume evolution of an isothermal sessile droplet with a moving contactline and with an initial contact angle of \(\theta = 90\,^\circ\) and the analytical solution,
\begin{equation}
D_i^2 - D^2 = \frac{8D(c_s -c_{\infty})}{\rho_l}\, t, \label{eq:isothermal-analytical}
\end{equation}
where $c_s$ is the vapor concentration at the sphere surface, $c_{\infty}$ is the vapor concentration at infinity, $D_i$ is the initial diameter of the spherical drop, and $D$ is the instantaneous drop diameter.
The sensitivity of the results with respect to the mesh resolution is assessed using three different mesh resolutions, where the initial contact radius of the droplet is resolved by 24, 32 and 40 cells.
% Simulations are performed until the droplet volume reaches approximately 65 \% of the initial value \unclear{\bf Is this true? Because below we show figures of droplets with $25 \ \%$ their initial volume.}.
% \aman{For the above figure 2, where the validation and mesh sensitivity has been shown, the simulations for all three resolutions are performed till the droplet volume reaches approximately 55\% (correction: not 65\%). For all the cases studied below the simulations were performed till 20\% volume of the initial droplet is left}.
As depicted in Figure \ref{fig:CCAIsothermal}, the results from all the numerical resolutions agree well with the analytical solution, demonstrating that the results from stage 1 are independent of the mesh resolution.

\begin{figure}
    \centering
		\includegraphics[width=0.45\textwidth]{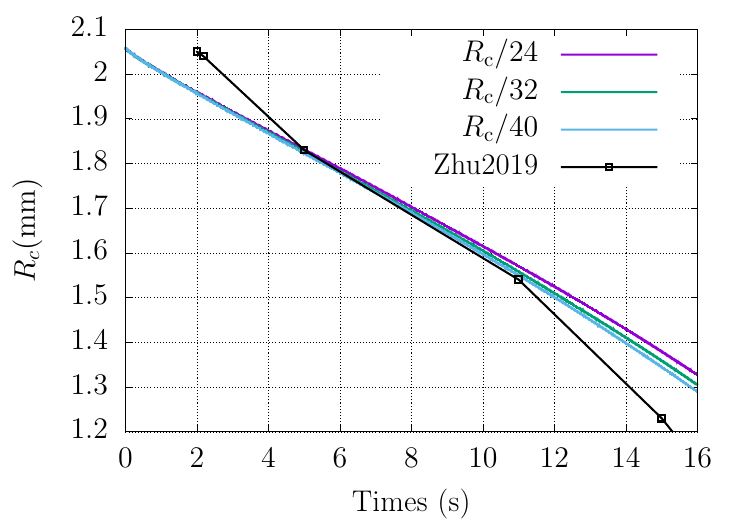}
			    \caption{Validation and mesh sensitivity study for the n-hexane evaporating droplet, comparing the numerical results of the contact radius evolution with three mesh resolutions with the experimental solution given by \citet{Zhu2019}.}
       \label{fig:ModelVerification}
\end{figure}

Further validation of the model is carried out with the experimental results of \citet{Zhu2019}, wherein an n-hexane droplet evaporating in CCA mode with contact angle \(\theta_c = 27.47\,^\circ \) and initial contact radius \(R_c = 2.1\) mm on a heated substrate with \(T_s = 35.33^{\, \circ}\)C is considered. Figure~\ref{fig:ModelVerification} shows the evolution of the contact radius predicted by the simulations at three different resolutions, compared against the experimental measurements of \citet{Zhu2019}. The simulation results are in good agreement with the experimental results.
Thus, for subsequent analyses, the lower mesh resolution, where the contact radius is resolved by 24 cells, is used. To validate the mesh independence of the fluid flow in the presence of Marangoni stresses, case C4 is simulated with different mesh resolution, where the contact angle is \(\theta_c = 90^\circ\) and Marangoni flow is present. The results, presented in Appendix A, demonstrate that the flow is independent of the mesh resolution even in the presence of Marangoni stresses.
% Various Types of particle force implementations has been validated in our previous work \citet{Jain2024}.

\section{Results of stage 1: Flow inside the droplet}
\label{Flow}

\begin{figure}
      \centering
     \begin{subfigure}[b]{0.48\textwidth}
		\includegraphics[width=\textwidth]{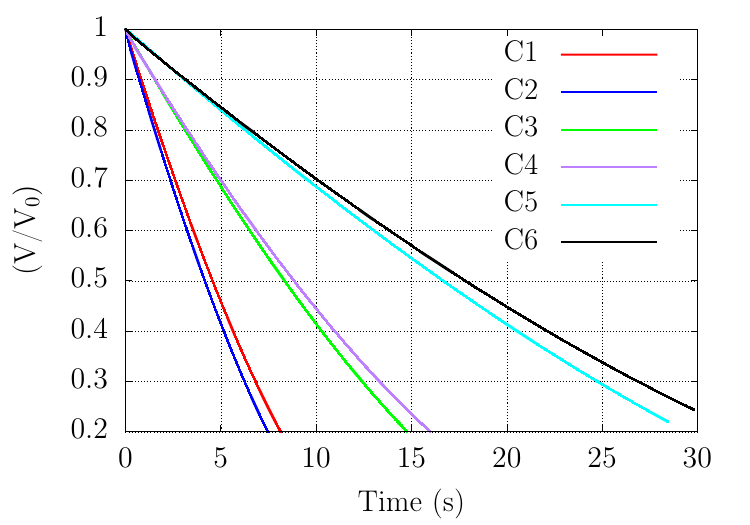}
			    \caption{}
       \label{fig:volumeEvolution}
       \end{subfigure}
       \begin{subfigure}[b]{0.48\textwidth}
		\includegraphics[width=\textwidth]{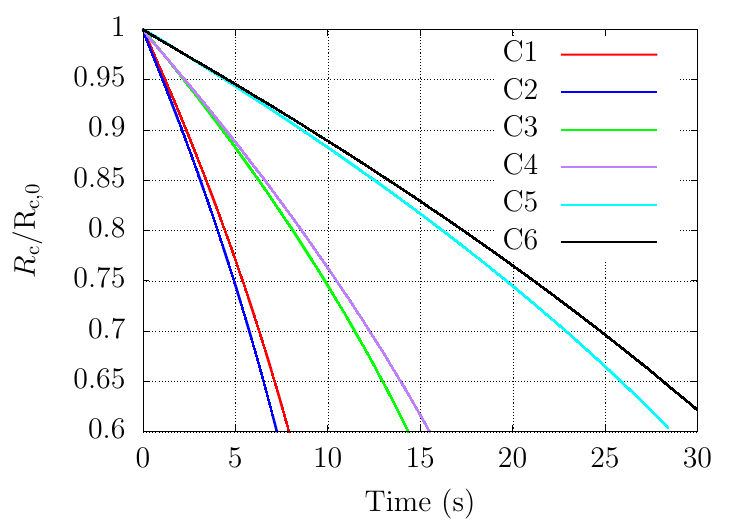}
			    \caption{}
       \label{fig:angleEvolution}
        \end{subfigure}
        \caption{The simulated normalized droplet volume \textbf{(a)} and the contact radius of the droplet \textbf{(b)} as a function of time for all 6 cases.}
\end{figure}

In the first stage of this study, the focus is on analyzing the evolution of the contact radius and the fluid flow inside the droplet, as well as the spatial distribution of the temperature and vapor concentration.
The temporal evolution of the droplet volume, normalized by the initial droplet volume, is shown in Figure \ref{fig:volumeEvolution}.
It is observed that increasing the contact angle results in a lower evaporation rate, resulting from the larger initial droplet volume.
The presence of Marangoni flow enhances the evaporation rate for cases C1 and C2 (\(\theta_c = 60^\circ\)), whereas it reduces the evaporation rate for cases C3-C6 (\(\theta_c = \{90^\circ, 120^\circ\}\)).
The evolution of the contact radius is shown in Figure \ref{fig:angleEvolution}, exhibiting an almost linear reduction and a trend similar to the volume of the droplet.

To analyze the fluid velocity and temperature fields inside the droplet, four specific time instances are discussed below, corresponding to times when 90\%, 75\%, 50\% and 25\% of the initial volume of the droplet is left, for three different contact angles.

\subsection{Contact angle of $\theta_c = 60\,^\circ$}

\begin{figure}
        \centering
       \begin{subfigure}{0.45\textwidth}
          \includegraphics[clip,trim = 0 4cm 0 0, width=\textwidth]{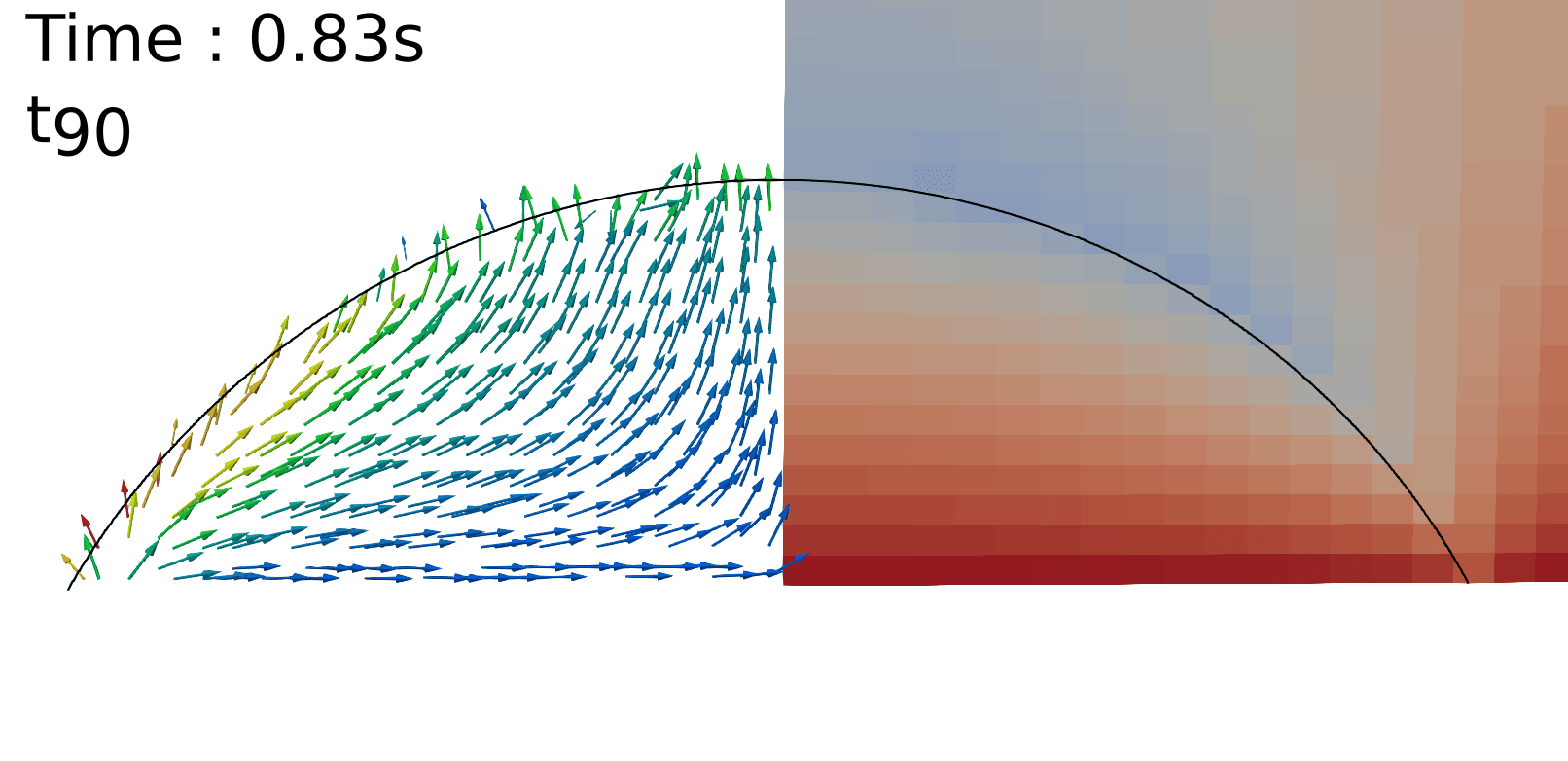}
                  \caption{}
         \label{fig:C1T1}
         \end{subfigure}
         \begin{subfigure}{0.45\textwidth}
          \includegraphics[clip,trim =0 4cm 0 0,width=\textwidth]{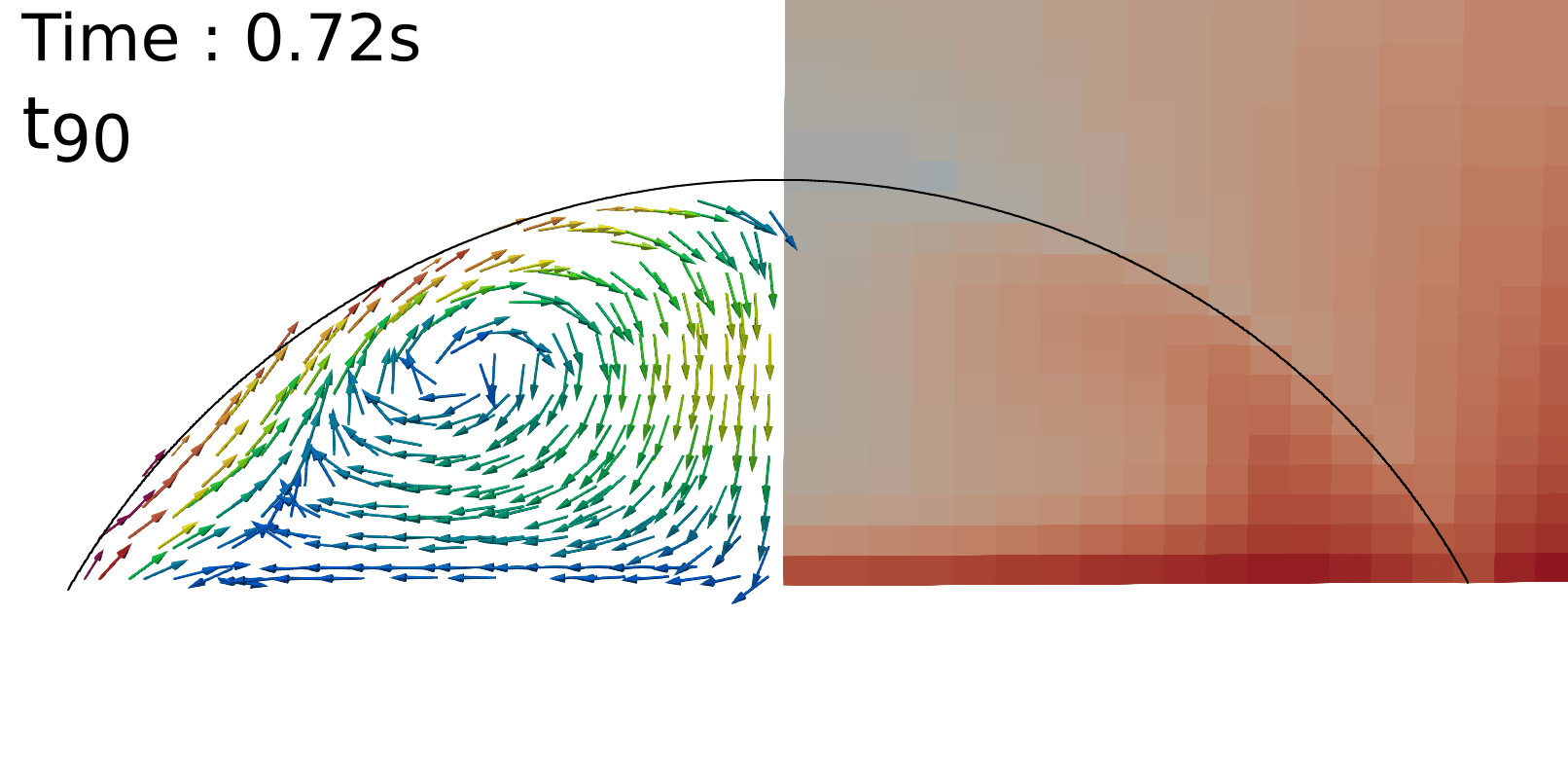}
                  \caption{}
         \label{fig:C2T1}
          \end{subfigure}
               \begin{subfigure}{0.45\textwidth}
          \includegraphics[clip,trim =0 4cm 0 0,width=\textwidth]{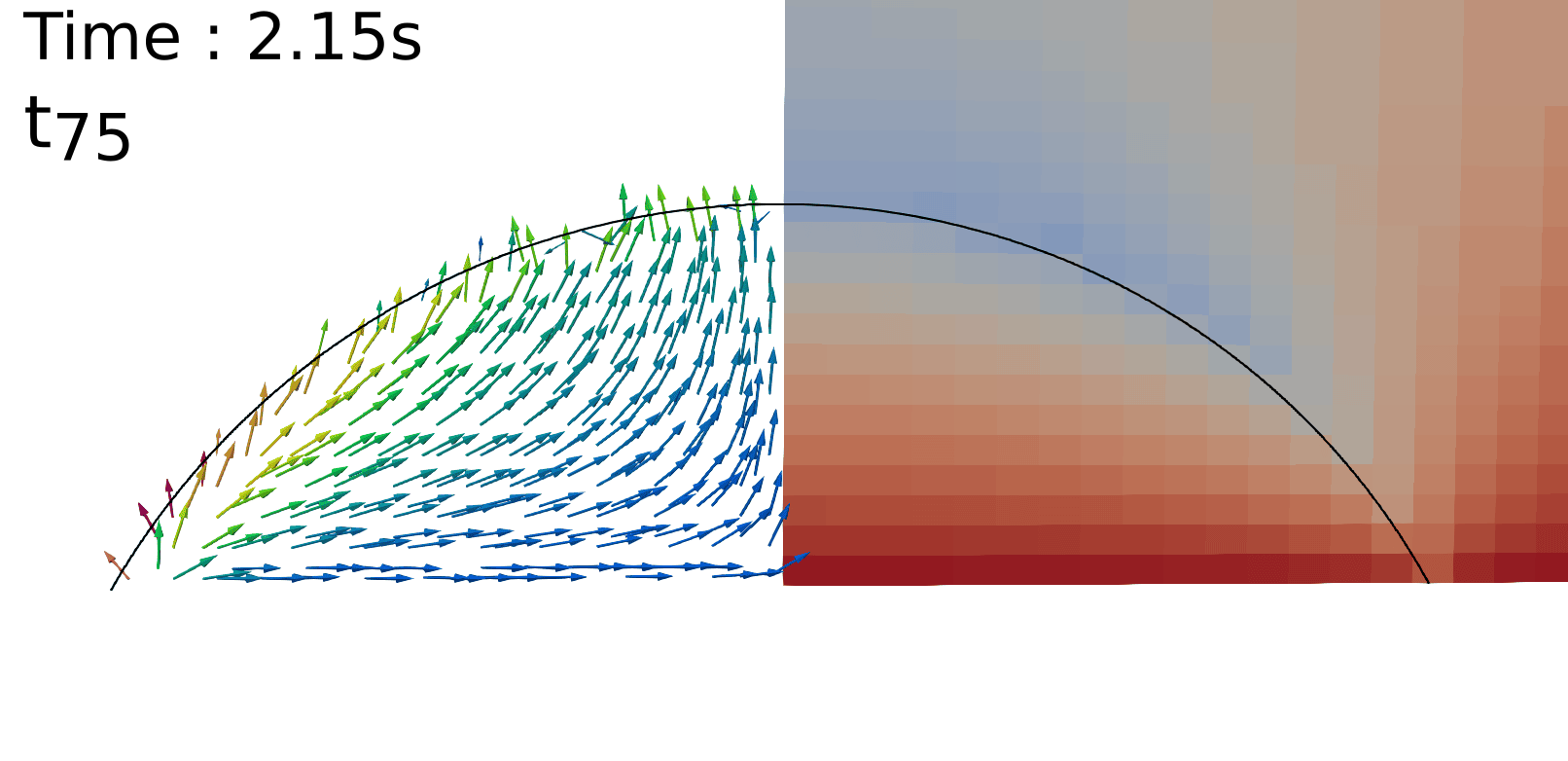}
                  \caption{}
         \label{fig:C1T2}
         \end{subfigure}
         \begin{subfigure}{0.45\textwidth}
          \includegraphics[clip,trim =0 4cm 0 0,width=\textwidth]{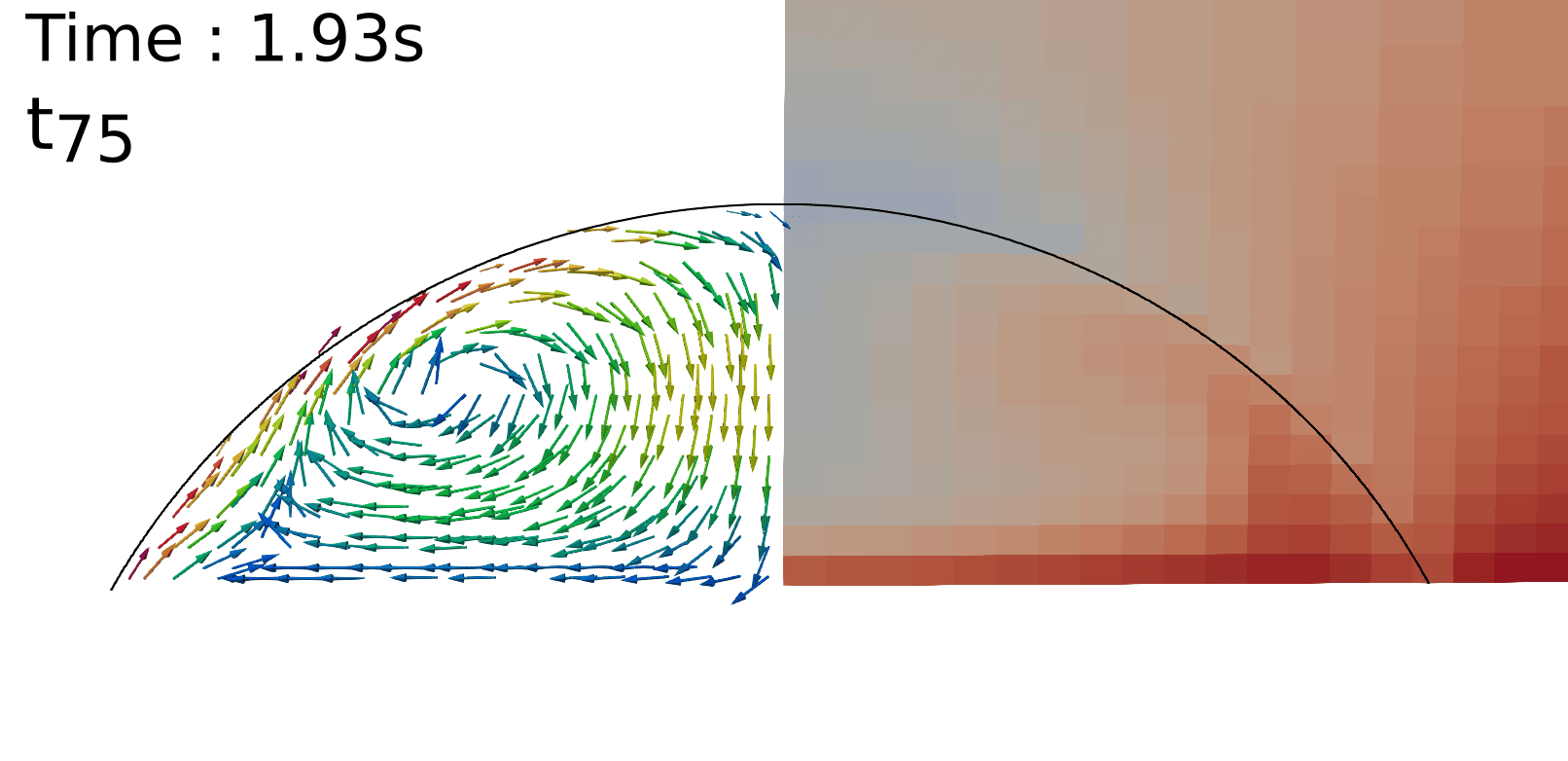}
                  \caption{}
         \label{fig:C2T2}
          \end{subfigure}
               \begin{subfigure}{0.45\textwidth}
          \includegraphics[clip,trim =0 4cm 0 0,width=\textwidth]{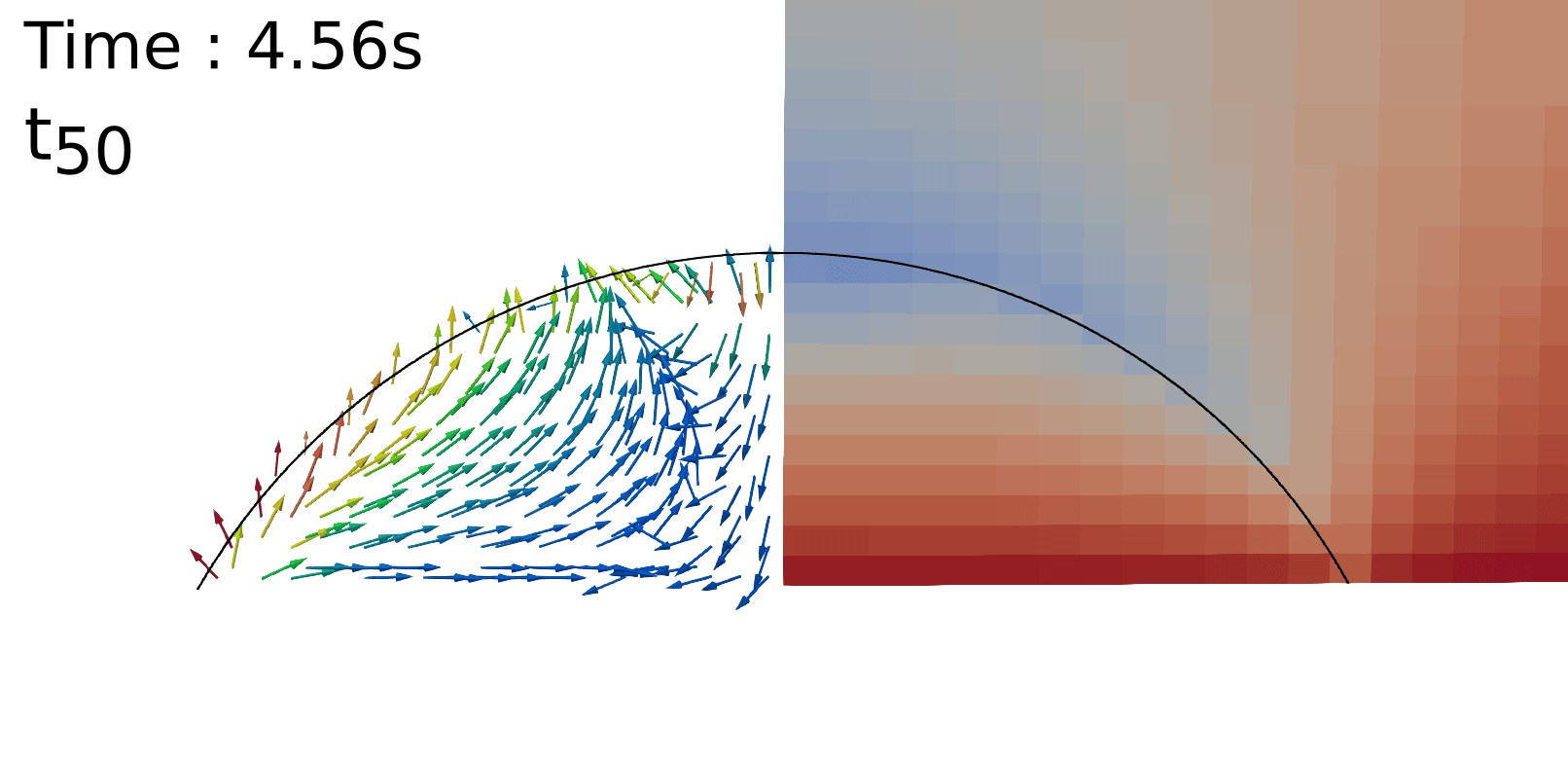}
                  \caption{}
         \label{fig:C1T3}
         \end{subfigure}
         \begin{subfigure}{0.45\textwidth}
          \includegraphics[clip,trim =0 4cm 0 0,width=\textwidth]{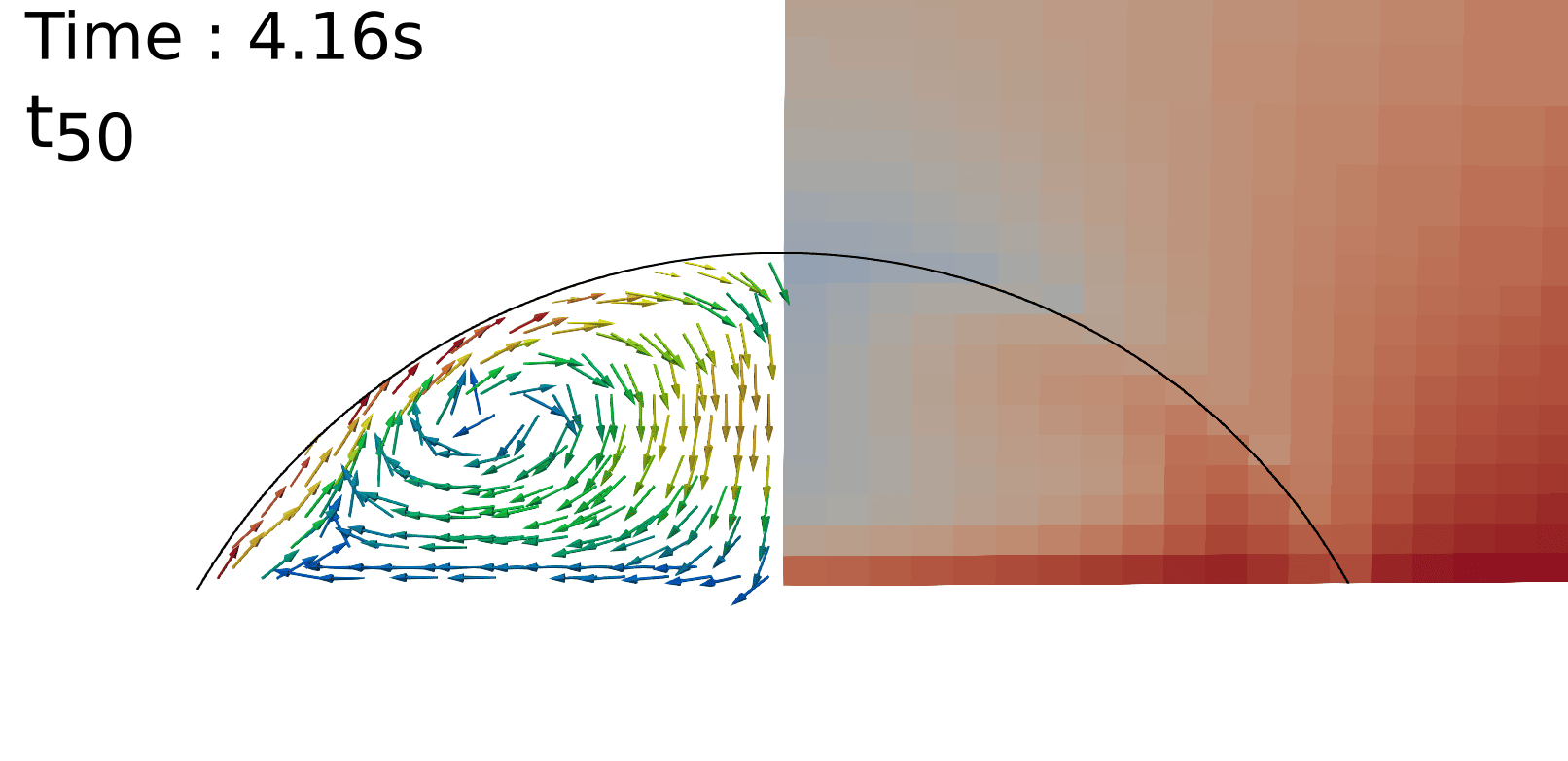}
                  \caption{}
         \label{fig:C2T3}
          \end{subfigure}
               \begin{subfigure}{0.45\textwidth}
          \includegraphics[clip,trim =0 4cm 0 0,width=\textwidth]{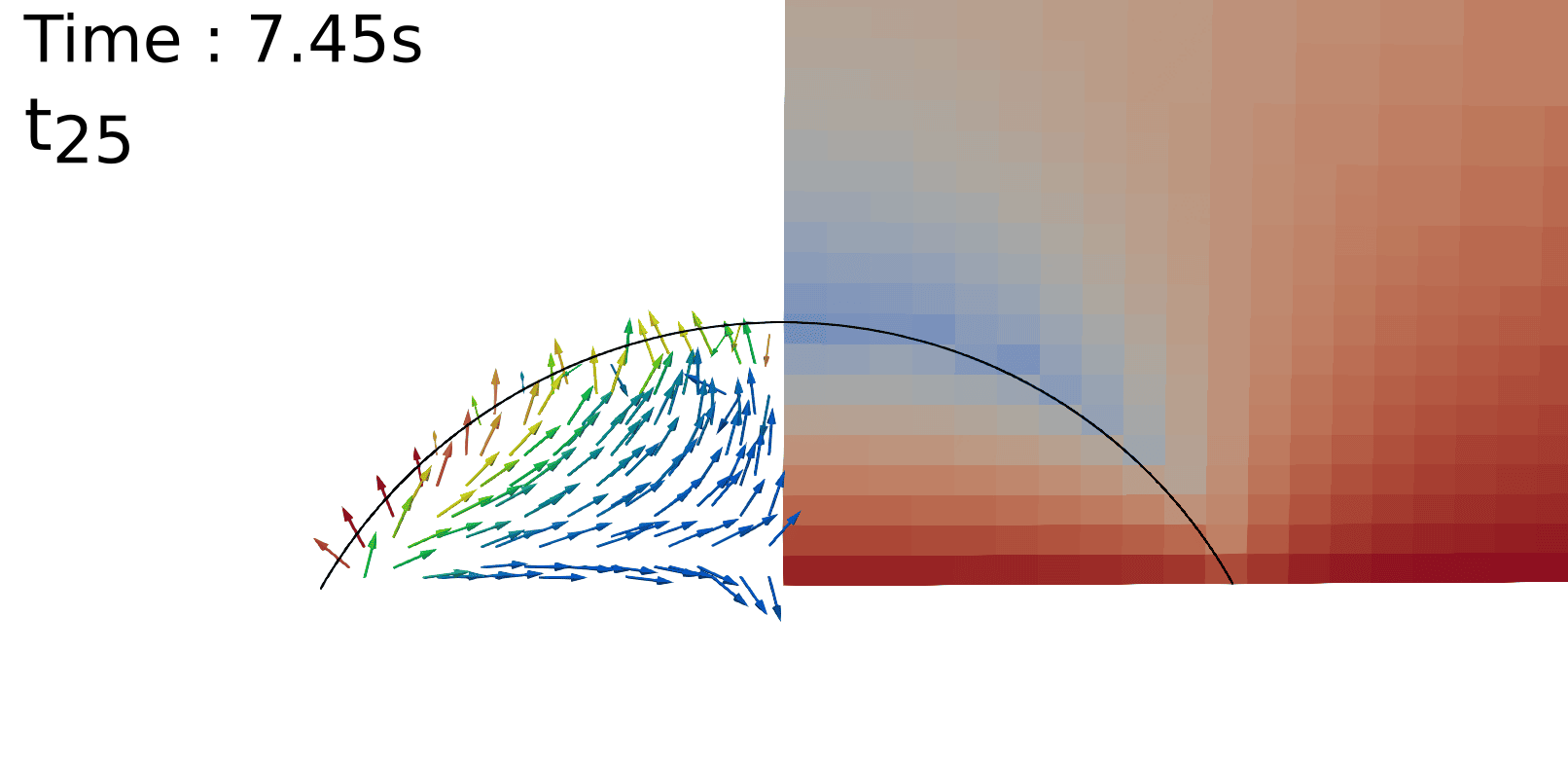}
                  \caption{}
         \label{fig:C1T4}
         \end{subfigure}
         \begin{subfigure}{0.45\textwidth}
          \includegraphics[clip,trim =0 4cm 0 0,width=\textwidth]{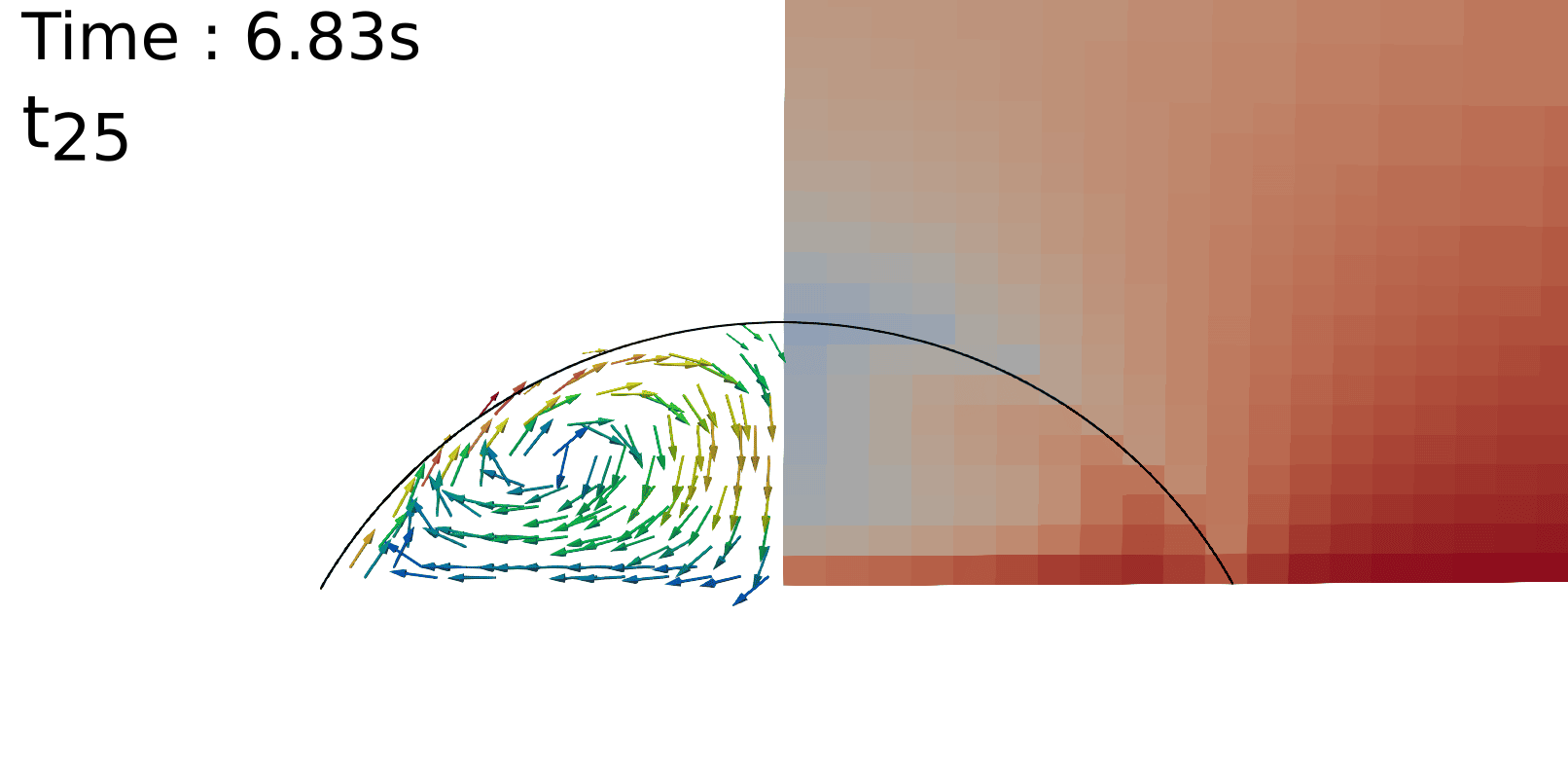}
                  \caption{}
         \label{fig:C2T4}
          \end{subfigure}
                       \begin{subfigure}{0.48\textwidth}
          \includegraphics[clip,trim =10cm 5cm 10cm 15cm,width=\textwidth]{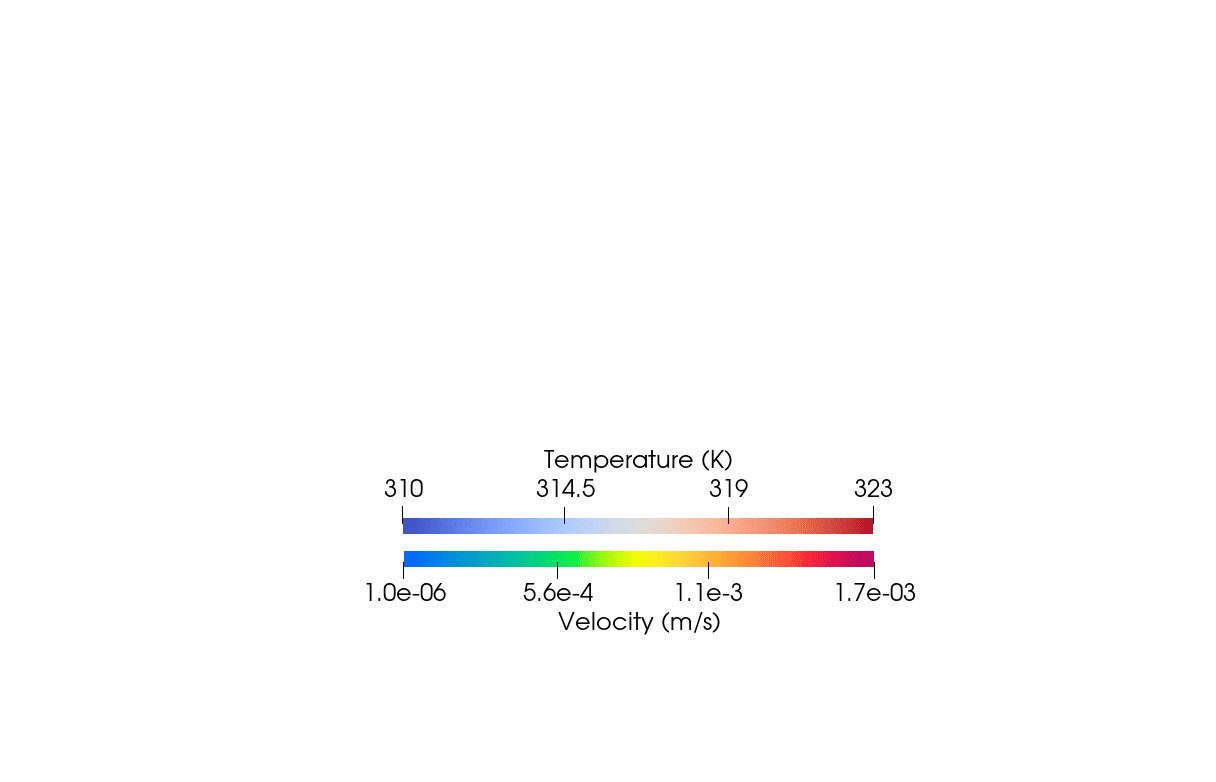}
         \end{subfigure}
         \begin{subfigure}{0.48\textwidth}
          \includegraphics[clip,trim =10cm 5cm 10cm 15cm,width=\textwidth]{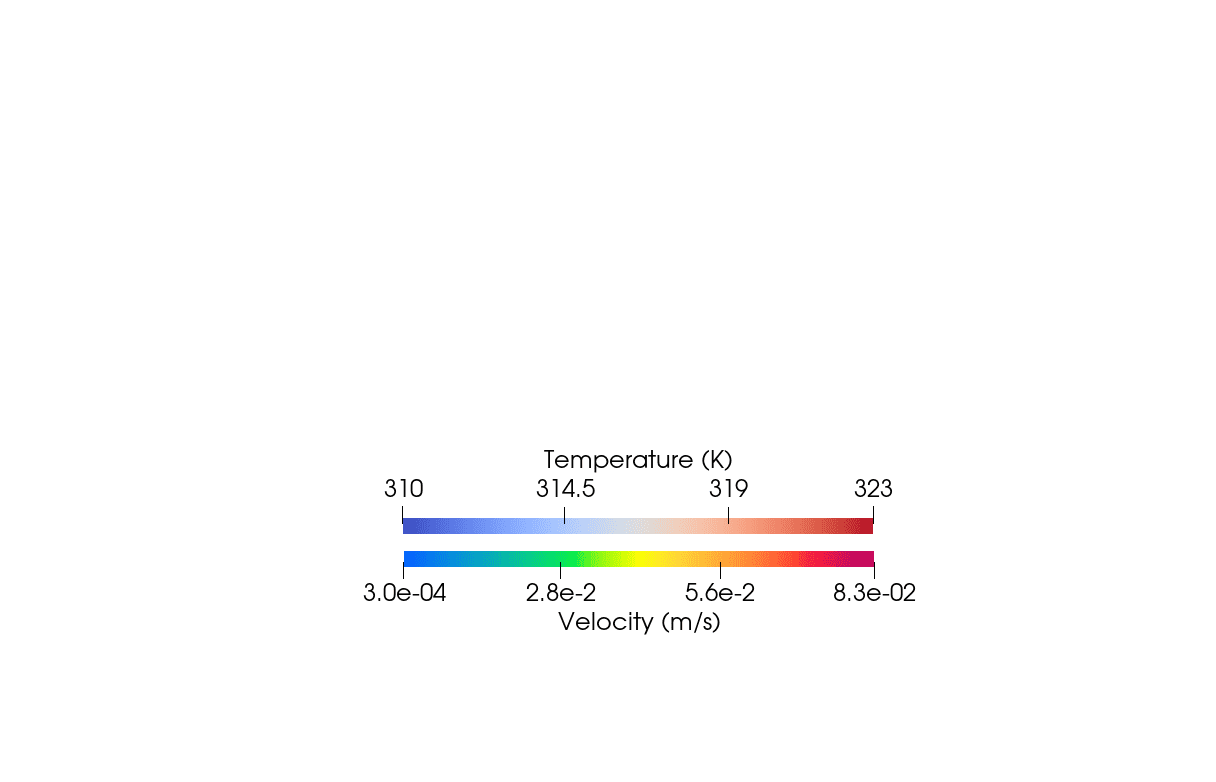}
          \end{subfigure}
                  \caption{Velocity vectors and temperature plots at four different time instances of the cases with a contact angle of $\theta_\mathrm{c}=60\, ^\circ$  without taking into account the Marangoni stresses, C1 (left column), and with Marangoni stresses, C2 (right column). Velocity vector plots are shown on the left, where the color indicates the velocity magnitude. Contour plots of the temperature are shown on the right. \textbf{(a)} C1, $t_{90}$, \textbf{(c)} C1, $t_{75}$, \textbf{(e)} C1, $t_{50}$, \textbf{(g)} C1, $t_{25}$.
                    \textbf{(b)} C2, $t_{90}$, \textbf{(d)} C2, $t_{75}$, \textbf{(f)} C2, $t_{50}$, \textbf{(h)} C2, $t_{25}$. }
                  \label{fig:fig4}
    \end{figure}
    
For case C1, where the contact angle is $\theta_c = 60\, ^\circ$ and the Marangoni stresses are neglected, the figures showing the velocity on the left and the temperature distribution inside the droplet on the right are presented for four different time instances in Figures \ref{fig:C1T1}, \ref{fig:C1T2}, \ref{fig:C1T3}, and \ref{fig:C1T4}.
A flow from the contact line to the apex of the droplet is observed, which is different from the well-known capillary flow pattern {\citep{Hu2005a, Barmi2014}} observed in the absence of Marangoni flow in pinned droplets, with fluid flowing from the apex of the droplet towards the contact line.
As evaporation continues and the contact radius decreases, the flow from the contact line to the apex remains consistent, with a higher velocity magnitude near the contact line and along the interface compared to the interior bulk of the droplet. The temperature distribution on the right shows that the temperature along the interface increases towards the contact line.

The velocity and temperature distributions inside the droplet for case C2, where the contact angle is $\theta_c = 60\,^\circ$ and the Marangoni
stresses are taken into account, are presented in Figures \ref{fig:C2T1}, \ref{fig:C2T2}, \ref{fig:C2T3}, and \ref{fig:C2T4}.
The flow velocity inside the droplet has a magnitude that is approximately ten times larger than that in case C1 (in which the Marangoni stresses neglected), driven by the Marangoni stresses. The Marangoni stresses move the liquid from the contact line to the apex of the droplet along the interface, forming a vortex inside the droplet.
As evaporation continues and the contact radius decreases, the influence of the Marangoni flow reduces as the temperature difference along the interface reduces.
A stagnation point is formed at the base of the droplet, which is slightly offset from the contact line. This is a result of the Marangoni flow pushing the liquid inside the droplet, whereas the opposing capillary flow pushes the liquid outward towards the contact line. As the evaporation progresses, the distance of the offset from the contact line decreases.
The results in Figure \ref{fig:fig4} show that, as a result of Marangoni flow, the temperature of the liquid in case C2 is larger along the interface near the contact line than in the droplet center.

\subsection{Contact angle of $\theta_c = 90\,^\circ$}

\begin{figure}
        \centering
       \begin{subfigure}{0.45\textwidth}
          \includegraphics[clip,trim = 0 0cm 0 0, width=\textwidth]{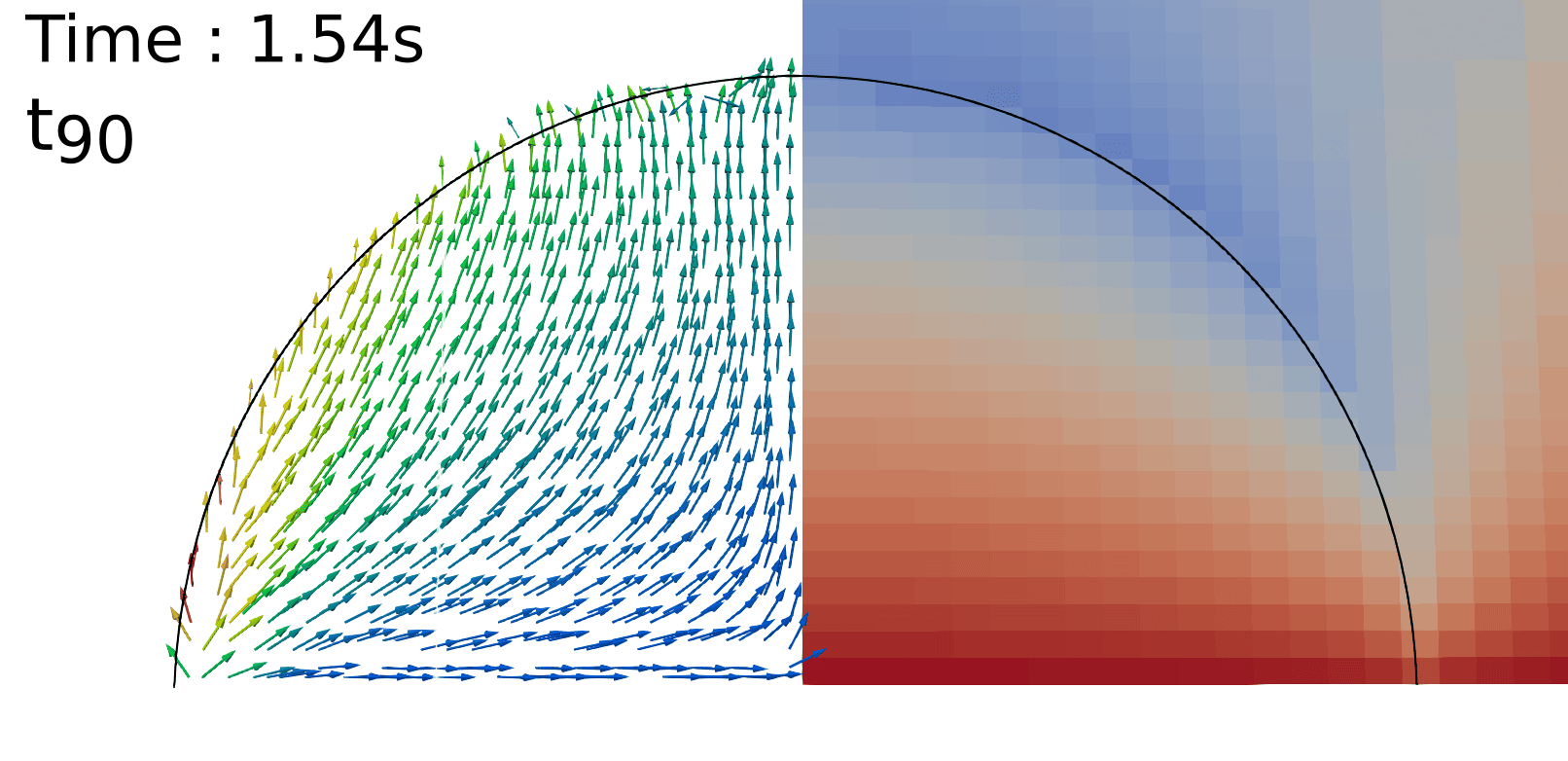}
                  \caption{}
         \label{fig:C3T1}
         \end{subfigure}
         \begin{subfigure}{0.45\textwidth}
          \includegraphics[clip,trim =0 0cm 0 0,width=\textwidth]{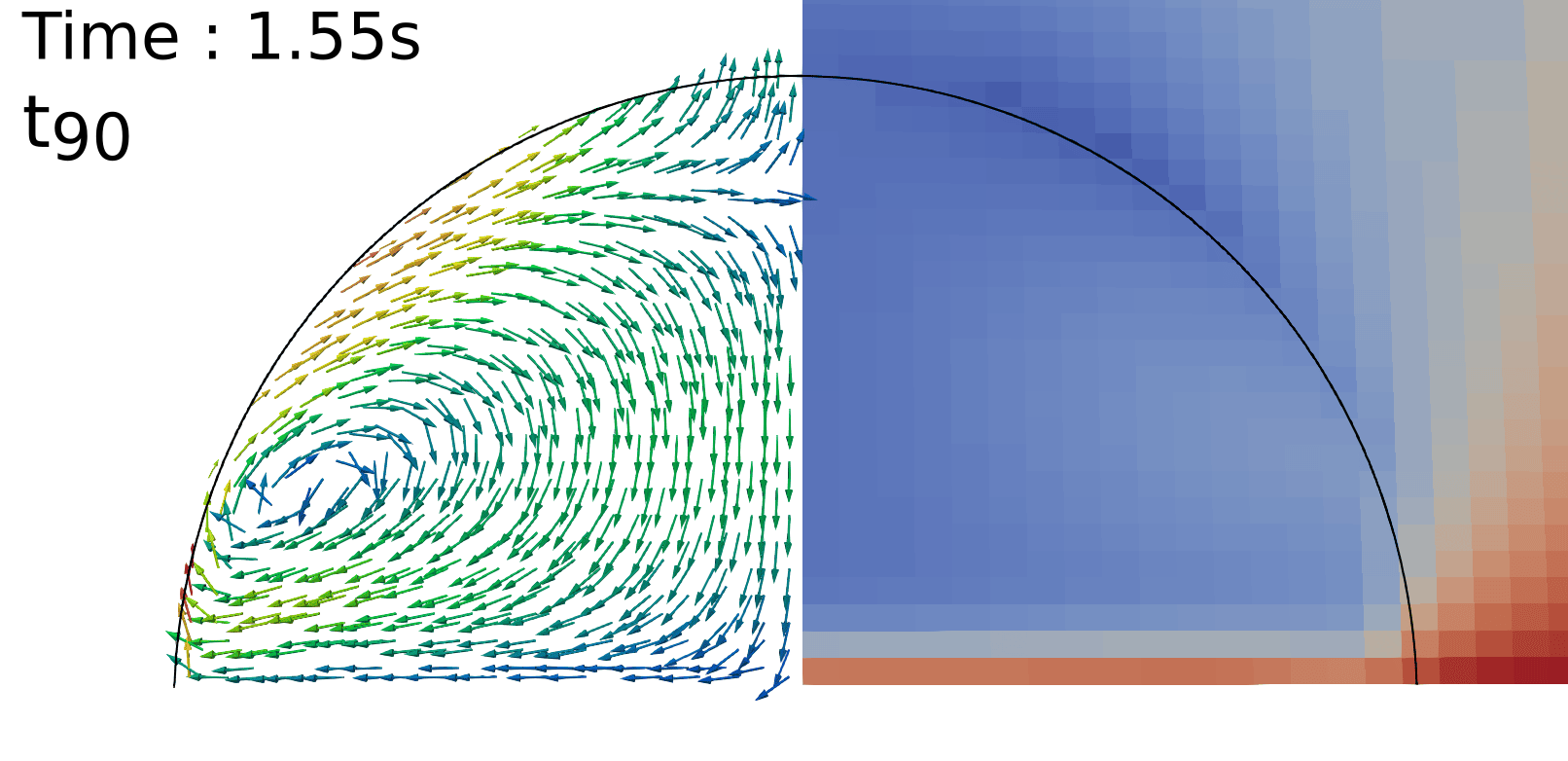}
                  \caption{}
         \label{fig:C4T1}
          \end{subfigure}
               \begin{subfigure}{0.45\textwidth}
          \includegraphics[clip,trim =0 0cm 0 0,width=\textwidth]{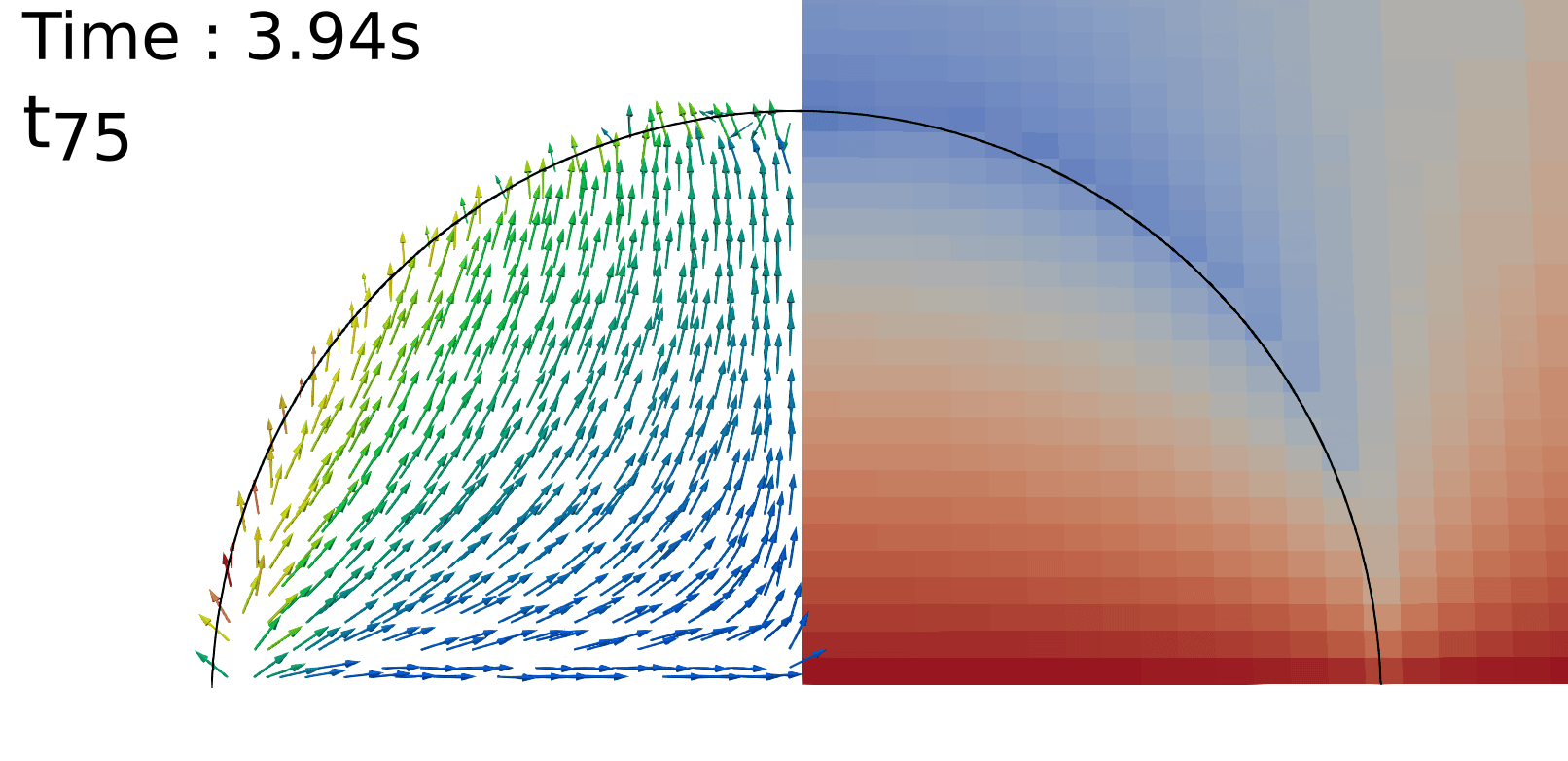}
                  \caption{}
         \label{fig:C3T2}
         \end{subfigure}
         \begin{subfigure}{0.45\textwidth}
          \includegraphics[clip,trim =0 0cm 0 0,width=\textwidth]{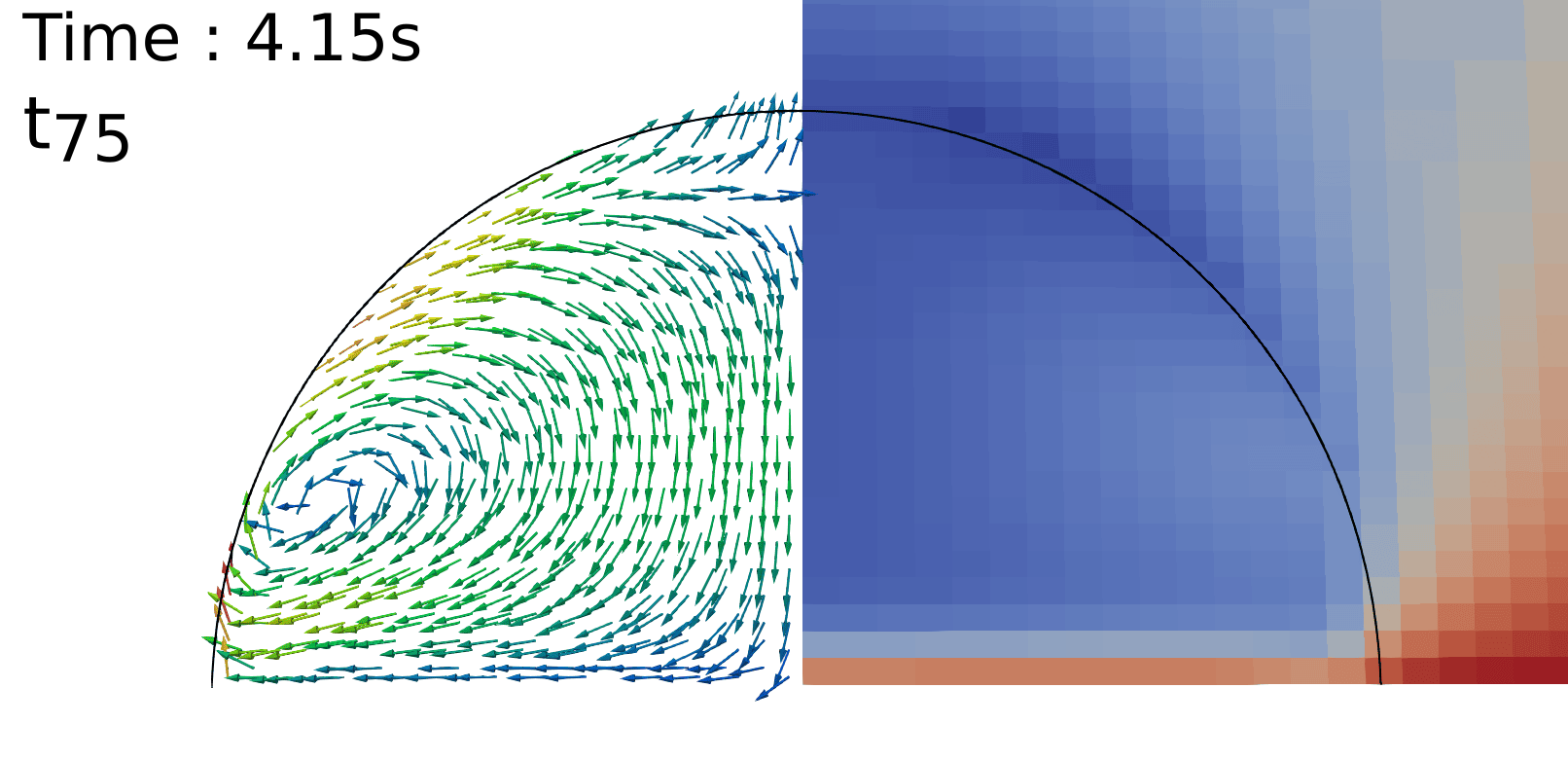}
                  \caption{}
         \label{fig:C4T2}
          \end{subfigure}
               \begin{subfigure}{0.45\textwidth}
          \includegraphics[clip,trim =0 0cm 0 0,width=\textwidth]{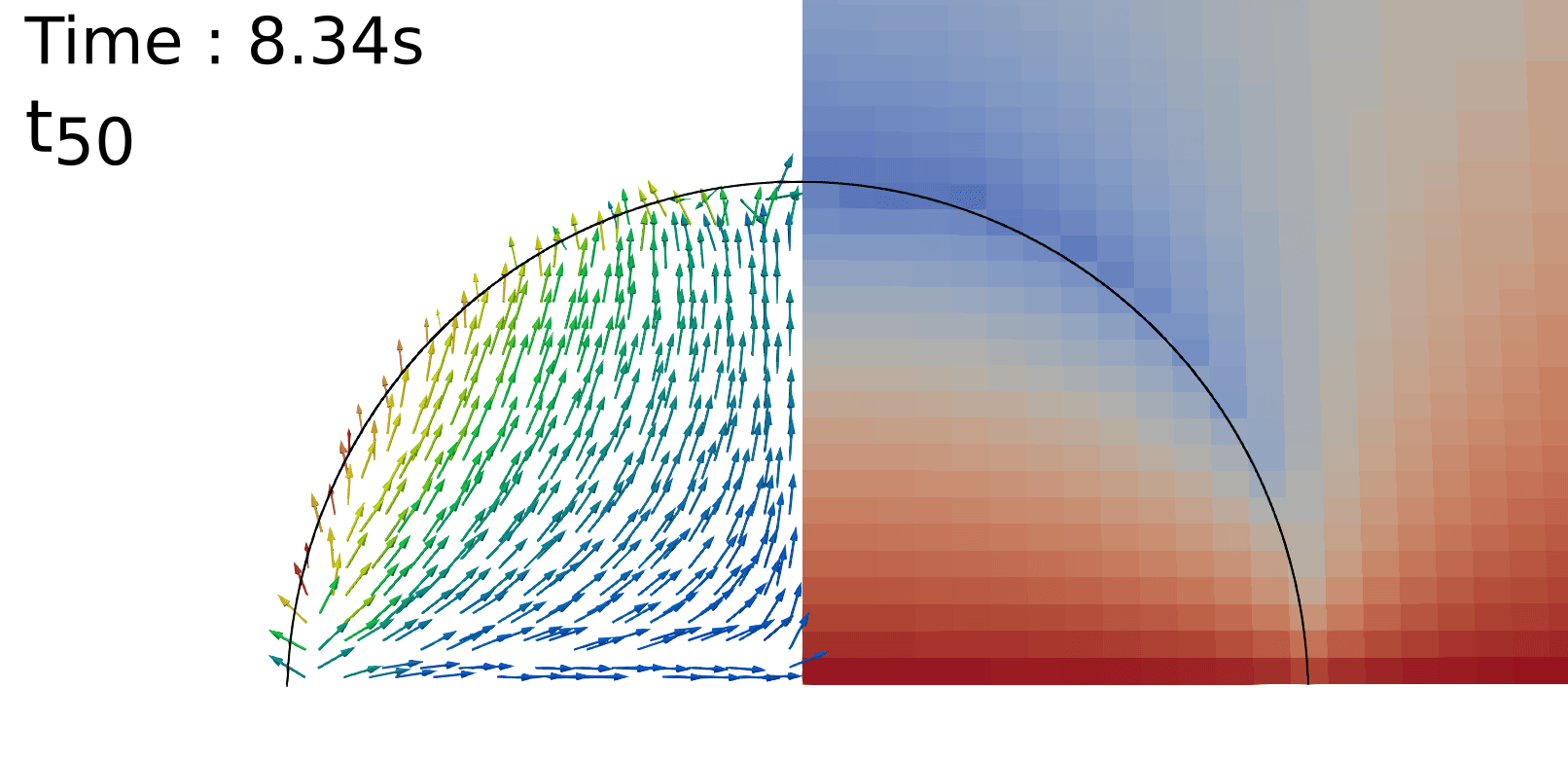}
                  \caption{}
         \label{fig:C3T3}
         \end{subfigure}
         \begin{subfigure}{0.45\textwidth}
          \includegraphics[clip,trim =0 0cm 0 0,width=\textwidth]{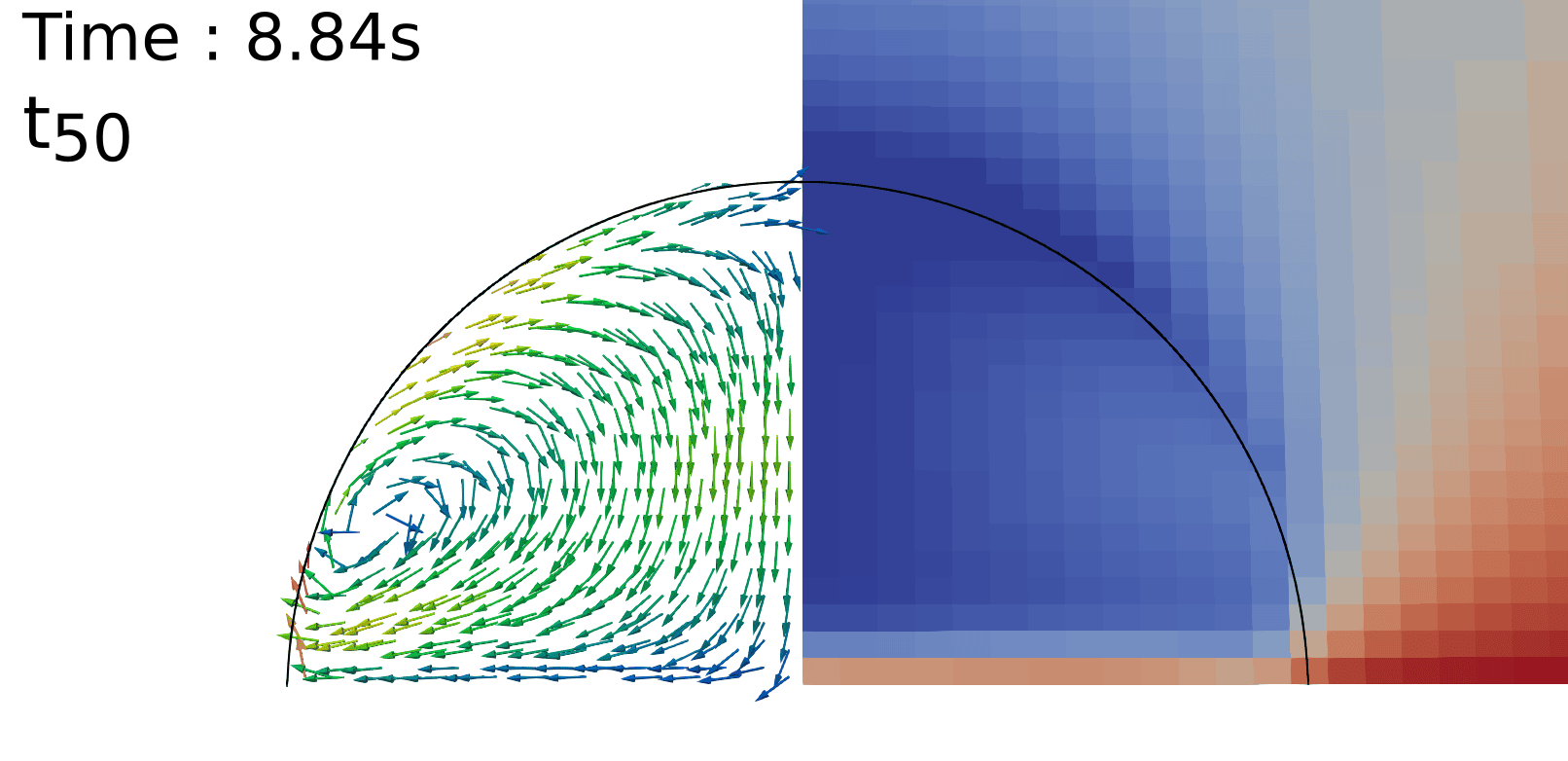}
                  \caption{}
         \label{fig:C4T3}
          \end{subfigure}
               \begin{subfigure}{0.45\textwidth}
          \includegraphics[clip,trim =0 0cm 0 0,width=\textwidth]{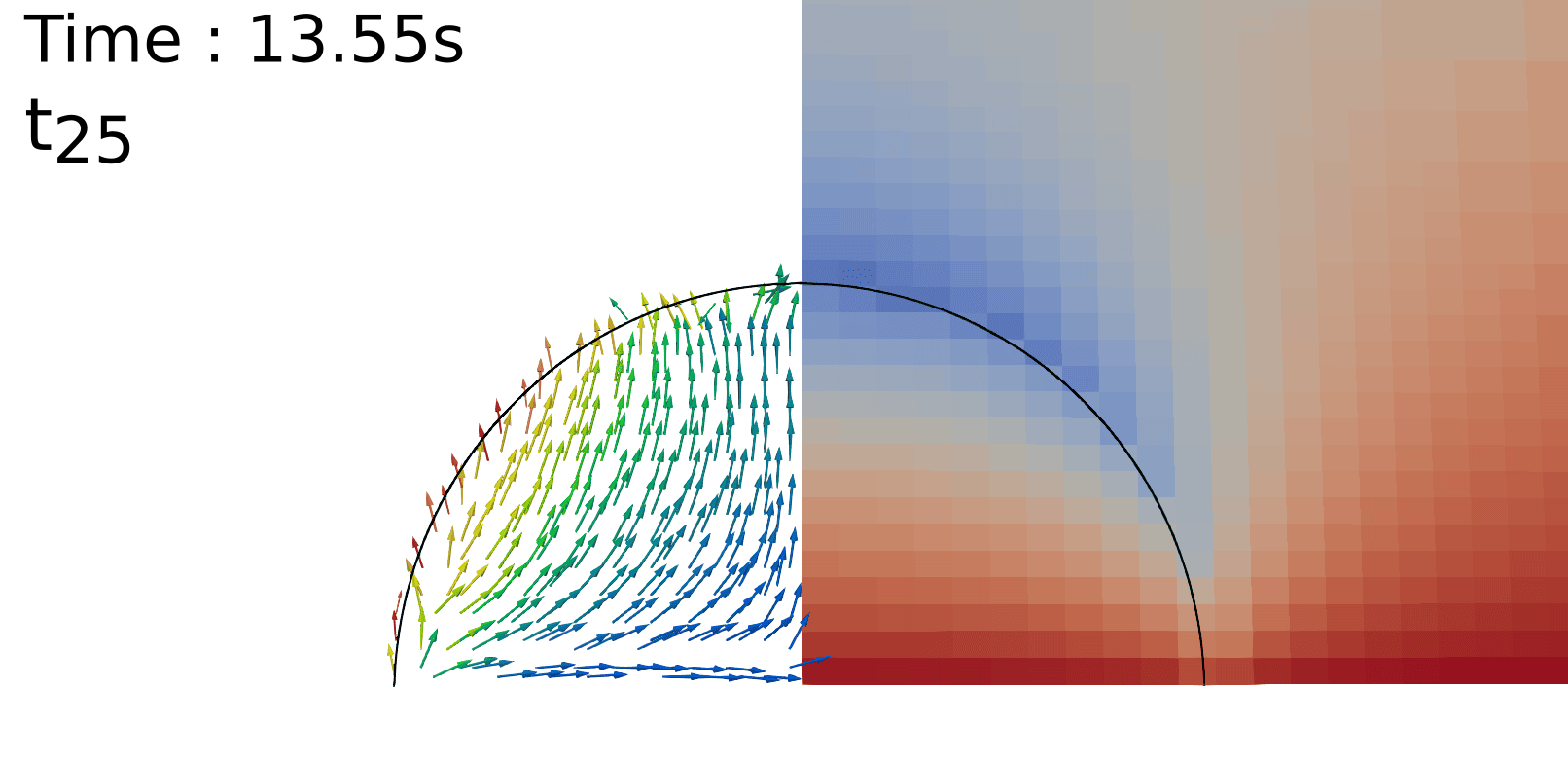}
                  \caption{}
         \label{fig:C3T4}
         \end{subfigure}
         \begin{subfigure}{0.45\textwidth}
          \includegraphics[clip,trim =0 0cm 0 0,width=\textwidth]{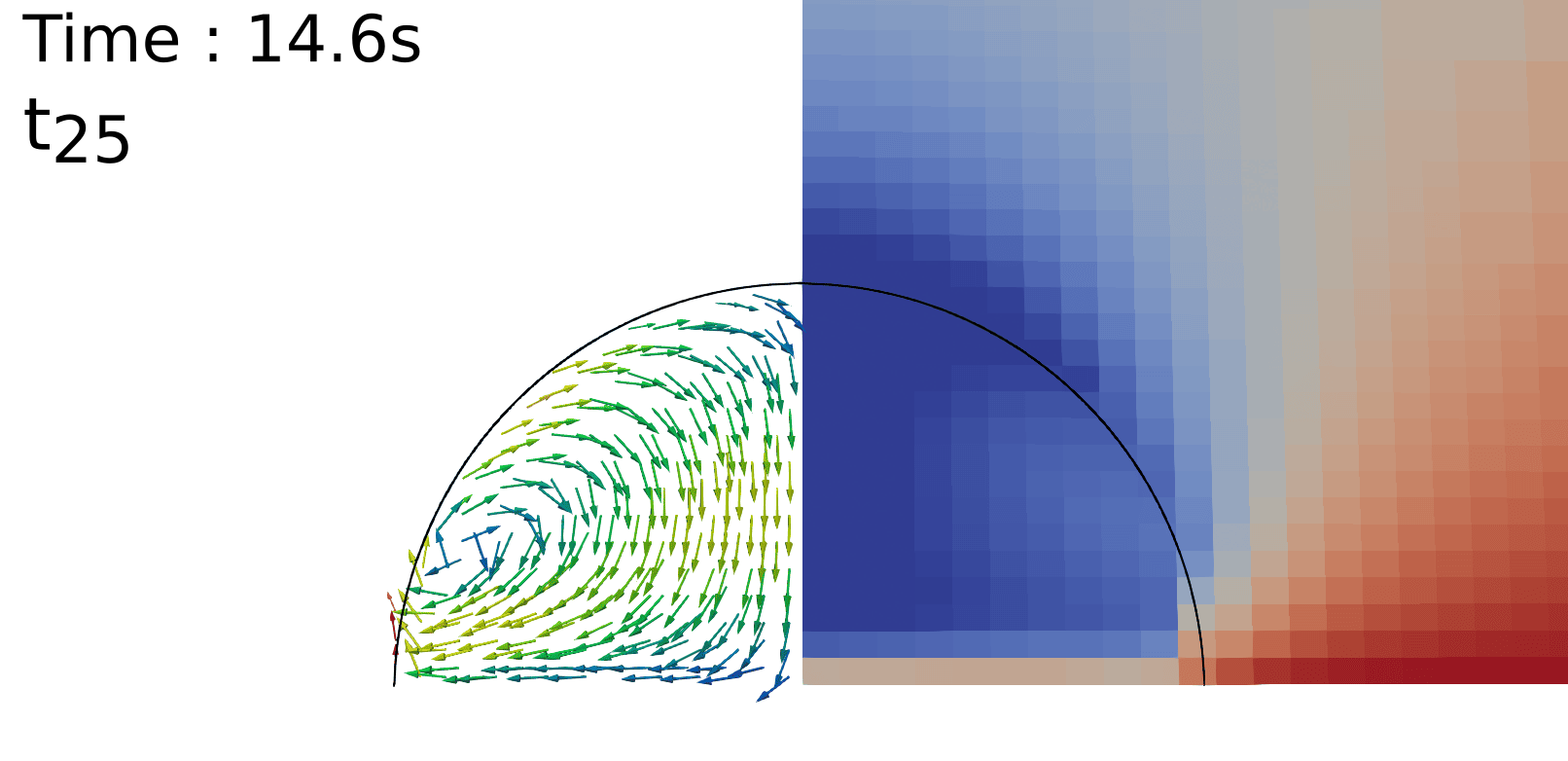}
                  \caption{}
         \label{fig:C4T4}
          \end{subfigure}
                       \begin{subfigure}{0.48\textwidth}
          \includegraphics[clip,trim =10cm 5cm 10cm 15cm,width=\textwidth]{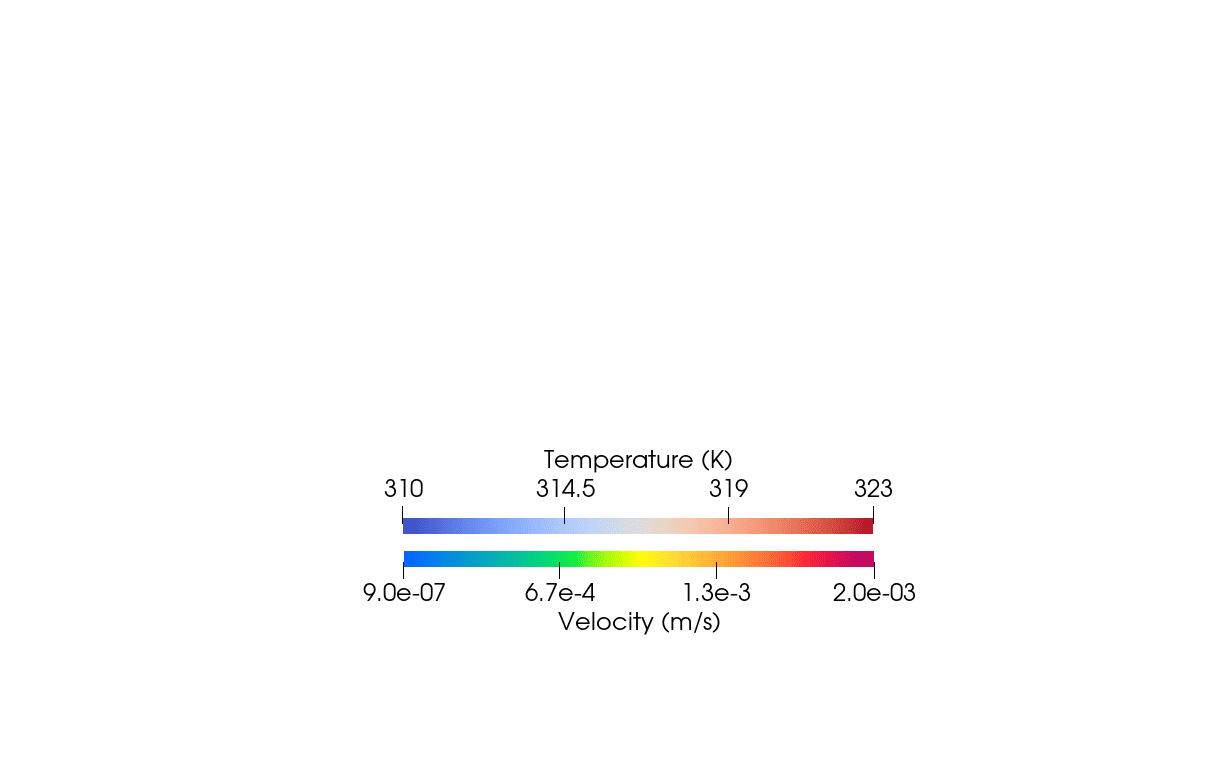}
         \end{subfigure}
         \begin{subfigure}{0.48\textwidth}
          \includegraphics[clip,trim =10cm 5cm 10cm 15cm,width=\textwidth]{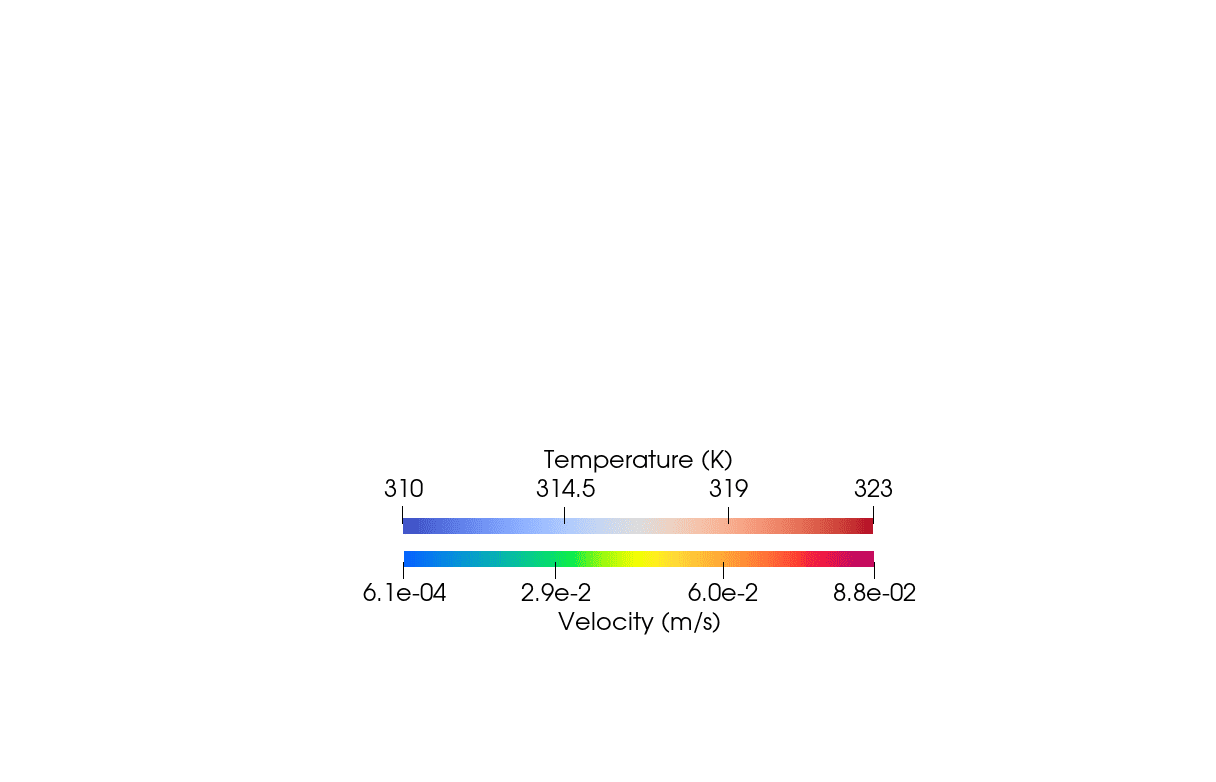}
          \end{subfigure}
                  \caption{Velocity vectors and temperature plots at four different time instances of the cases with contact angle  $\theta_\mathrm{c}=90\, ^\circ$  without taking into account the Marangoni stresses, C3 (left column), and with Marangoni stresses, C4 (right column).  Velocity vector plots are shown on left, where the color indicates the velocity magnitude. Contour plots of the temperature are shown on the right. \textbf{(a)} C3, $t_{90}$, \textbf{(c)} C3, $t_{75}$, \textbf{(e)} C3, $t_{50}$, \textbf{(g)} C3, $t_{25}$.
                    \textbf{(b)} C4, $t_{90}$, \textbf{(d)} C4, $t_{75}$, \textbf{(f)} C4, $t_{50}$, \textbf{(h)} C4, $t_{25}$.}
                  \label{fig:fig5}
    \end{figure}

In Figure \ref{fig:fig5}, similar plots as in the previous section are shown for cases C3 and C4, for four different time instances, in which the contact angles are $\theta_c = 90\, ^\circ$.
The figures showing the fluid velocity field for case C3, where the Marangoni stresses are absent, show similar flow patterns as for case C1; the fluid flows from the contact line to the apex of the droplet and the magnitude of the flow remains approximately constant. In addition, the temperature profiles observed in case C3 show a consistent distribution over time, with the lowest temperature at the apex of the droplet and a higher temperature in the interior of the droplet than near the interface.

The figure showing the velocity field for the first depicted time instance for case C4 (Figure \ref{fig:C4T1}), shows that the Marangoni flow is stronger than the capillary flow and the flow induced by the receding contact line, resulting in a branching of the flow.
A part of the flow moves towards the bottom bulk of the droplet, while the other part moves towards the apex of the droplet. 
This branching effect of the Marangoni flow is absent in the case of smaller contact angle in case C2, as seen in Figure \ref{fig:C2T1}. 
As the evaporation progresses, the magnitude of the Marangoni stresses and the resulting magnitude of the Marangoni flow reduce as a result of the decreasing temperature difference along the interface.
Unlike in case C2, the vortex formed by the Marangoni flow is much closer to the contact line of the droplet and does not exhibit a stagnation point near the substrate.

The figures showing the temperature for case C4 shows a lower temperature at the apex of the droplet than the bulk of the droplet owing to evaporative cooling. The Marangoni flow enhances convective heat transfer and mixing considerably, increasing the temperature along the interface compared to the bulk of the droplet.
Contrary to case C2, the temperature in the bulk of the droplet decreases, as evaporative cooling dominates, and the Marangoni-driven convective mixing affects the heat transfer. This is due to the fact that for a droplet with larger contact angle,
the thermal resistance is higher and the solid-liquid contact area is smaller.

\subsection{Contact angle of $\theta_c = 120\,^\circ$}

\begin{figure}
        \centering
       \begin{subfigure}{0.45\textwidth}
          \includegraphics[clip,trim = 0 0cm 0 0, width=\textwidth]{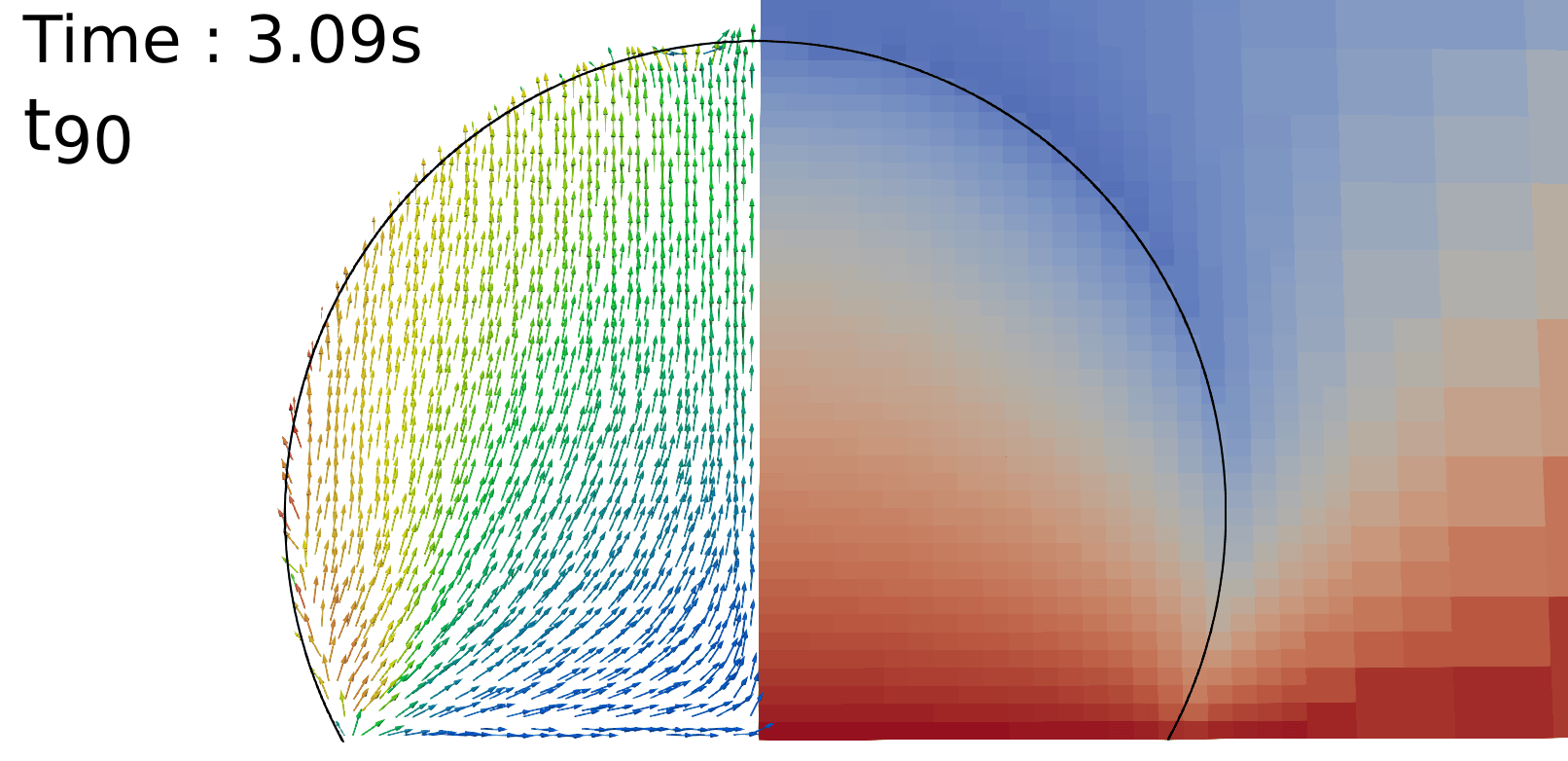}
                  \caption{}
         \label{fig:C5T1}
         \end{subfigure}
         \begin{subfigure}{0.45\textwidth}
          \includegraphics[clip,trim =0 0cm 0 0,width=\textwidth]{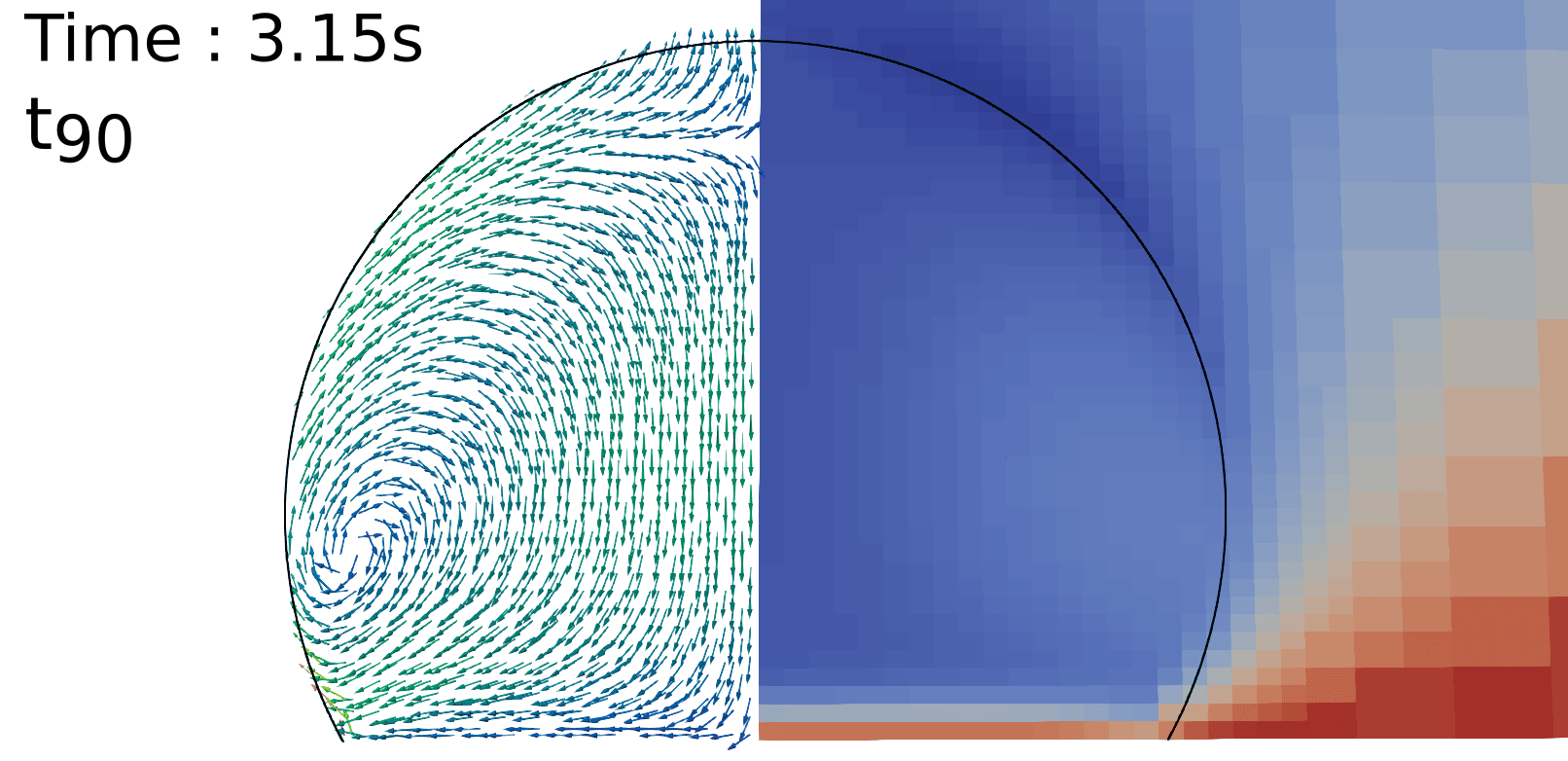}
                  \caption{}
         \label{fig:C6T1}
          \end{subfigure}
               \begin{subfigure}{0.45\textwidth}
          \includegraphics[clip,trim =0 0cm 0 0,width=\textwidth]{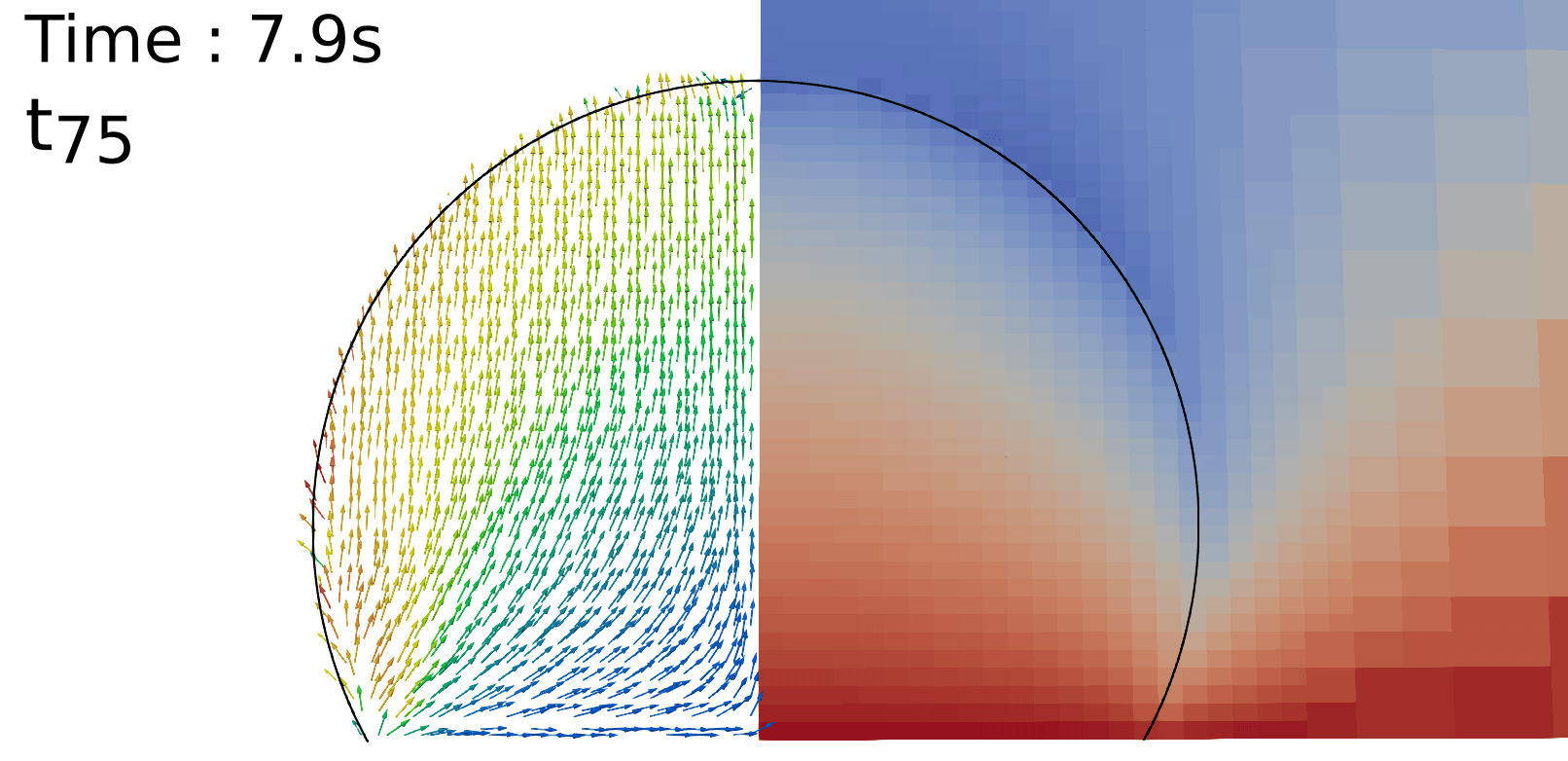}
                  \caption{}
         \label{fig:C5T2}
         \end{subfigure}
         \begin{subfigure}{0.45\textwidth}
          \includegraphics[clip,trim =0 0cm 0 0,width=\textwidth]{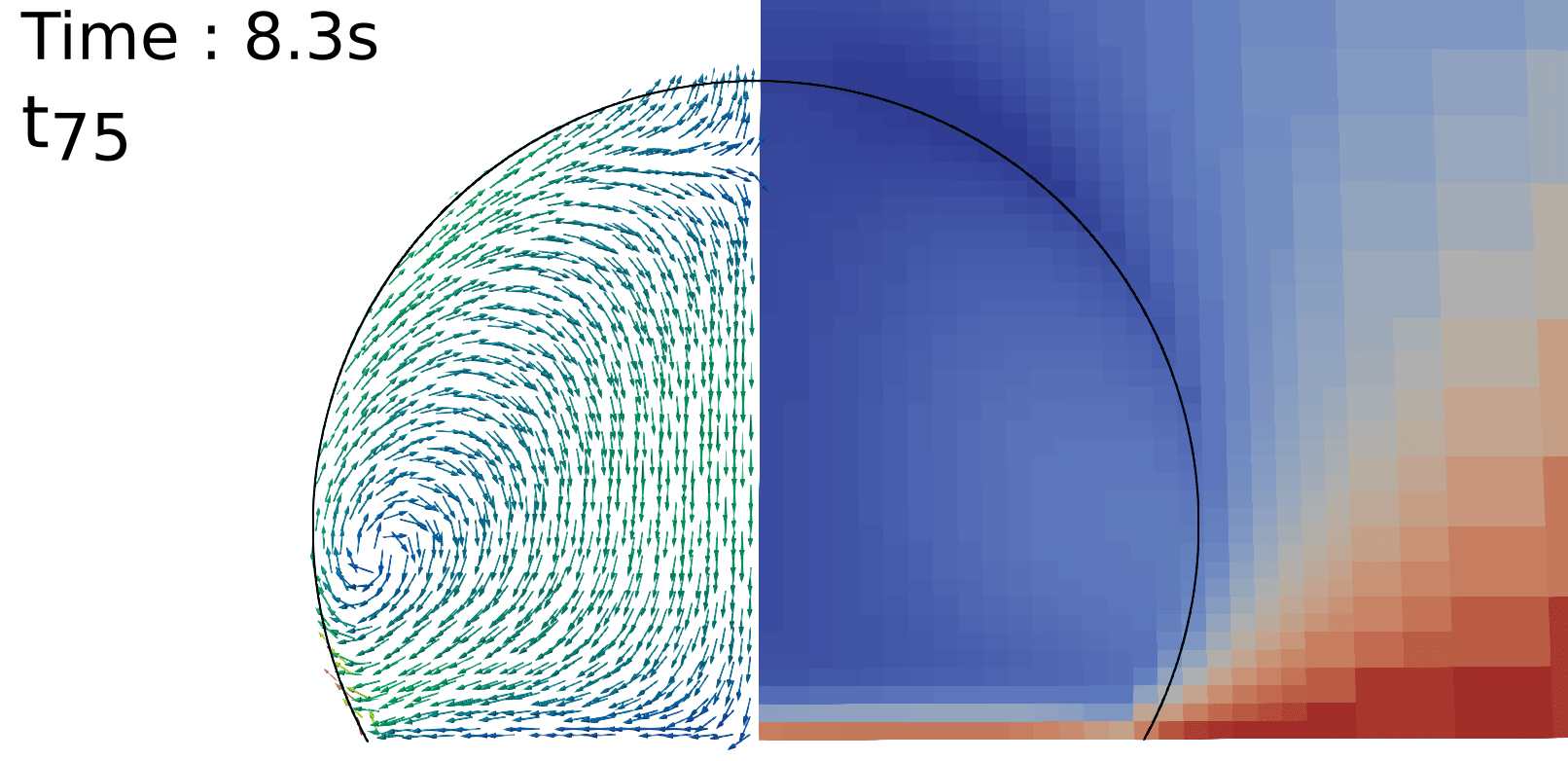}
                  \caption{}
         \label{fig:C6T2}
          \end{subfigure}
               \begin{subfigure}{0.45\textwidth}
          \includegraphics[clip,trim =0 0cm 0 0,width=\textwidth]{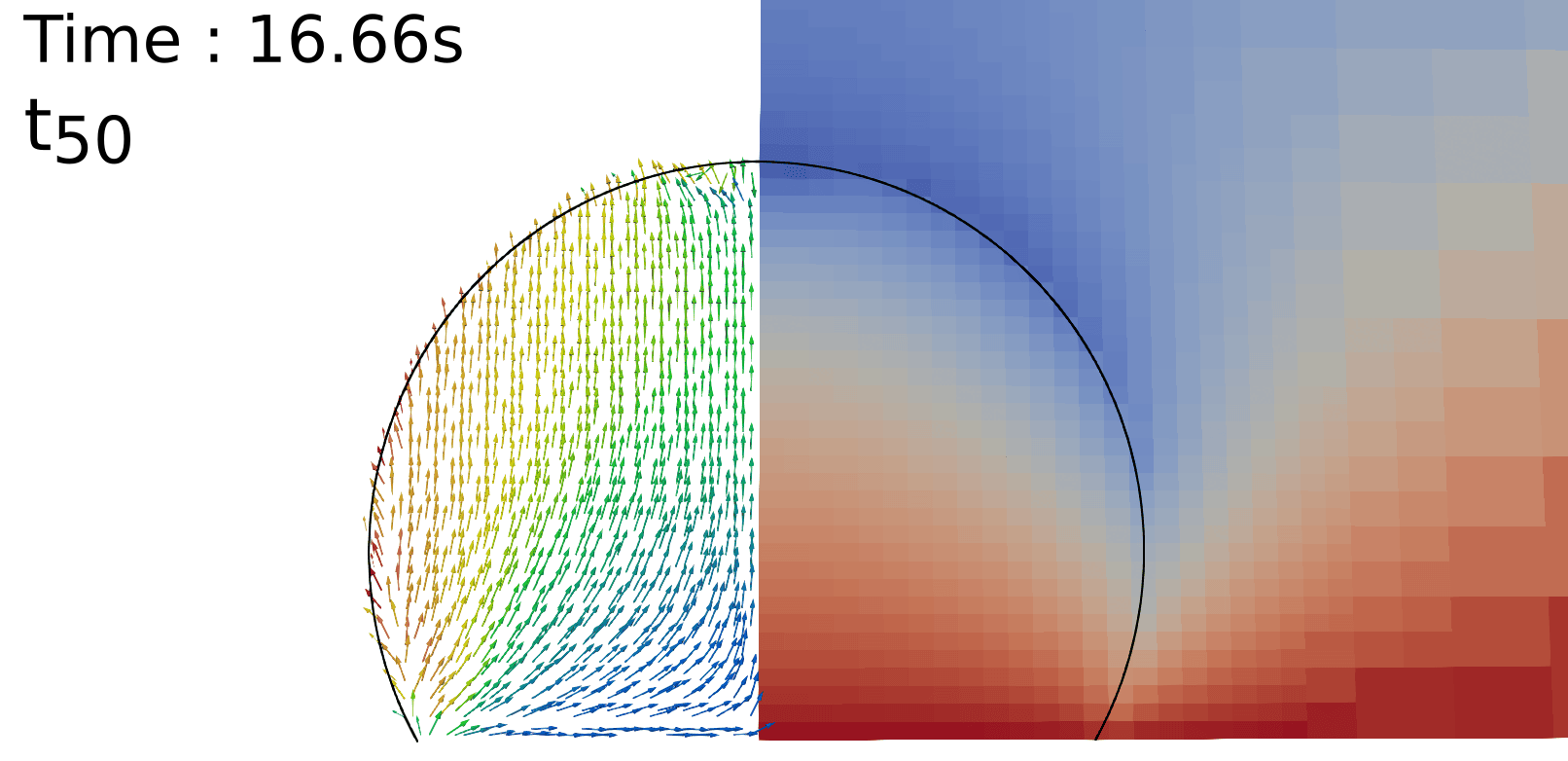}
                  \caption{}
         \label{fig:C5T3}
         \end{subfigure}
         \begin{subfigure}{0.45\textwidth}
          \includegraphics[clip,trim =0 0cm 0 0,width=\textwidth]{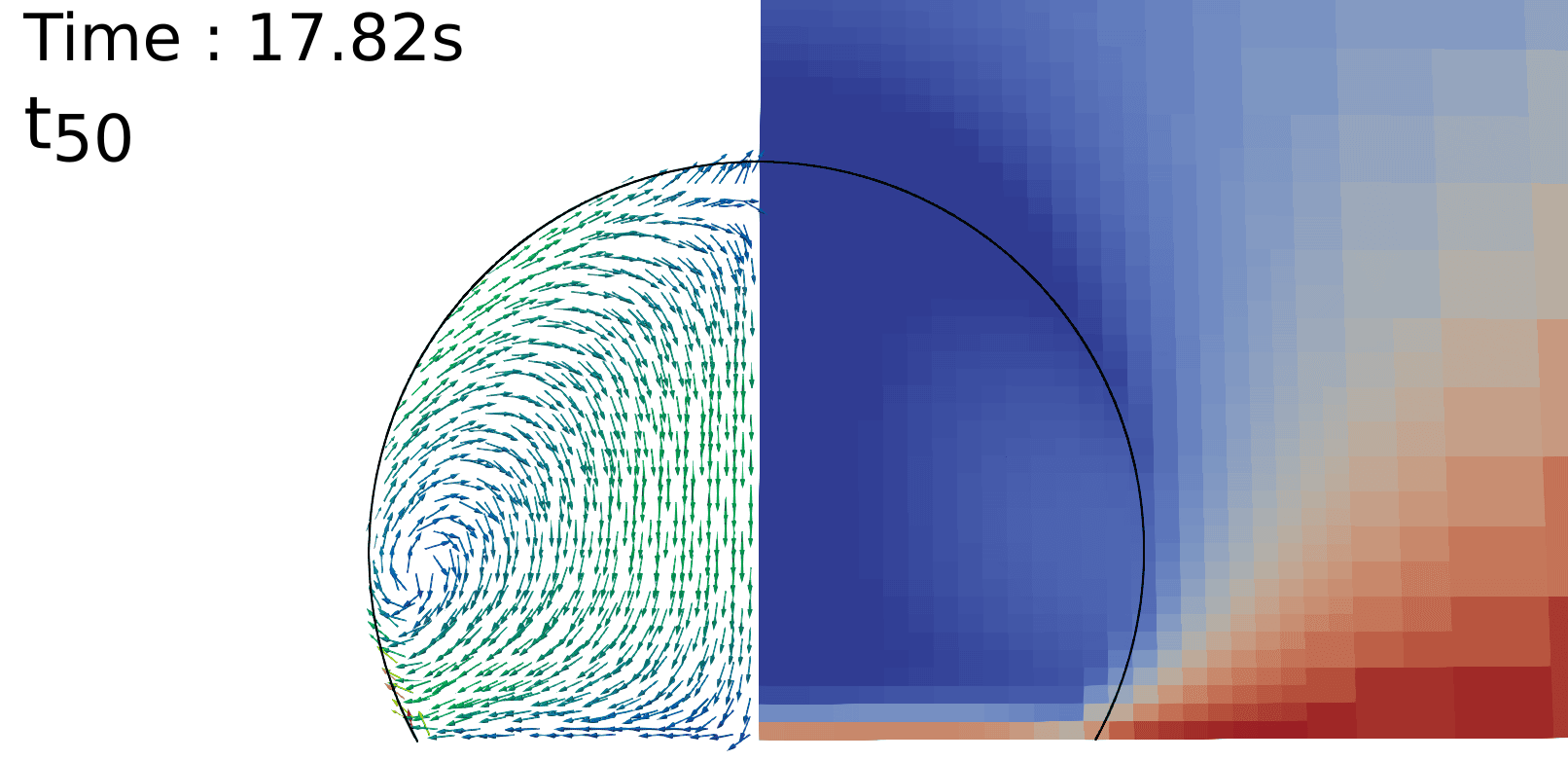}
                  \caption{}
         \label{fig:C6T3}
          \end{subfigure}
               \begin{subfigure}{0.45\textwidth}
          \includegraphics[clip,trim =0 0cm 0 0,width=\textwidth]{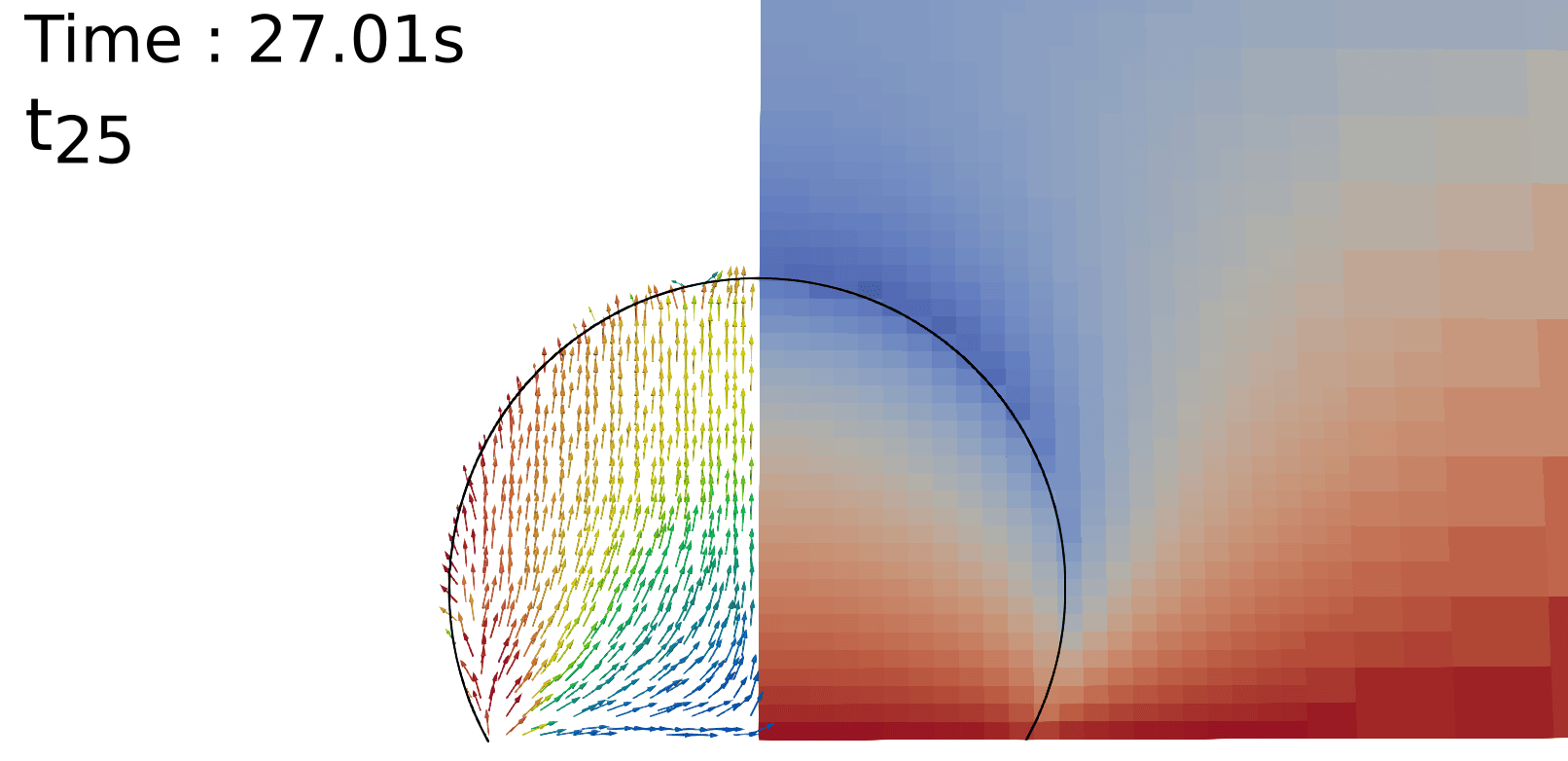}
                  \caption{}
         \label{fig:C5T4}
         \end{subfigure}
         \begin{subfigure}{0.45\textwidth}
          \includegraphics[clip,trim =0 0cm 0 0,width=\textwidth]{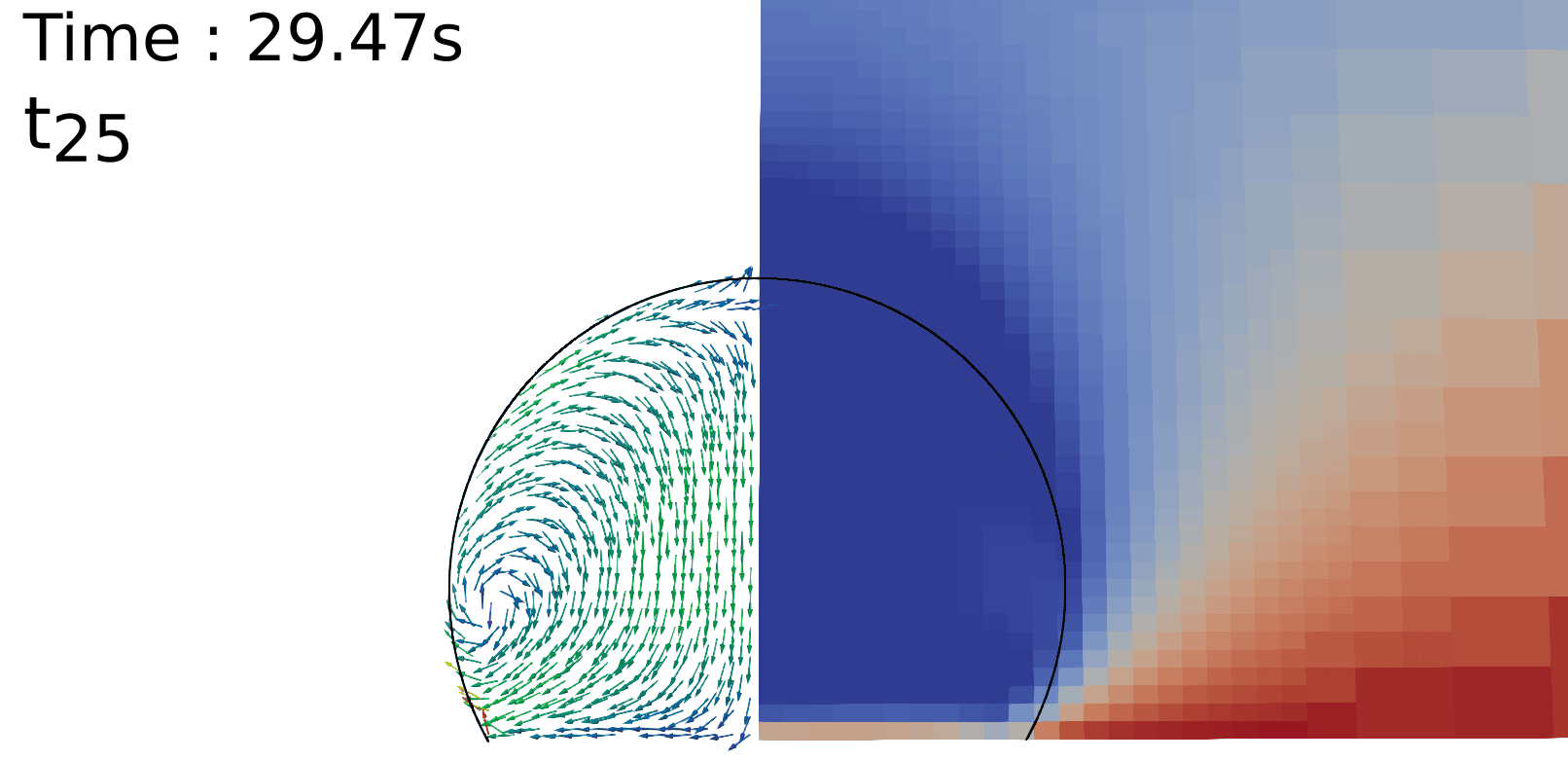}
                  \caption{}
         \label{fig:C6T4}
          \end{subfigure}
                       \begin{subfigure}{0.48\textwidth}
          \includegraphics[clip,trim =10cm 5cm 10cm 15cm,width=\textwidth]{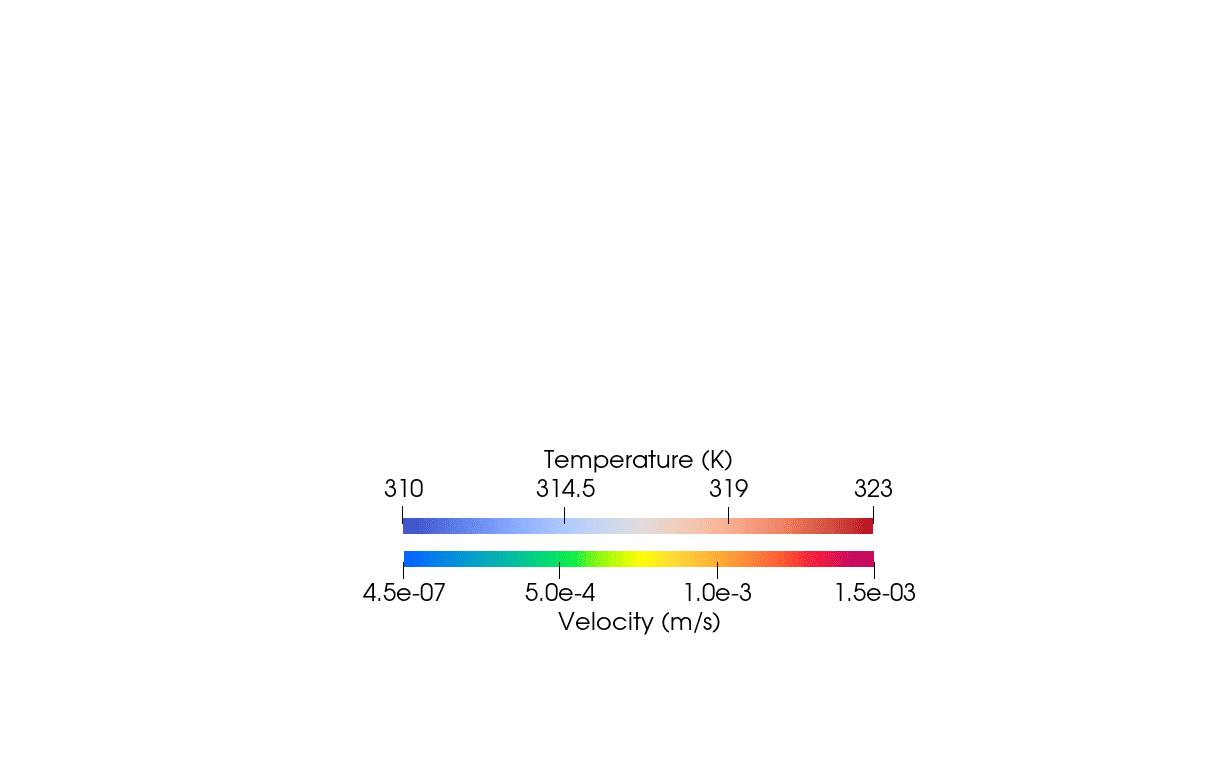}
         \end{subfigure}
         \begin{subfigure}{0.48\textwidth}
          \includegraphics[clip,trim =10cm 5cm 10cm 15cm,width=\textwidth]{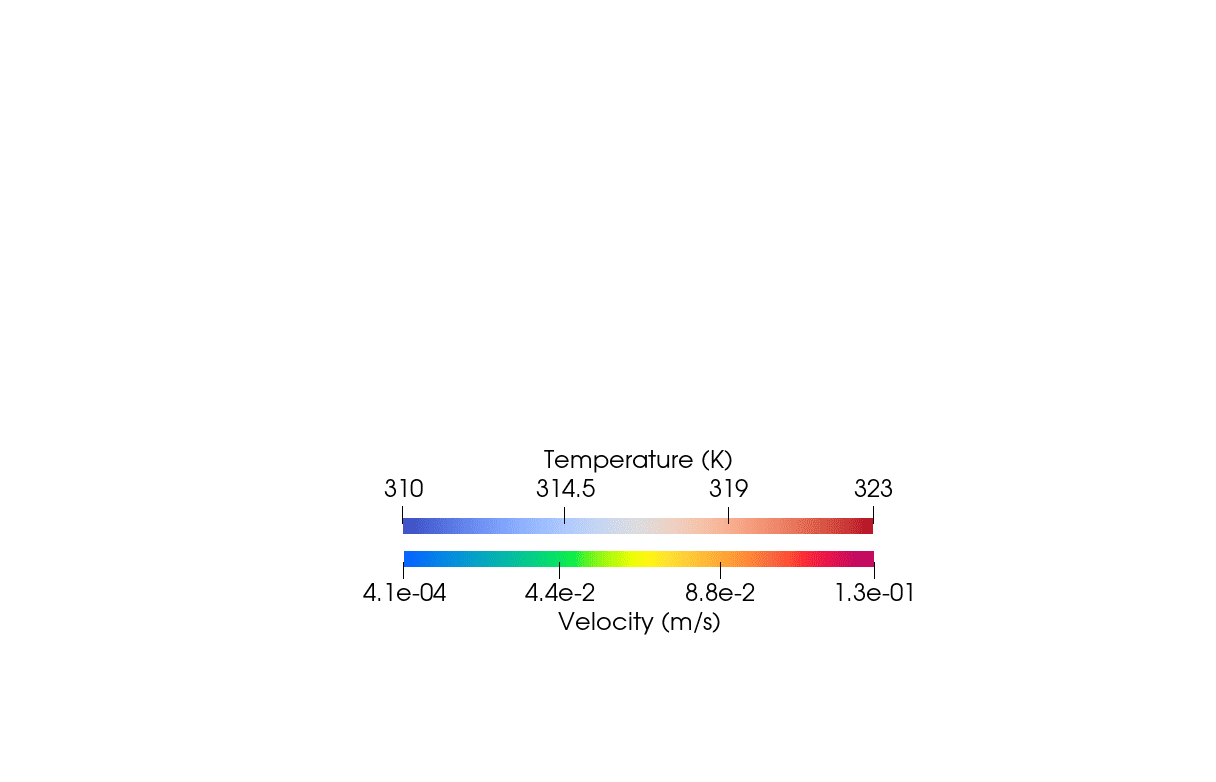}
          \end{subfigure}
                  \caption{Velocity vectors and temperature plots at four different time instances of the cases with contact angle  $\theta_\mathrm{c}=120\, ^\circ$  without taking into account the Marangoni stresses, C5 (left column), and with Marangoni stresses, C6 (right column). Velocity vector plots are shown on the left, where the color indicates the velocity magnitude. Contour plots of the temperature on shown on the right. \textbf{(a)} C5, $t_{90}$, \textbf{(c)} C5 $t_{75}$, \textbf{(e)} C5, $t_{50}$, \textbf{(g)} C5, $t_{25}$.
                    \textbf{(b)} C6, $t_{90}$, \textbf{(d)} C6, $t_{75}$, \textbf{(f)} C6, $t_{50}$, \textbf{(h)} C6, $t_{25}$.}
                  \label{fig:fig6}
    \end{figure}

Figure \ref{fig:fig6} shows the velocity and temperature fields for cases C5 and C6, at four different time instances, in which the contact angles are $\theta_c = 120\, ^\circ$.
The figures showing the fluid velocity for case C5, where the Marangoni stresses are absent, show similar flow patterns as are observed in case C3. The fluid flows from the contact line to the apex of the droplet. However, the magnitude of the fluid flow increases during the course of the evaporation process. In addition, the temperature distribution of case C5 exhibits a monotonic distribution along the interface, with the lowest temperature  at the apex of the droplet. The temperature in the interior of the droplet is higher than in the proximity of the interface, similar as is observed in cases C1 and C3.

Comparing the velocity fields of cases C5 and C6 suggests that the Marangoni flow is stronger than the capillary flow and the flow induced by the receding contact line, resulting in a branching of the flow in case C6, similar to the branching observed in case C4.
The vortex formed by the Marangoni flow is close to the interface and does not show any stagnation point near the substrate. The temperature distribution in case C6 is similar in nature to case C4, showing a lower temperature at the apex of the droplet as a result of evaporative cooling and the large distance of the apex from the heated substrate.

\subsection{Discussion}
Marangoni stresses play a crucial role for the fluid flow and temperature distribution inside the droplet. For cases in which the Marangoni stresses are neglected, irrespective of the contact angle, the flow induced by the moving contact line dominates the capillary flow, with the fluid flowing from the contact line to the apex of the droplet. 
This stands in contrast to the explanation provided by \citet{Bhardwaj2018}, in which it is suggested that the fluid flows from the contact line to the apex of droplets with moving contact if the contact angle is $\theta_c > 90^\circ$. \citet{Bhardwaj2018} also suggests that the evaporation at the apex of the droplet is higher than at the contact line. However, our results demonstrate that, irrespective of the contact angle, the flow inside a droplet resting on the heated substrate and evaporating in the CCA mode is dominated by the contact line motion rather than by evaporation.
If Marangoni stresses are present, the flow induced by these Marangoni stresses dominates the flow inside the droplet, leading to a classical single-vortex flow.

\begin{figure}
      \centering
      \begin{subfigure}[b]{0.48\textwidth}
		\includegraphics[clip,trim=3cm 2cm 4cm 0cm, width=\textwidth]{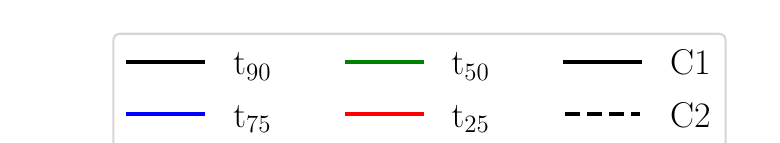}
       \end{subfigure}
                    \begin{subfigure}[b]{0.48\textwidth}
		\includegraphics[clip,trim=3cm 2cm 4cm 0cm,width=\textwidth]{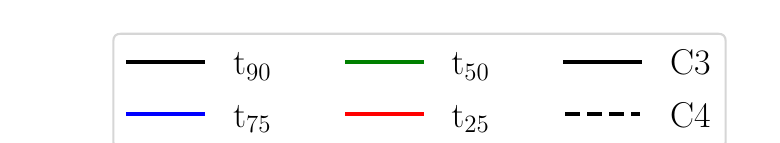}
       \end{subfigure}
     \begin{subfigure}[b]{0.48\textwidth}
		\includegraphics[width=\textwidth]{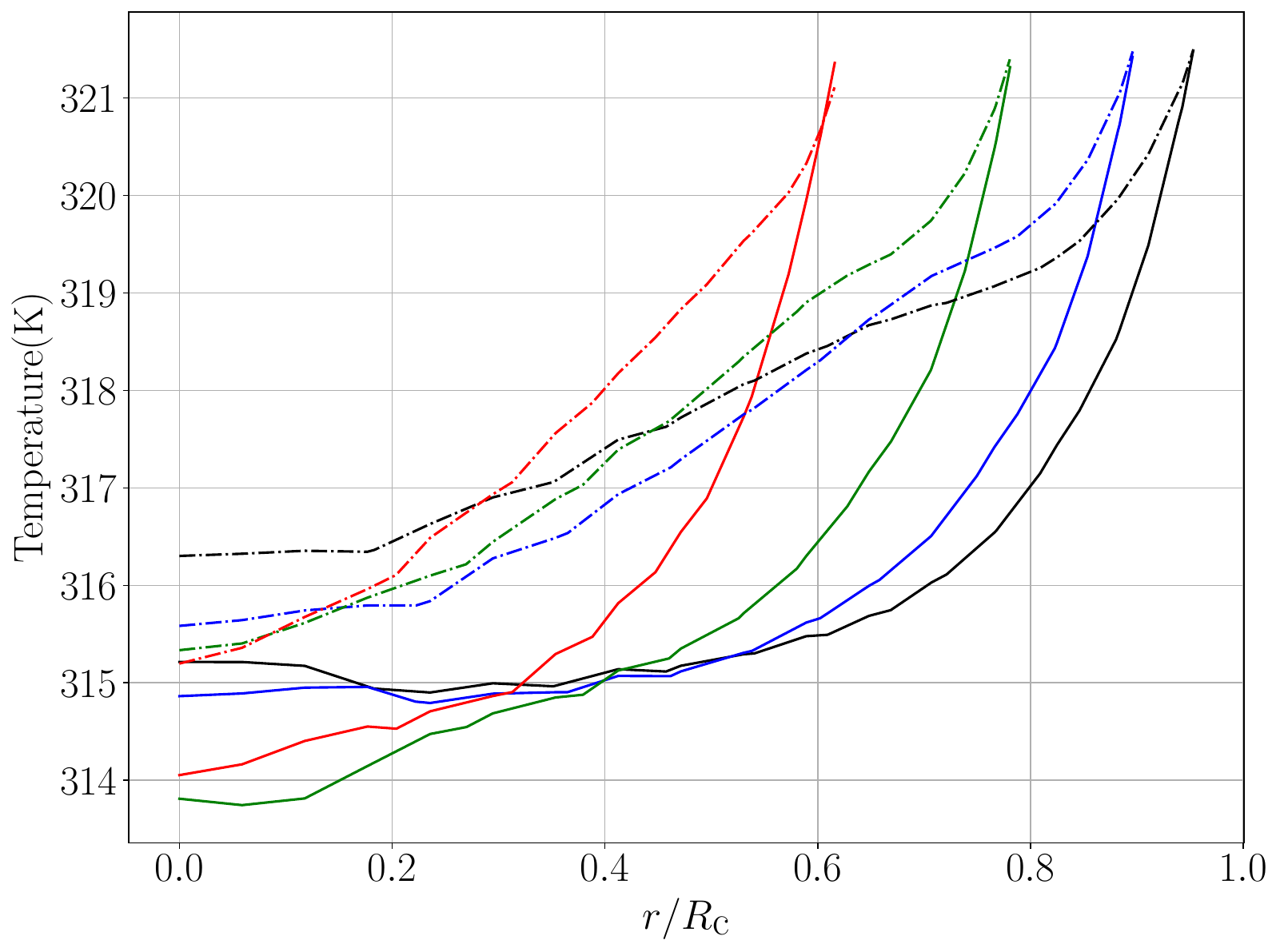}
			    \caption{}
       \label{fig:IntTempC1C2}
       \end{subfigure}
       \begin{subfigure}[b]{0.47\textwidth}
		\includegraphics[width=\textwidth]{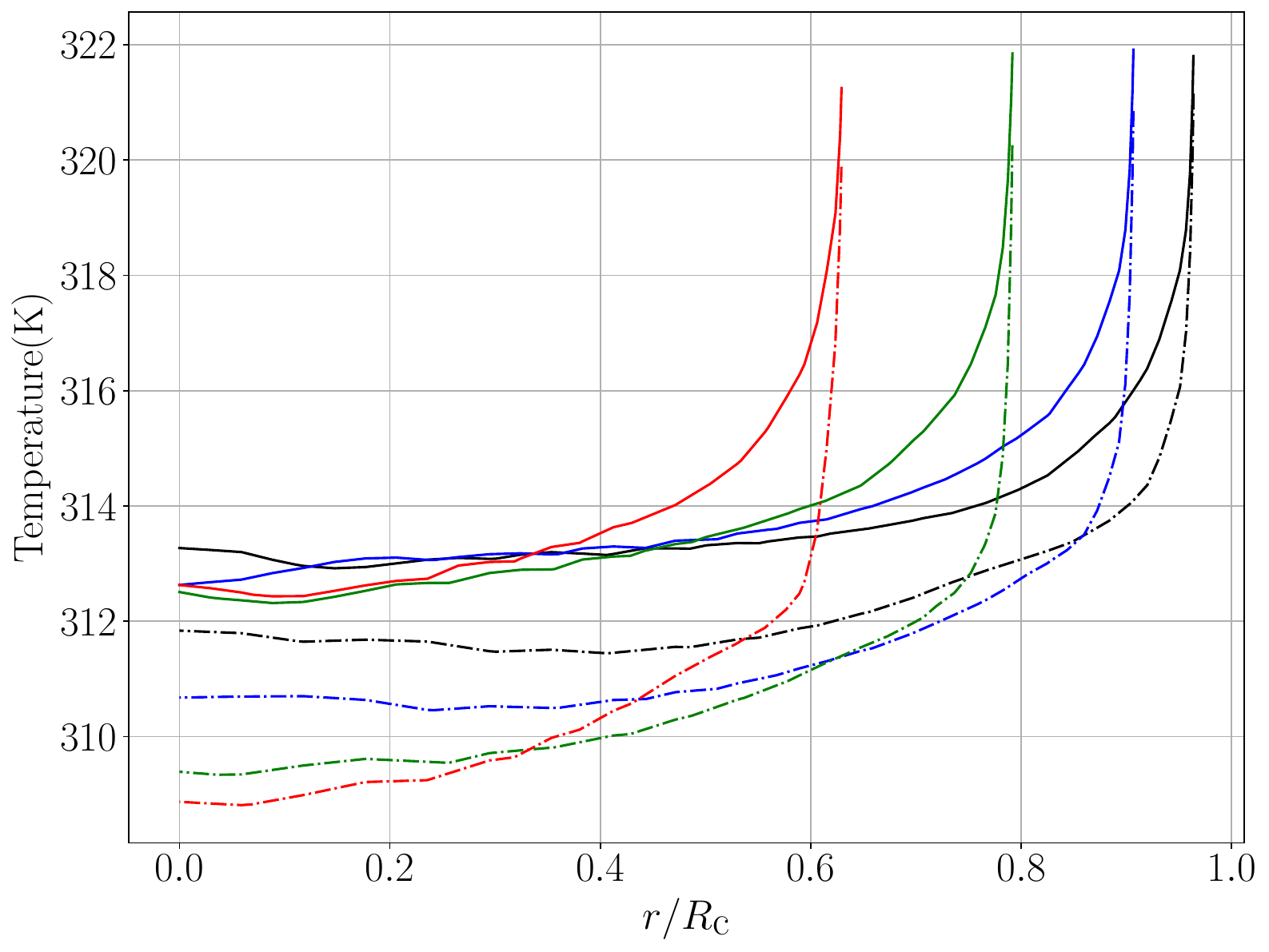}
			    \caption{}
       \label{fig:IntTempC3C4}
\end{subfigure}\\
       \begin{subfigure}[b]{0.48\textwidth}
        \includegraphics[clip,trim=3cm 2cm 4cm 0cm,width=\textwidth]{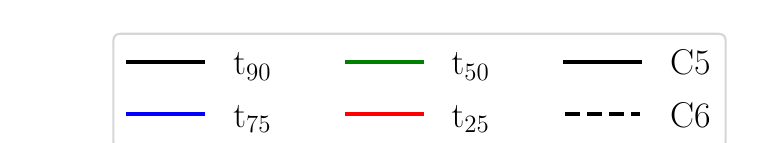}
\end{subfigure}\\
\begin{subfigure}[b]{0.47\textwidth}
        \includegraphics[width=\textwidth]{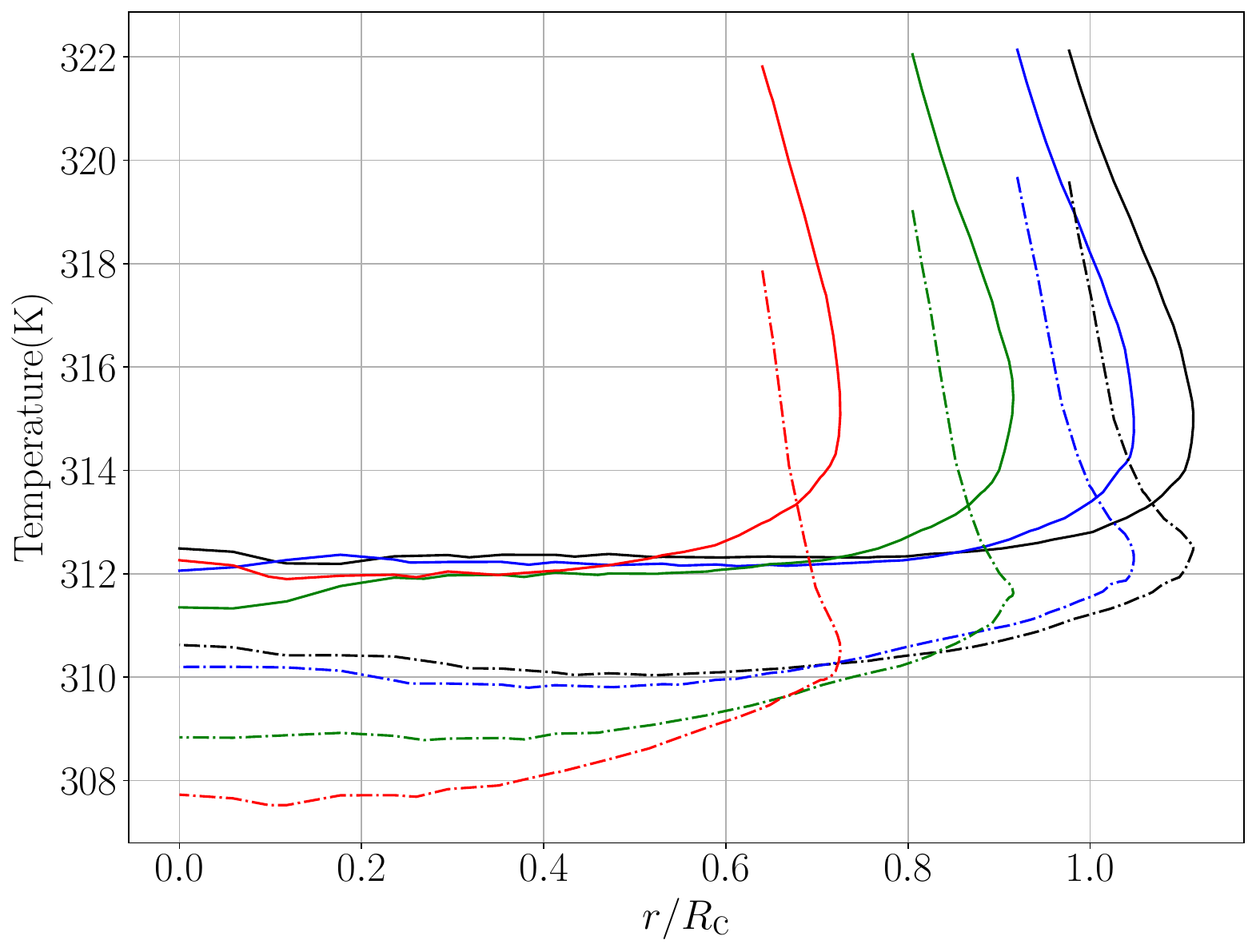}
                    \caption{}
\label{fig:IntTempC5C6}
\end{subfigure}
        \caption{Temperature distribution along the interface for different time instances. \textbf{(a)} $\theta_\mathrm{c}=60\, ^\circ$,  neglecting the Marangoni stresses, C1 (solid lines), and including the Marangoni stresses, C2 (dashed lines).  \textbf{(b)}  $\theta_\mathrm{c}=90\, ^\circ$, neglecting the Marangoni stresses, C3 (solid lines), and including the Marangoni stresses, C4 (dashed lines).  \textbf{(c)} $\theta_\mathrm{c}=120\, ^\circ$,  neglecting the Marangoni stresses, C5 (solid lines), and including the Marangoni stresses, C6 (dashed lines).}
        \label{fig:fig7}
\end{figure}

The temperature distribution along the interface is shown in Figure \ref{fig:fig7}. Figure \ref{fig:IntTempC1C2} shows the interface temperature for cases C1 and C2, where the contact angle is $60\, ^\circ$. 
The interface temperatures are shown for four different time instances with droplet volumes of 90\%, 75\%, 50\%, and 25\% of the initial volume of the droplet.
For case C1, in which the Marangoni stresses are neglected, the interface temperatures are lower compared to those in case C2, which takes into account the Marangoni stresses.  This suggests that the convective mixing due to Marangoni flow increases the interface temperature.
Figure \ref{fig:IntTempC3C4} illustrates the interface temperature for cases C3 and C4, where the contact angle is $90\, ^\circ$.
The overall interface temperature for case C4, considering the effect of the Marangoni stresses, is lower than that for case C3, where the effect of the Marangoni stresses is neglected, unlike the cases where the contact angle is $60\, ^\circ$. In case C3, the interface temperature converges towards a constant value at the apex. However, for case C4, the temperature at the apex continuously decreases, with a lower temperature than in case C3.
Similar results are obtained for cases C5 and C6, shown in Figure \ref{fig:IntTempC5C6}, where the presence of Marangoni stresses result in a lower interface temperature in case C6 compared to the interface temperature of case C5. In addition, the apex temperature continuously decreases in case C6, whereas in case C5, the interface region near the apex maintains a constant temperature.

\begin{figure}
      \centering

        \begin{subfigure}[b]{0.48\textwidth}
		\includegraphics[clip,trim=3cm 2cm 4cm 0cm,width=\textwidth]{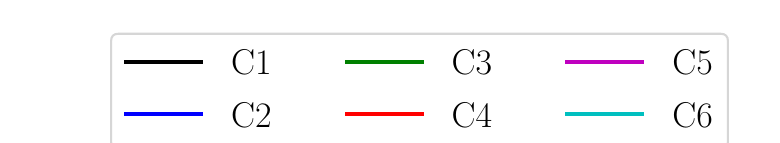}
       \end{subfigure}
       \\
       \begin{subfigure}[b]{0.47\textwidth}
		\includegraphics[width=\textwidth]{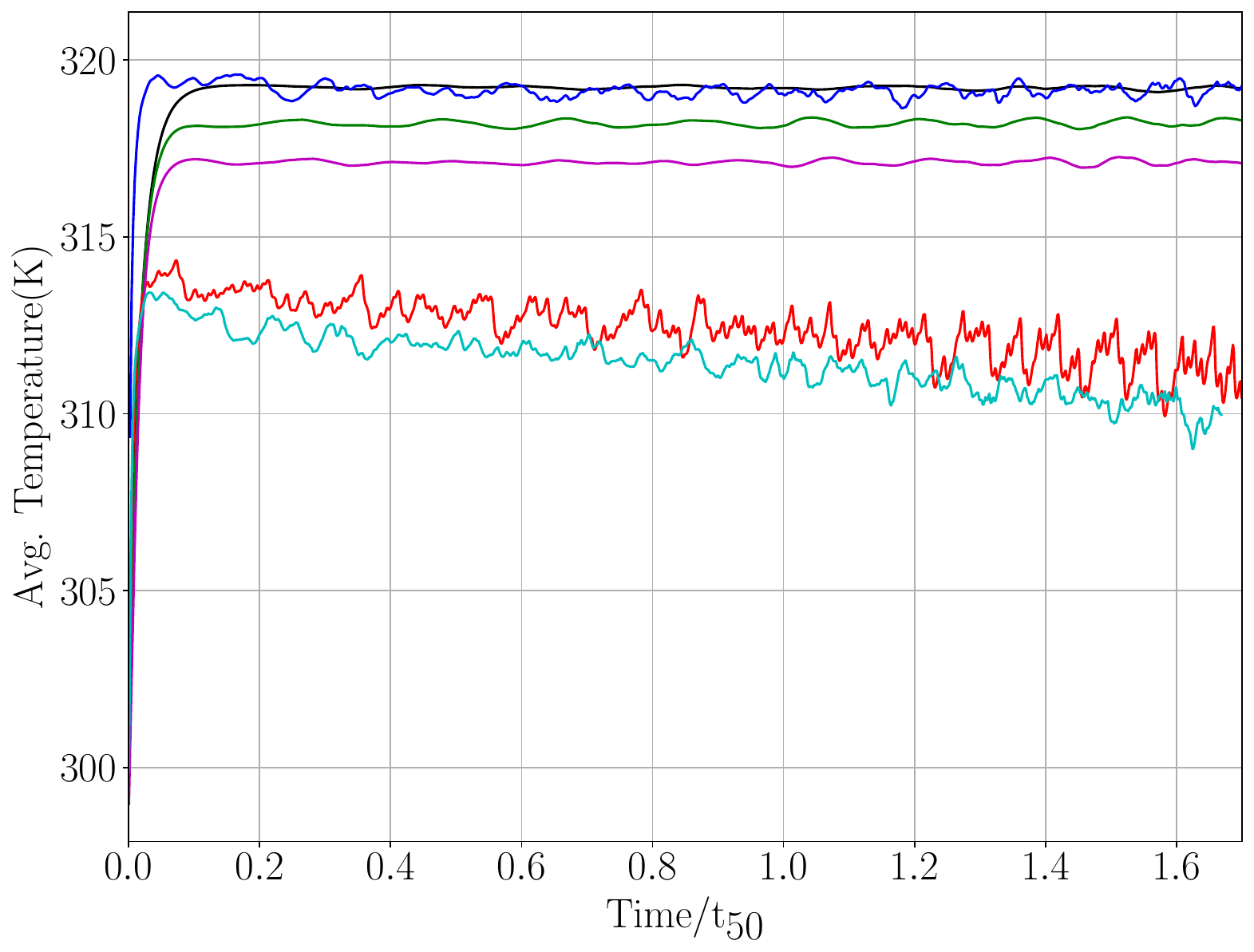}
			    \caption{}
       \label{fig:AvgTemp}
        \end{subfigure}
        \caption{Evolution of the average temperature of the droplet for all the 6 cases. Time is normalized by the time when half of the droplet has evaporated.}
        \label{fig:fig9}
\end{figure}

Further analysis of the average temperature of the droplet accounting for both the bulk and interface temperatures is shown in Figure \ref{fig:fig9}. 
In cases C1 and C2, where the contact angle is $60 ^\circ$, the average temperature reaches almost the same equilibrium temperature and is maintained for the later stages of evaporation. In case C2, the equilibrium temperature is reached earlier than in the case without Marangoni stress, case C1, due to effective thermal mixing.
For cases in which Marangoni stresses are absent and where the contact angles are $90 ^\circ$ and $120 ^\circ$, the droplet reaches an equilibrium temperature and maintains it for the later stages of the evaporation process.
For cases without Marangoni stresses, droplets with larger contact angles exhibit smaller average temperatures, owing to a higher thermal resistance and a smaller solid-liquid contact area.
For cases with Marangoni stresses and contact angles $90 ^\circ$ and $120 ^\circ$, the average temperature of the droplet decreased over time, with case C6 having a comparatively lower average temperature than case C4. This is due to the fact that as the evaporation progresses, the solid-liquid contact area decreases. Thus, evaporative cooling dominates for droplets with contact angles $\theta_c \geq 90 ^\circ$ for cases with Marangoni stresses. 
Similar observations were reported by~\citet{Paul2023}, where the authors numerically showed that for water droplets evaporating in a depinned state with the Marangoni flow, the evaporative cooling dominates for large contact angles, leading to a smaller average temperature of the droplet.

\section{Results of stage 2: Dispersion of particles in the evaporating droplet}
\label{Particles}

In the second simulation stage, multiple simulations with particles are carried out based on the flow fields resulting from the simulations carried out in stage 1.
The location of the particles is analyzed for four different time instances, corresponding to the volume of the droplet left being 90\%, 75\%, 50\% and 25\% of the initial volume of the droplet, for three different contact angles.

\subsection{Contact angle of $\theta_c = 60\,^\circ$}

\begin{figure}
        \centering
                \begin{subfigure}[b]{0.6\textwidth}
                    \includegraphics[clip,trim=1cm 2cm 0cm 0cm,width=\textwidth]{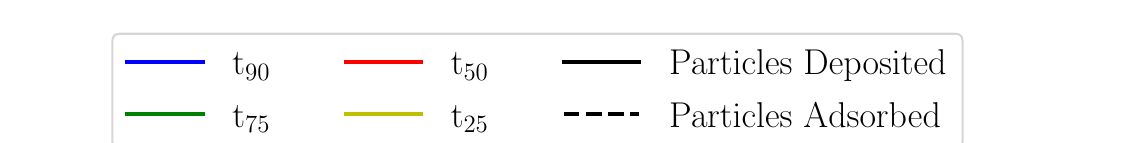}
           \end{subfigure}
           \\
       \begin{subfigure}{0.45\textwidth}
          \centerline{\includegraphics[clip,trim = 0 0cm 0 0, height=0.7\textwidth]{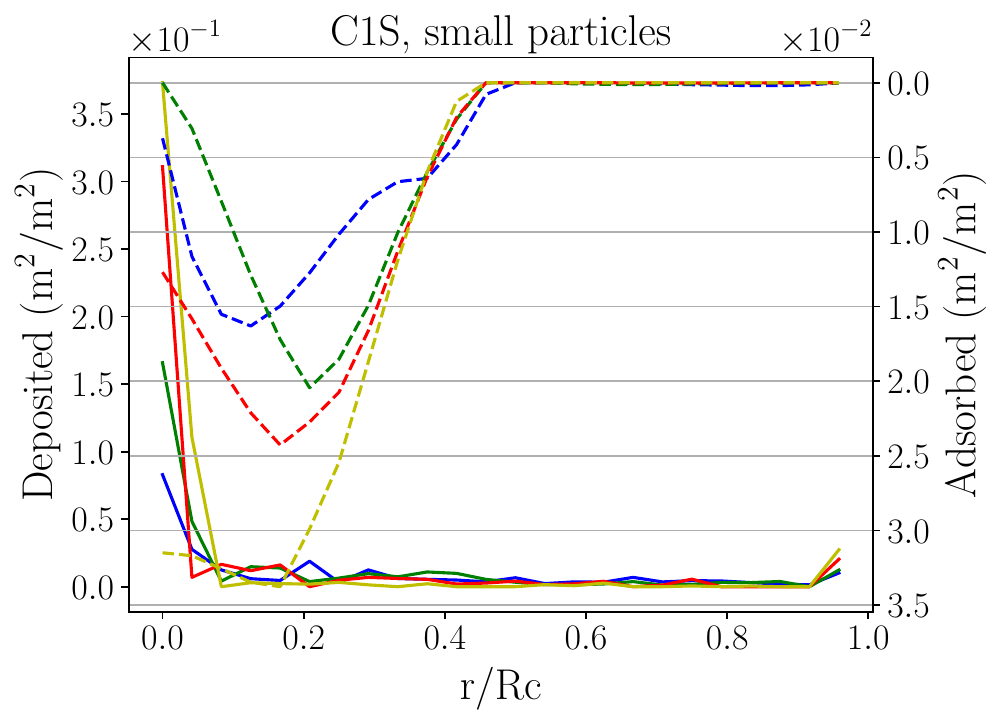}}
                  \caption{}
         \label{fig:C1ST1}
         \end{subfigure}
         \begin{subfigure}{0.45\textwidth}
                \centerline{\includegraphics[clip,trim =0 0cm 0 0,height=0.7\textwidth]{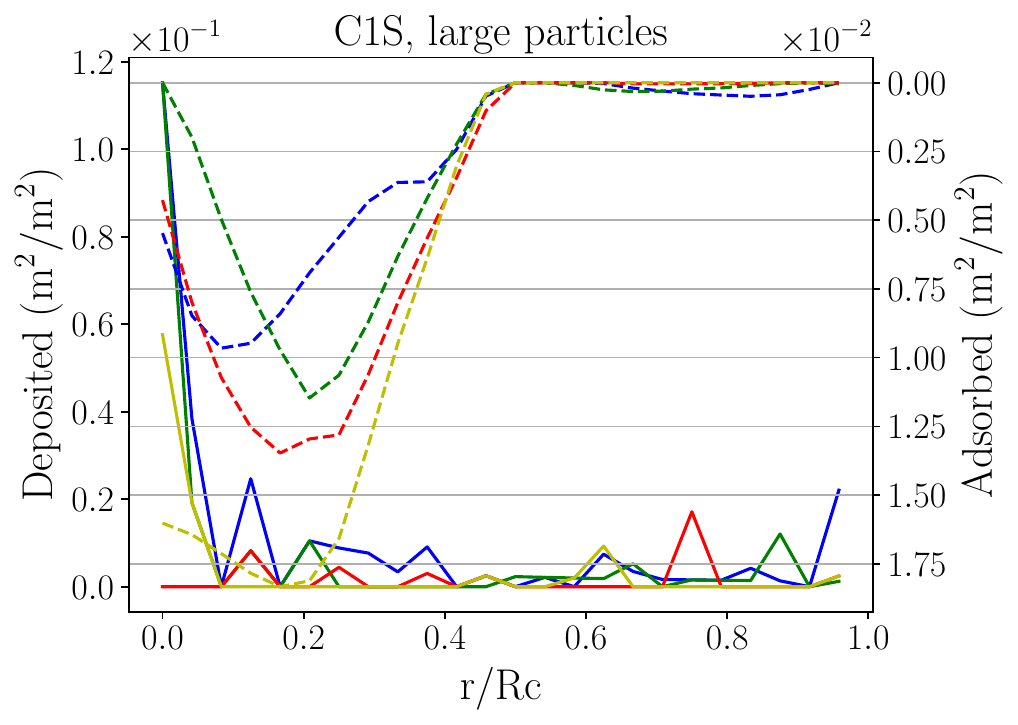}}
                  \caption{}
         \label{fig:C1ST2}
          \end{subfigure}
               \begin{subfigure}{0.45\textwidth}
                \centerline{\includegraphics[clip,trim =0 0cm 0 0,height=0.7\textwidth]{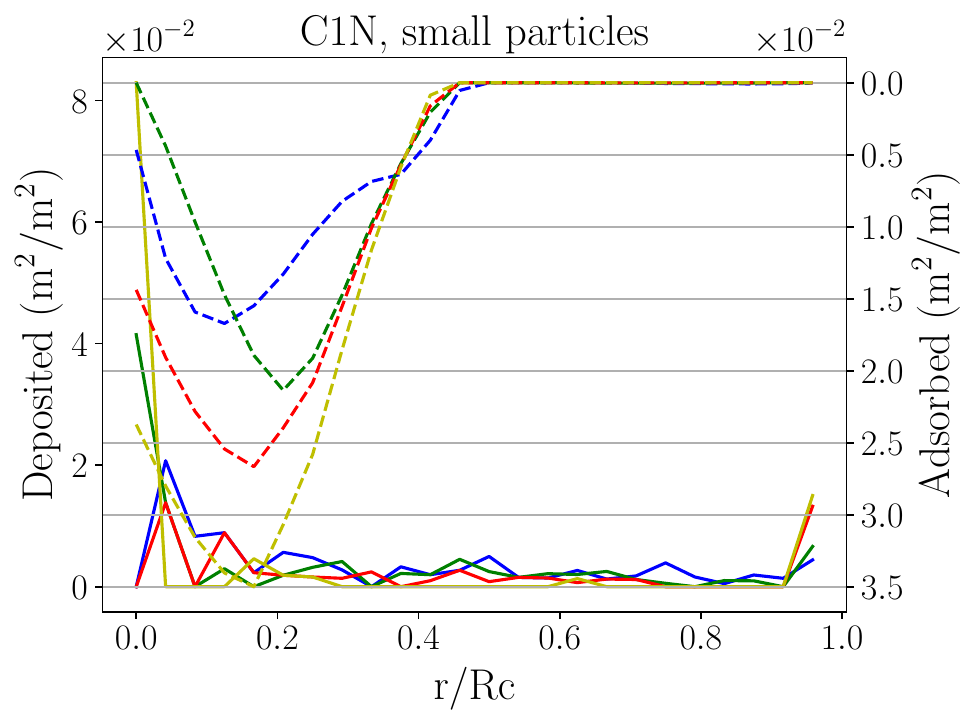}}
                  \caption{}
         \label{fig:C1NT1}
         \end{subfigure}
         \begin{subfigure}{0.45\textwidth}
                \centerline{\includegraphics[clip,trim =0 0cm 0 0,height=0.7\textwidth]{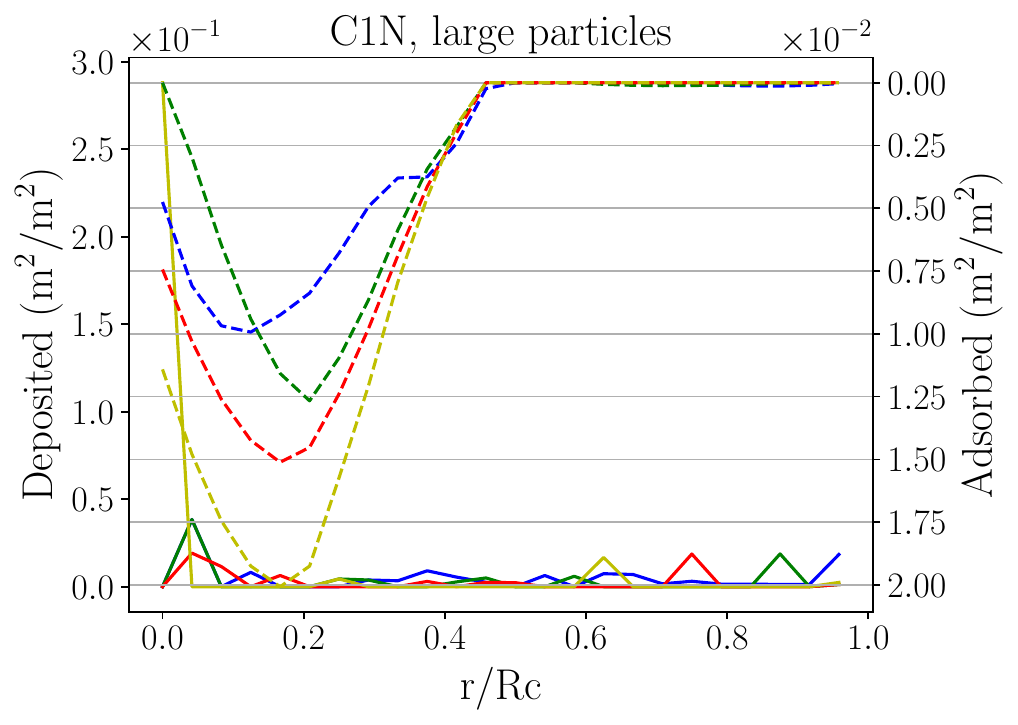}}
                  \caption{}
         \label{fig:C1NT2}
          \end{subfigure}
                  \caption{Evolution of surface concentration of particles adsorbed at the interface (dashed lines, right $y$-axes) and deposited at the substrate (solid lines, left $y$-axes), for cases without the Marangoni stresses and the contact angle $\theta_\mathrm{c} = 60\, ^\circ$. \textbf{(a)} Small silica particles of case C1S, \textbf{(b)} large silica particles of case C1S, \textbf{(c)} small neutrally buoyant particles of case C1N, and \textbf{(d)} large neutrally buoyant particles of case C1N.}
                  \label{fig:fig11}
    \end{figure}

    \begin{figure}
        \centering
                \begin{subfigure}[b]{0.6\textwidth}
                    \includegraphics[clip,trim=1cm 2cm 0cm 0cm,width=\textwidth]{LegendParticles.pdf}
           \end{subfigure}
           \\
       \begin{subfigure}{0.45\textwidth}
        \centerline{\includegraphics[clip,trim = 0 0cm 0 0, height=0.7\textwidth]{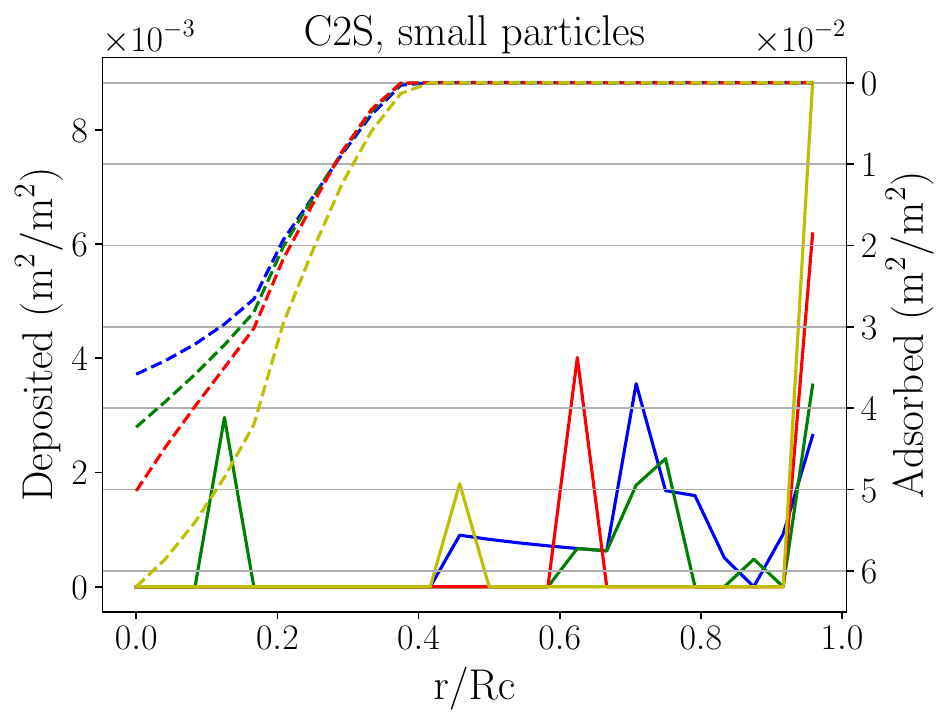}}
                  \caption{}
         \label{fig:C2ST1}
         \end{subfigure}
         \begin{subfigure}{0.45\textwidth}
                \centerline{\includegraphics[clip,trim =0 0cm 0 0,height=0.7\textwidth]{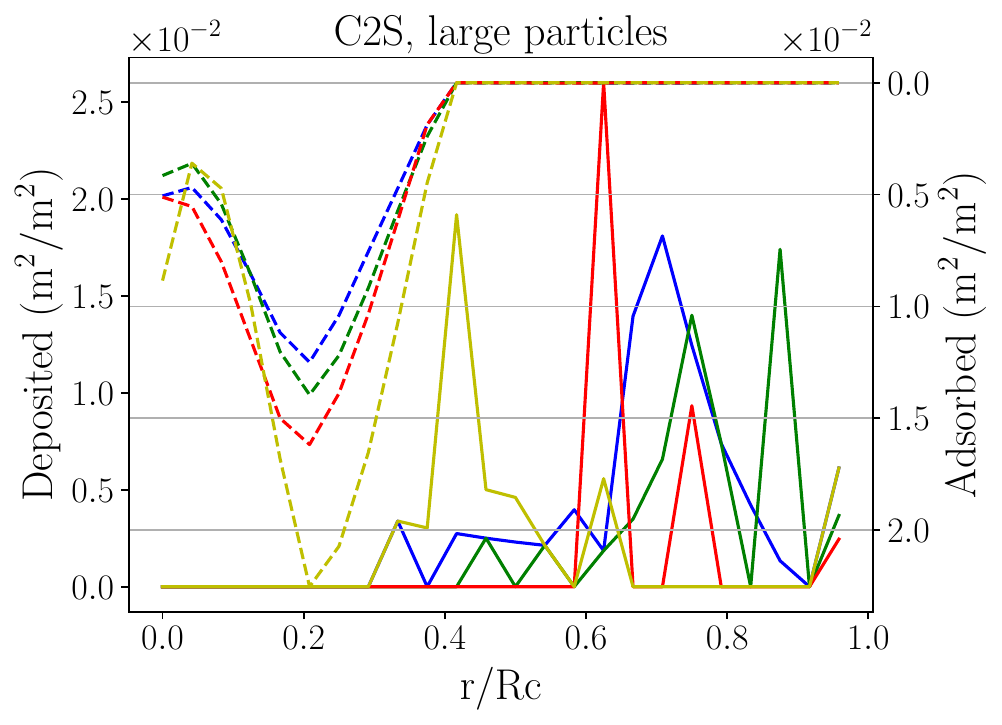}}
                  \caption{}
         \label{fig:C2ST2}
          \end{subfigure}
               \begin{subfigure}{0.45\textwidth}
                \centerline{\includegraphics[clip,trim =0 0cm 0 0,height=0.7\textwidth]{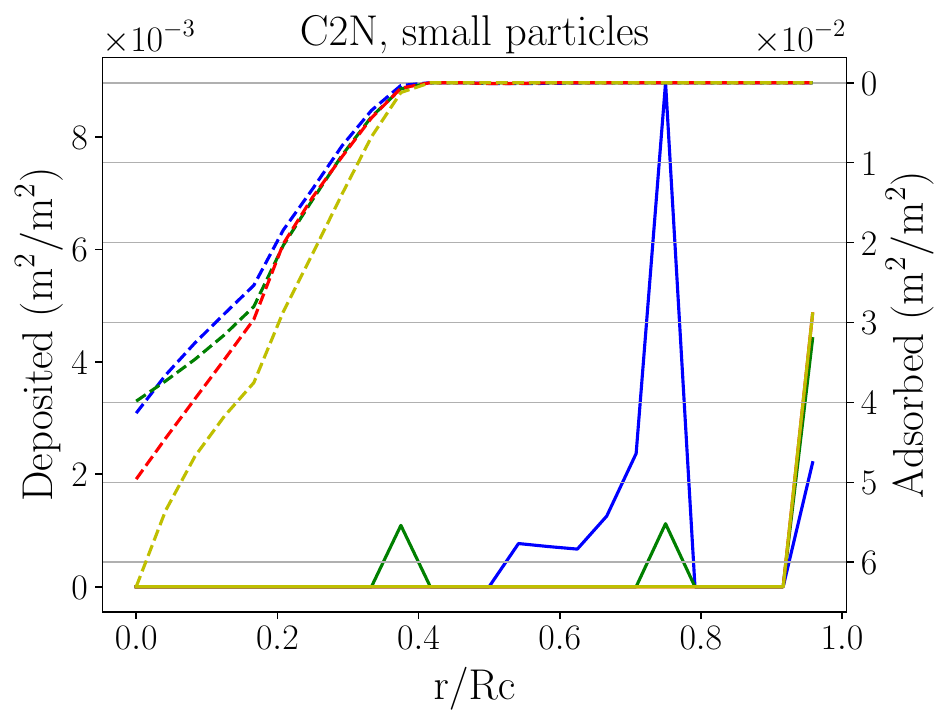}}
                  \caption{}
         \label{fig:C2NT1}
         \end{subfigure}
         \begin{subfigure}{0.45\textwidth}
                \centerline{\includegraphics[clip,trim =0 0cm 0 0,height=0.7\textwidth]{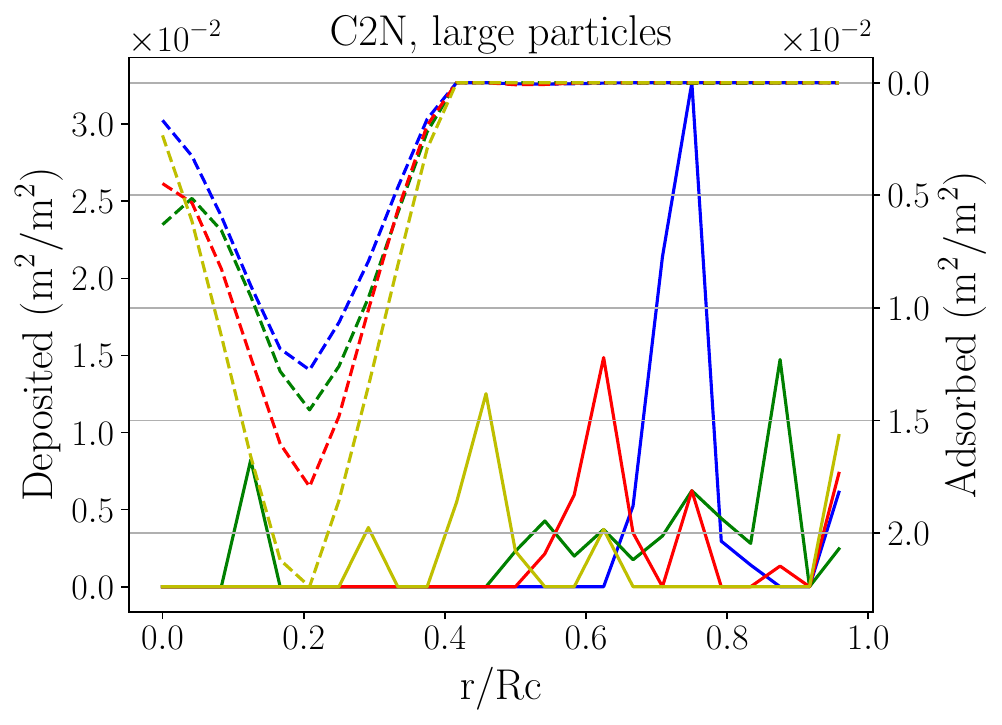}}
                  \caption{}
         \label{fig:C2NT2}
          \end{subfigure}
                  \caption{Evolution of surface concentration of particles adsorbed at the interface (dashed lines, right $y$-axes) and deposited at the substrate (solid lines, left $y$-axes) for cases with the Marangoni stresses and the contact angle $\theta_\mathrm{c} = 60\, ^\circ$. \textbf{(a)} smaller silica particles of case C2S \textbf{(b)} larger silica particles of case C2S \textbf{(c)} smaller neutrally buoyant particles of case C2N \textbf{(d)} larger  neutrally buoyant particles of case C2N.}
                  \label{fig:fig12}
    \end{figure}

In Figure \ref{fig:fig11}, the surface concentrations of the small and the large particles that are adsorbed at the interface or deposited on the substrate are shown for case C1, where the contact angle is $60^\circ$ and Marangoni stresses are absent.
Figures \ref{fig:C1ST1} and \ref{fig:C1ST2} show the surface concentrations of silica particles, Figures \ref{fig:C1NT1} and \ref{fig:C1NT2} show the surface concentrations of neutrally buoyant particles.
% Particles deposited on the substrate are represented by solid lines, and the surface concentration is projected on the left side of the plot.
% Particles adsorbed on the interface are represented by dashed lines, and the surface concentration is projected on the right side of the plot.
% From fig \ref{fig:C1ST1} and \ref{fig:C1ST2}, it is observed that 
Both the small and the large particles occupy the central region of the droplet, whether they are deposited on the substrate or adsorbed at the interface.
For the small particles, the surface concentration of the deposited particles is nearly three times larger than for the larger particles. We attribute this preferential deposition of small particles to the stronger influence of the adhesive van der Waals forces.
In addition, the surface concentration for the small particles adsorbed at the interface is nearly double the surface concentration of the large particles. The concentration of particles of both sizes gradually increases at the droplet apex.
Similar observations can be made for both types of particles with the same density as the fluid, shown in Figures \ref{fig:C1NT1} and \ref{fig:C1NT2}.
The only visible difference is that the surface concentration of the neutrally buoyant particles deposited on the substrate is higher for the large particles than for the small particles.
The positions of both the small and the large particles are at the same location, suggesting that no significant segregation is seen.

Similar plots are shown in Figure \ref{fig:fig12} for the surface concentrations of smaller and larger particles that are adsorbed and deposited in droplet in case C2, where the contact angle is $60^\circ$ and the Marangoni stresses are taken into account. The small particles adsorbed at the interface of the droplet move towards the apex of the droplet forming a core, as observed in Figures  \ref{fig:C2ST1} and \ref{fig:C2ST2}, while the larger particles are adsorbed around the core of the smaller particles.
As the evaporation progresses, the concentration of particles adsorbed at the interface increases, and the segregation of the small and large particles becomes more prominent.
The small particles are deposited on the substrate mostly near the stagnation point, which moves towards the center of the droplet as evaporation progresses. 
The surface concentration of the large particles on the substrate is nearly three times larger than the surface concentration of the small particles, and it can be seen that the majority of the large particles are deposited at the stagnation point, which later moves towards the center of the droplet.
Similar observations can be made for both types of particles with the same density as the fluid of the droplet, as shown in Figures \ref{fig:C2NT1} and \ref{fig:C2NT2}.

Overall, we can summarize that the trend of particle adsorption and deposition is mostly affected by the presence of the Marangoni flow. In the presence of Marangoni flow, we observe a separation of small and large particles. The density of particles relative to the density of the fluid has, however, no significant influence.

\subsection{Contact angle of $\theta_c = 90\,^\circ$}

\begin{figure}
    \centering
            \begin{subfigure}[b]{0.6\textwidth}
		\includegraphics[clip,trim=1cm 2cm 0cm 0cm,width=\textwidth]{LegendParticles.pdf}
       \end{subfigure}
       \\
   \begin{subfigure}{0.45\textwidth}
        \centerline{\includegraphics[clip,trim = 0 0cm 0 0, height=0.7\textwidth]{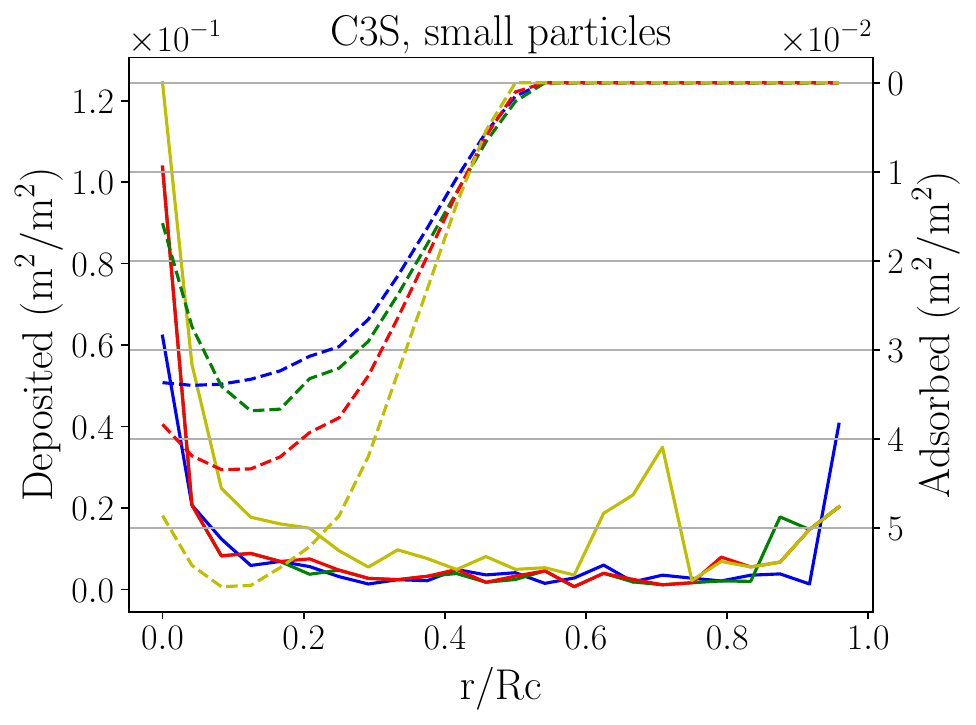}}
              \caption{}
     \label{fig:C3ST1}
     \end{subfigure}
     \begin{subfigure}{0.45\textwidth}
        \centerline{\includegraphics[clip,trim =0 0cm 0 0,height=0.7\textwidth]{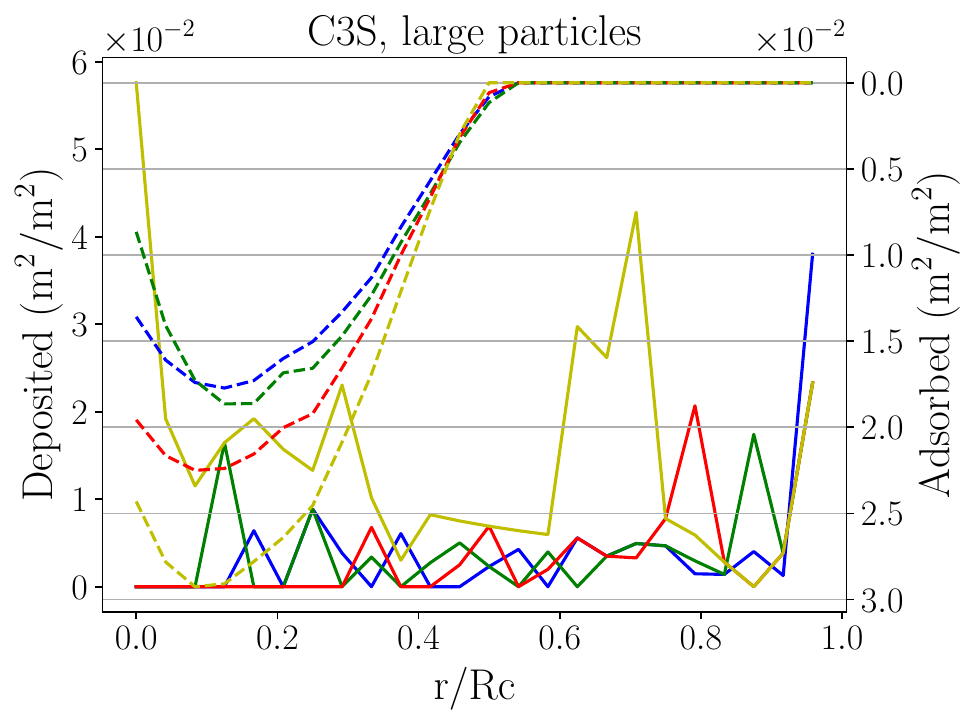}}
              \caption{}
     \label{fig:C3ST2}
      \end{subfigure}
           \begin{subfigure}{0.45\textwidth}
                \centerline{\includegraphics[clip,trim =0 0cm 0 0,height=0.7\textwidth]{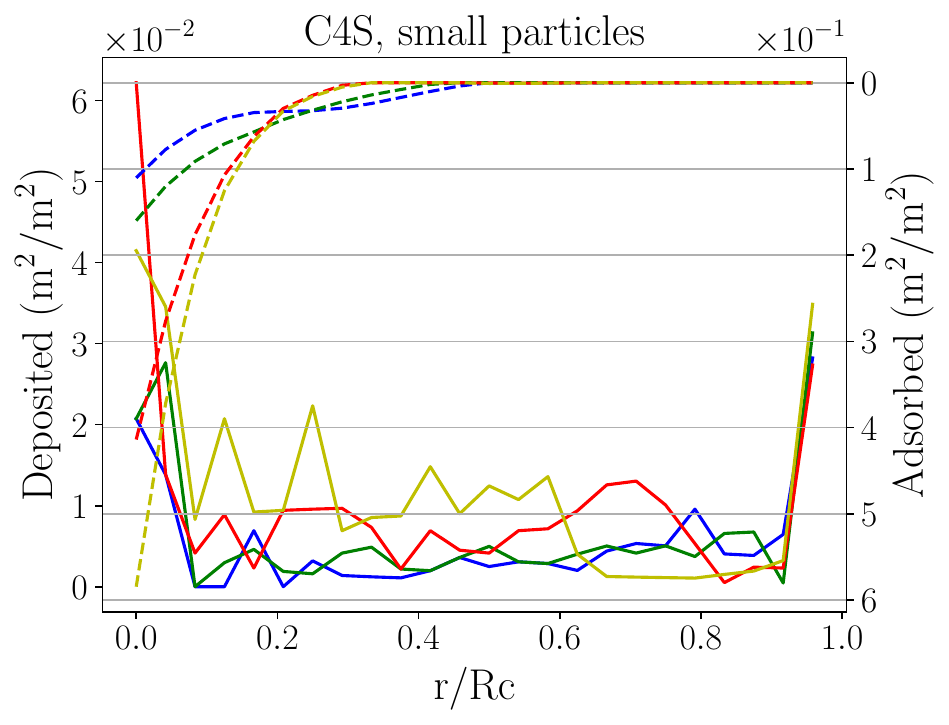}}
              \caption{}
     \label{fig:C4ST1}
     \end{subfigure}
     \begin{subfigure}{0.45\textwidth}
        \centerline{\includegraphics[clip,trim =0 0cm 0 0,height=0.7\textwidth]{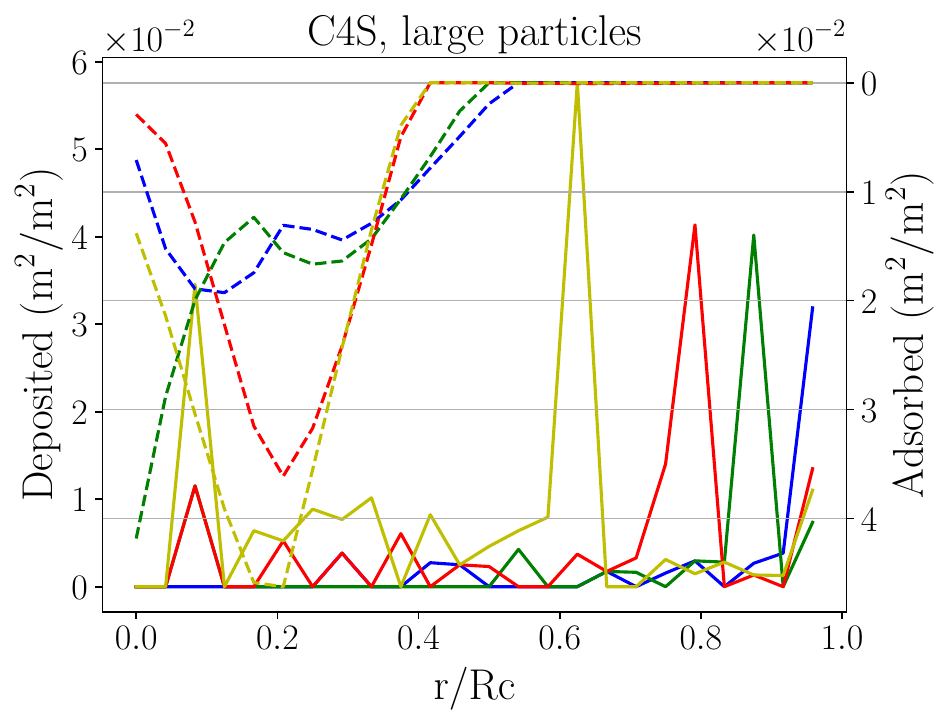}}
              \caption{}
     \label{fig:C4ST2}
      \end{subfigure}
              \caption{Evolution of surface concentration of particles adsorbed at the interface (dashed lines, right $y$-axes) and deposited at the substrate (solid lines, left $y$-axes), for cases with the contact angle $\theta_\mathrm{c} = 90\, ^\circ$. \textbf{(a)} Small silica particles of case C3S, \textbf{(b)} large silica particles of case C3S, \textbf{(c)} small neutrally buoyant particles of case C4S, and \textbf{(d)} large neutrally buoyant particles of case C4S.}
              \label{fig:fig13}
\end{figure}

Figure \ref{fig:fig13} shows the results of the deposition and adsorption of the silica particle for cases C3 and C4, where the contact angle is $90^\circ$ and Marangoni stresses are absent and present, respectively.
Comparing the plots in Figures \ref{fig:C3ST1} and \ref{fig:C3ST2} for case C3, where the Marangoni flow is absent and the fluid flows from the contact line to the apex of the droplet, both the small and the large particles are adsorbed at the apex of the droplet, where the surface concentration gradually increases. For the small particles that are deposited on the substrate, the surface concentration is higher at the center of the droplet for all four shown time instances. On the other hand, the large particles are only deposited near the contact line, which gradually moves towards the center of the droplet.

Figures \ref{fig:C4ST1} and \ref{fig:C4ST2} shows the positions of the particles for case C4, where the Marangoni flow is present. 
A arrangement of particles similar to the one seen for case C2 can be observed in Figures \ref{fig:C4ST1} and \ref{fig:C4ST2} for case C4: The small particles adsorbed at the apex of the droplet form a core and the large particles adsorbed at the interface surround the core of the small particles. The surface concentration of the adsorbed particles gradually increases as the evaporation proceeds. For the deposited particles, small particles are mainly found at the center of the droplet. In contrast, the large particles primarily occupy the region near the contact line and a few large particles surround the core of the small particles deposited on the substrate.

Overall, we see a similar scenario for cases C3 and C4 as for cases C1 and C2, where the Marangoni flow has a significant effect on the segregation of the small and large particles, with small particles moving towards the center, where they form a core of adsorbed and deposited particles, and the large particles surround them.

\subsection{Self-sorting mechanisms}
\label{Mechanism}
\begin{figure}
      \centering
     \begin{subfigure}[b]{0.48\textwidth}
		\includegraphics[width=\textwidth]{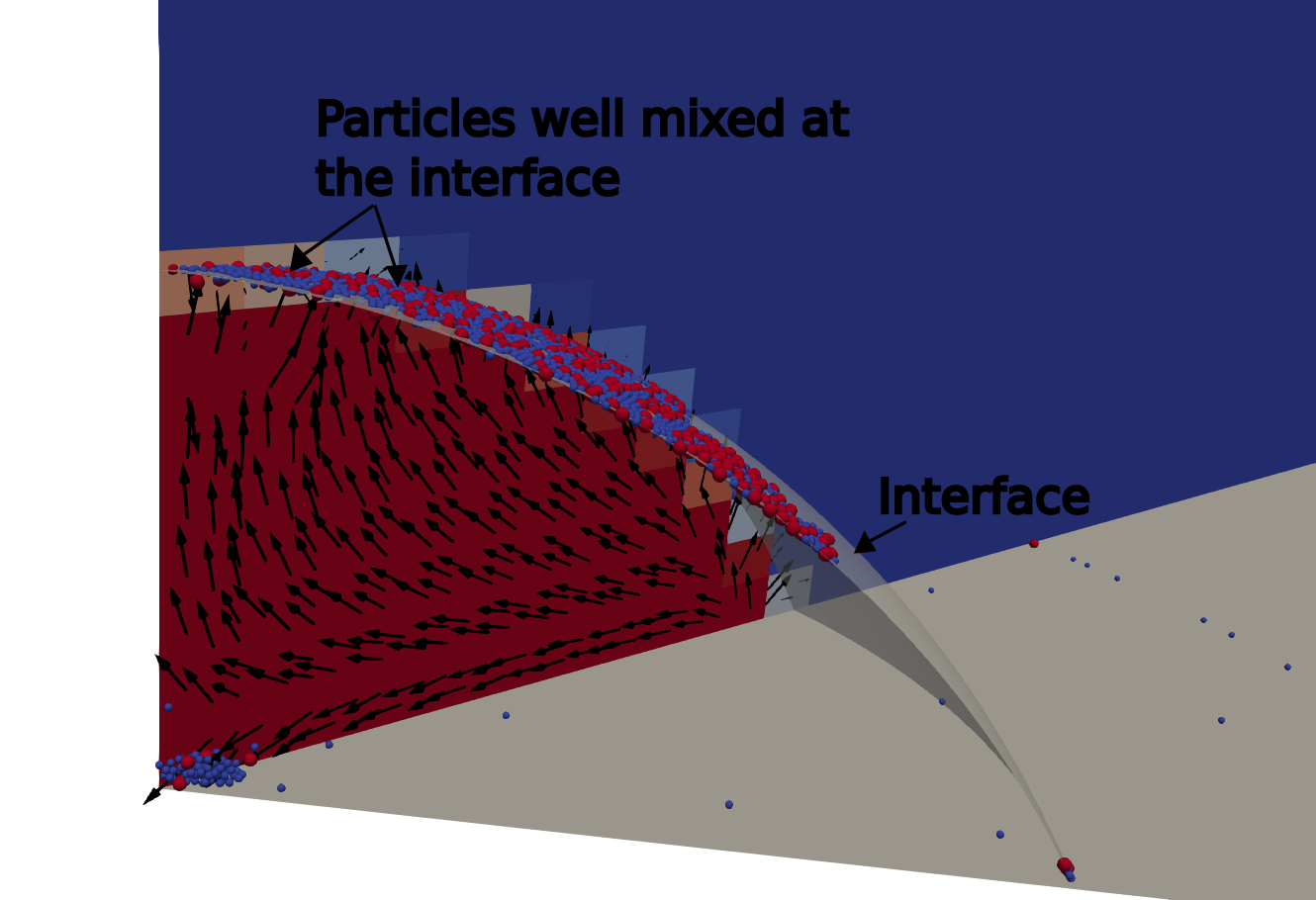}
			    \caption{}
       \label{fig:Mechanism1}
       \end{subfigure}
       \begin{subfigure}[b]{0.48\textwidth}
		\includegraphics[width=\textwidth]{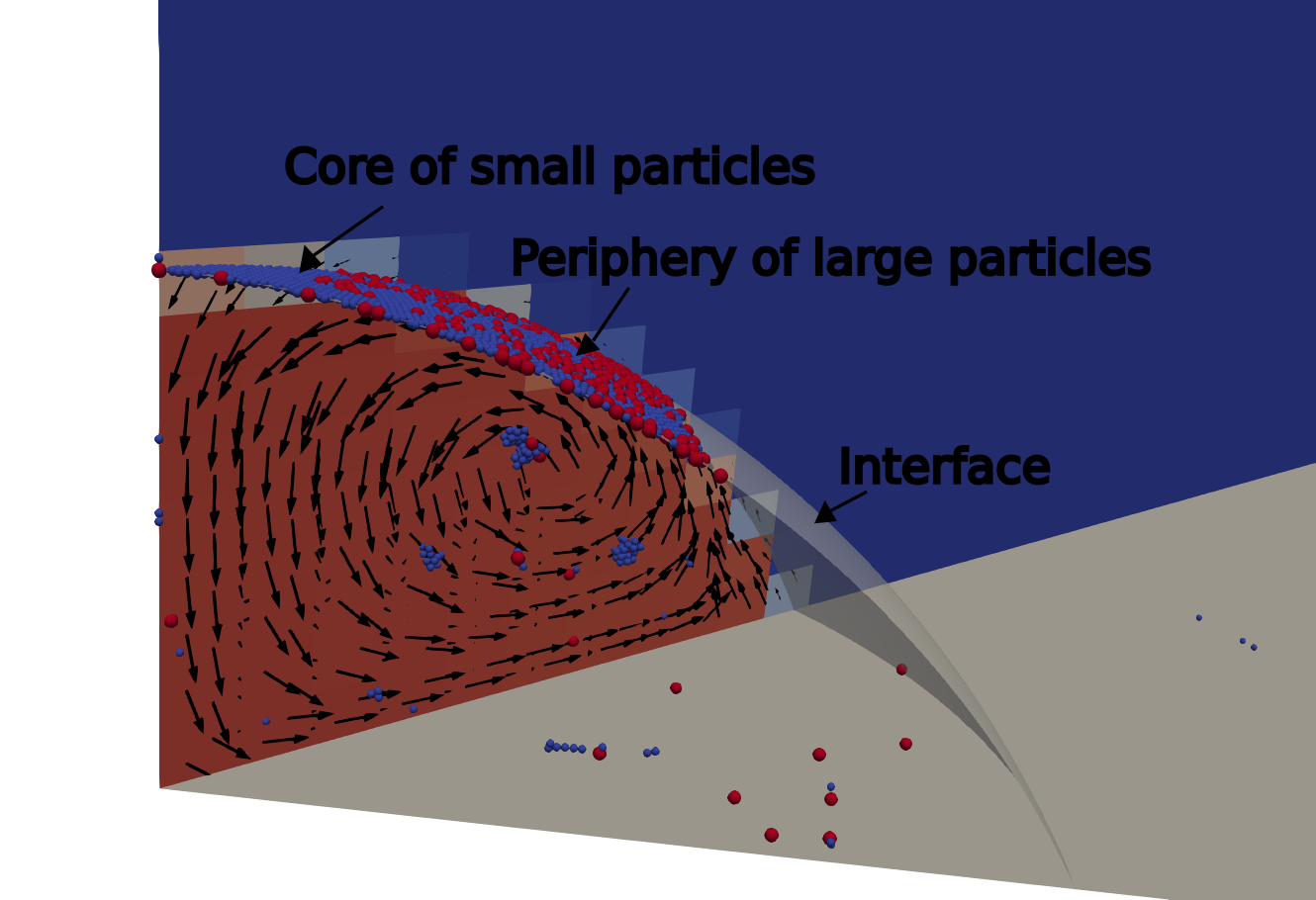}
			    \caption{}
       \label{fig:Mechanism2}
        \end{subfigure}
        \caption{Dispersion of particles in an evaporating ethanol droplet with contact angle $\theta_r = 60 \, ^\circ$, where Marangoni stresses are \textbf{(a)} neglected and \textbf{(b)} considered.}
\end{figure}

In most cases considered in this study, the particles classify at the interface; the smaller particles are sorted from the larger ones.
This self-sorting of the particles in an evaporating sessile droplet is governed by the flow inside the droplet.
For a sessile droplet with a moving contact line, three types of flow together form the resulting fluid flow: (i) the flow induced by the moving contact line, (ii) the capillary flow, generated by evaporation, and (iii) the Marangoni flow, generated by the Marangoni stresses associated with a local change in surface tension due to the temperature gradient along the interface.
While the capillary flow moves fluid from the apex of the droplet to the contact line, where the evaporation rate is typically highest, the motion of the receding contact line drives fluid from the contact line to the apex.
% Marangoni flow is generated along the interface owing to the surface tension gradient resulting from the thermal gradient along the interface. Capillary flow is generated by the evaporative flux distribution moving the fluid from the apex of the droplet to the contact line for a sessile droplet evaporating on a heated substrate.
% The flow induced by the receding motion of the contact line moves the fluid from the contact line to the apex of the droplet.
For the considered evaporating ethanol droplets, the flow induced by the receding motion of the contact line supersedes the capillary flow, leading to a resultant fluid flow from the contact line to the apex of the droplet in the absence of Marangoni stresses, as is observed in Figure~\ref{fig:Mechanism1}.
In contrast, in the presence of Marangoni stresses, the Marangoni flow dominates over the other two flows, resulting in a classical single vortex flow, as seen in Figure~\ref{fig:Mechanism2}.

In Figure \ref{fig:Mechanism1}, the dispersion of particles in case C1 is shown, where the contact angle \(\theta_c = 60^\circ\) and Marangoni flow is neglected. 
% The figure shows the streamlines of the velocity inside the droplet with the liquid volume fraction represented on a plane equidistant to the x and y axes.
The fluid flow induced by the receding motion of the contact line moves the particles from the bulk of the droplet to the apex of the droplet, where the particles adsorb to the interface. Since the flow in this droplet is oriented predominantly in the direction normal to the interface, the adsorbed particles retain their position at the interface. Hence, there is no mechanism present that promotes a size-dependent sorting of the particles adsorbed at the interface.
In contrast, if Marangoni stresses are considered, the resulting flow along the interface promotes particle interactions, which in turn creates gaps between the particles that are preferentially filled by the small particles, see Figure \ref{fig:Mechanism2}. As a result, the small particles agglomerate near the stagnation point of the flow at the apex of the droplet, with the large particles gathering around this agglomeration of small particles. We can, therefore, conclude that the dominant flow contribution associated with the thermocapillary Marangoni stresses is responsible for the size-based sorting of particles adsorbed at the gas-liquid interface.

\section{Conclusions}
\label{conclusions}

In this study, we have investigated the behavior of bidisperse particles in evaporating sessile droplets with a moving contact line. To this end, we have used a finite-volume method to model the gas-liquid fluid flow, in conjunction with Lagrangian particle tracking to model the behavior of the particles. Our study has focused on the influence of thermocapillary Marangoni stresses and the contact angle of the droplet on the fluid flow and the particle dispersion inside an ethanol droplet evaporating on a heated substrate.

Several interesting conclusions can be drawn from the results of this study.
In the absence of Marangoni flow, the flow induced by the receding contact line moves the fluid from the contact line to the apex of the droplet and dominates the flow to the contact line driven by evaporation, irrespective of the contact angle of the droplet.
The presence of Marangoni flow promotes evaporation when the contact angle is less than $90^\circ$, due to thermal convective mixing and low thermal resistance. However, when the contact angle exceeds $90^\circ$, the impact of the Marangoni flow decreases, resulting in a slower evaporation rate. This happens because of the increased thermal resistance, which is caused by the tall shape of the droplet and the decrease in the liquid-solid contact area, leading to a diminished effective thermal conductivity.
The Marangoni flow and the flow induced by the receding contact line both move the particles towards the apex of the droplet, although along different paths.
With respect to the particles, the flow originating from Marangoni stresses has been identified as the mechanism driving the self-sorting of the bidisperse particles adsorbed at the interface.
As a result of this self-sorting process, the small particles agglomerate at the apex of the droplet and the large particles surround this agglomeration of small particles.

These findings provide insights into controlling the dispersion and agglomeration of particles in evaporating sessile droplets by manipulating the surface wettability, heating conditions, and particle characteristics. This understanding has potential applications in fields such as inkjet printing, microfabrication, and coatings, where uniform or patterned particle deposition is desired.

\section*{Data Availability Statement}
\noindent The data that support the findings of this study are reproducible and data is openly available in the repository with DOI 
10.5281/zenodo.15114147, available at \href{https://doi.org/10.5281/zenodo.15114147}{https://doi.org/10.5281/zenodo.15114147}
\section*{Acknowledgments}

\noindent This research was funded by the Deutsche Forschungsgemeinschaft (DFG, German Research Foundation), grant number 452916560.

\section*{Appendix A. Mesh-independence study}

\begin{figure}
    \centering
		\includegraphics[width=0.45\textwidth]{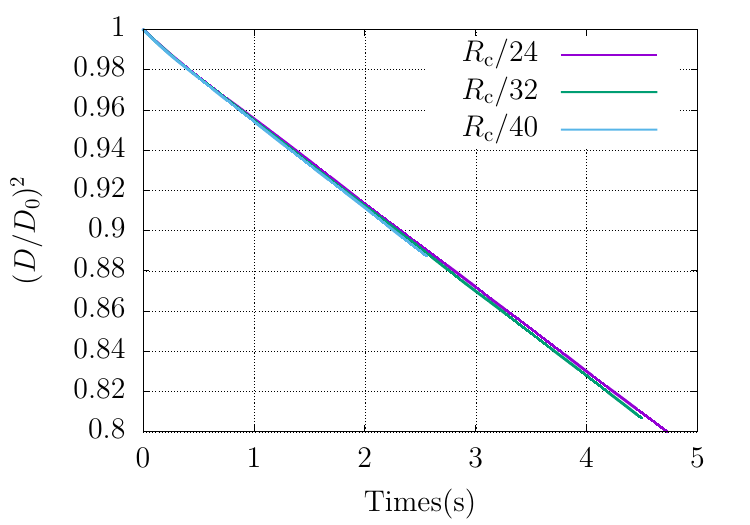}
			    \caption{Volume evolution for sessile droplet in case C4 with Marangoni flow and contact angle $\theta_c = 90^{\circ}$ for three different resolutions.}
       \label{fig:C4VolumeValidation}
\end{figure}

\begin{figure}
        \centering
       \begin{subfigure}{0.45\textwidth}
          \includegraphics[width=\textwidth]{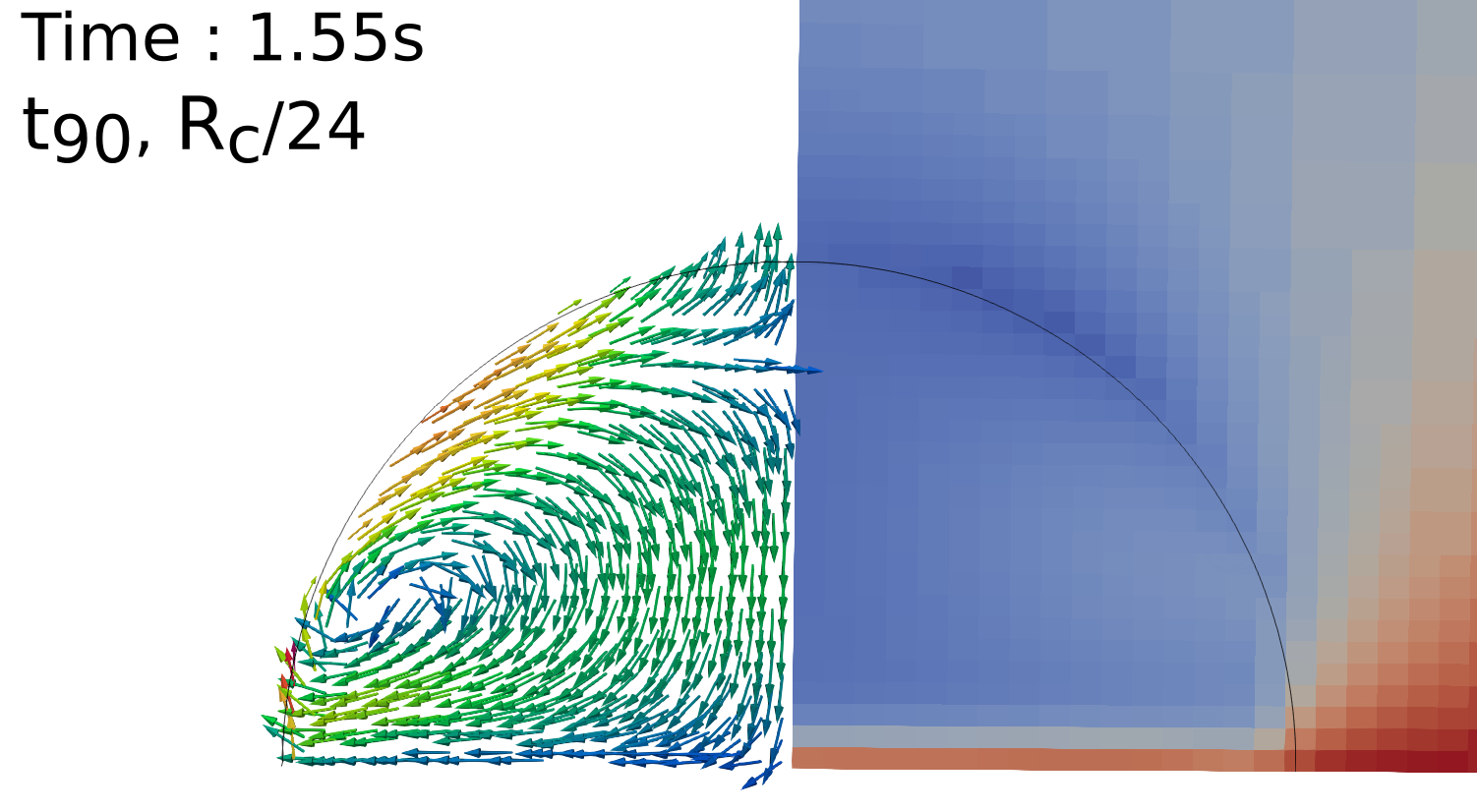}
                  \caption{}
         \label{fig:R24t90}
         \end{subfigure}
         \begin{subfigure}{0.45\textwidth}
          \includegraphics[width=\textwidth]{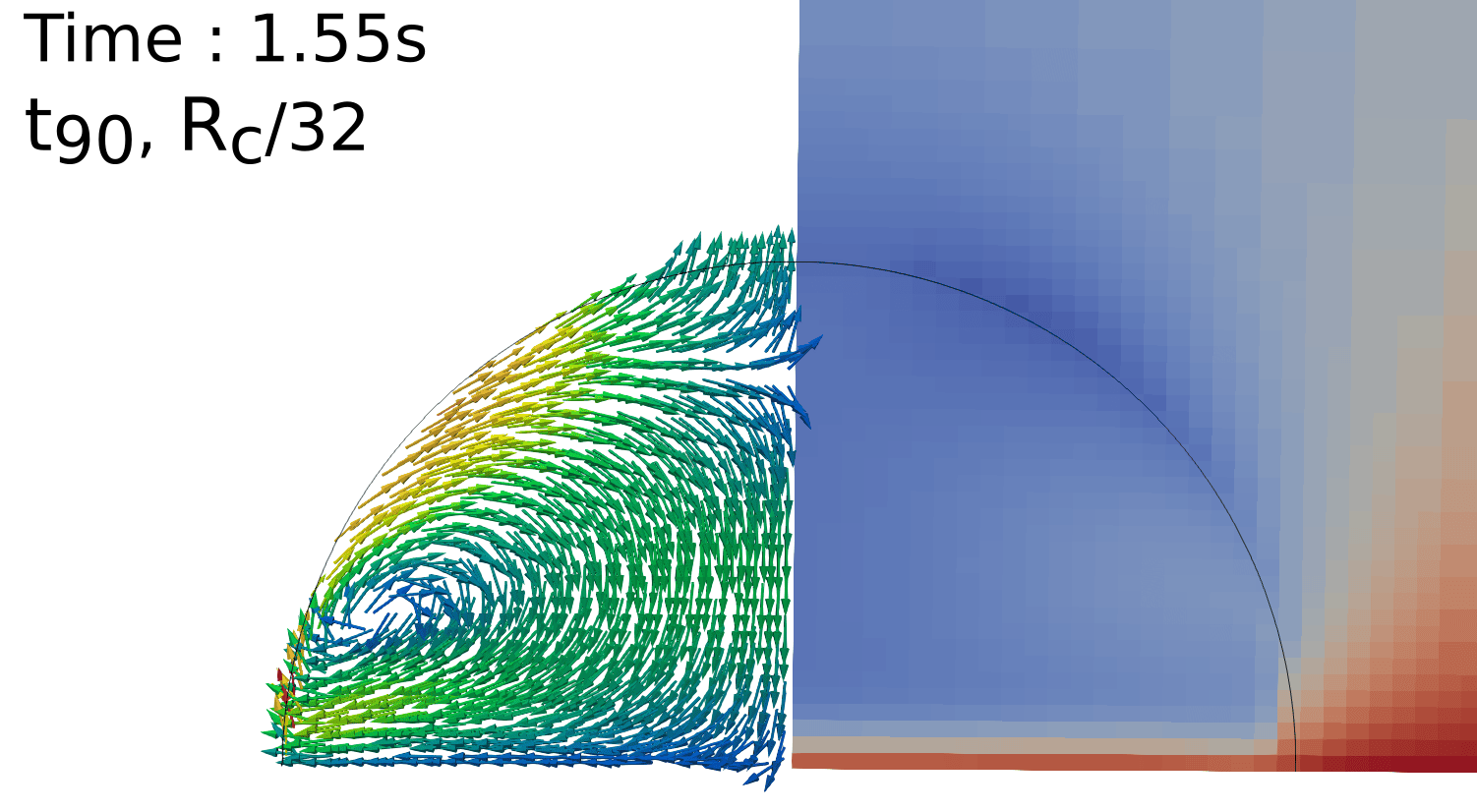}
                  \caption{}
         \label{fig:R32t90}
          \end{subfigure}
          \\
               \begin{subfigure}{0.45\textwidth}
          \includegraphics[width=\textwidth]{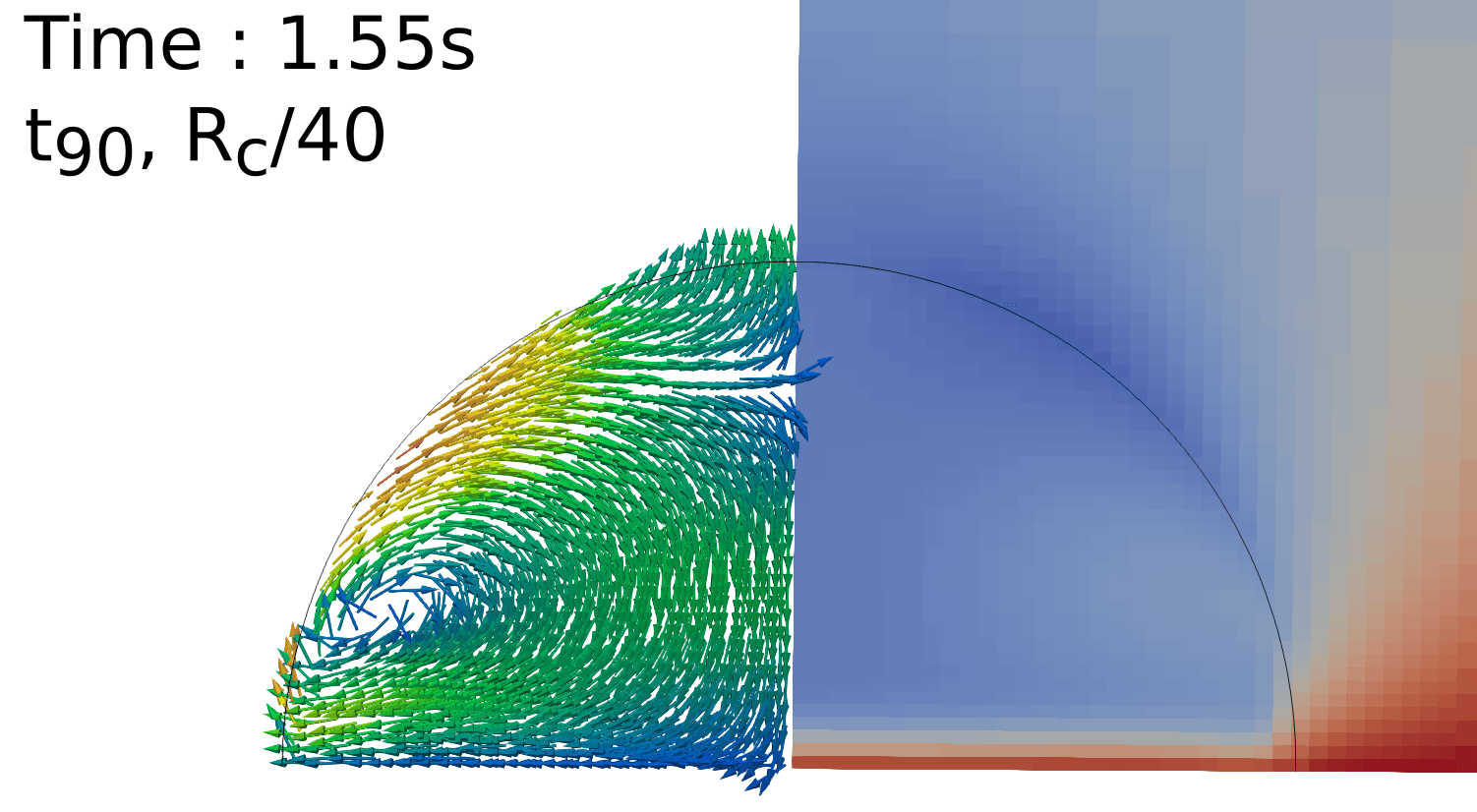}
                  \caption{}
         \label{fig:R40t90}
         \end{subfigure}
        \\
        \begin{subfigure}{0.5\textwidth}
         \includegraphics[clip,trim =10cm 5cm 10cm 15cm,width=\textwidth]{C4ColorMap.png}
         \end{subfigure}
    
                  \caption{Velocity and temperature plots for the cases including the Marangoni stresses, C4, where $\theta_\mathrm{c}=90\, ^\circ$, with different mesh resolutions. The left side of each figure shows the velocity vectors, of which the lengths are scaled with the color function and the color of the vectors shows their magnitude. The right side of each figure shows the temperature. \textbf{(a)} C4, $R_c/24$, $t_{90}$,\textbf{(b)} C4, $R_c/32$, $t_{90}$,\textbf{(c)} C4, $R_c/40$, $t_{90}$}
                  \label{fig:fig1}
    \end{figure}

A mesh-independence study for case C4 with Marangoni flow, substrate temperature $T_s = 50 \, ^{\circ}$C  and contact angle $\theta_c = 90^\circ$ is carried out, by resolving the initial contact radius by 24, 32, and 40 cells.
The results of the evolution of the droplet volume for the three different resolutions, shown in Figure~\ref{fig:C4VolumeValidation}, exhibit excellent agreement, indicating that the evaporation dynamics are mesh independent. In addition, Figure~\ref{fig:fig1} shows the velocity and temperature distribution inside the droplet. Both the velocity and temperature are in qualitatively very good agreement for the three mesh resolutions.

\section*{Appendix B. Slip length}

\begin{figure}
        \centering
         \begin{subfigure}{0.45\textwidth}
          \includegraphics[width=\textwidth]{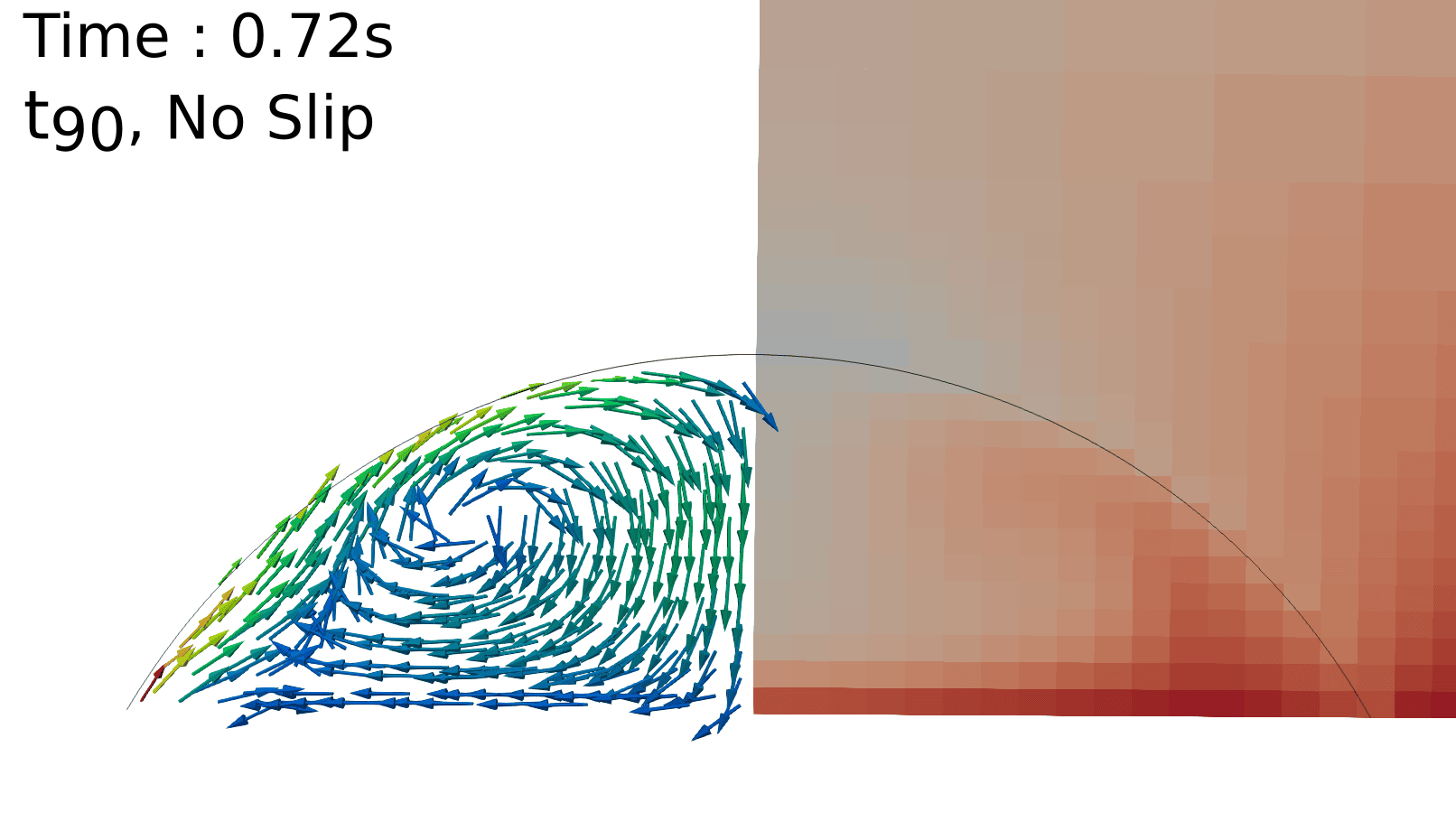}
                  \caption{}
         \label{fig:T1NoSlip}
          \end{subfigure}
        \begin{subfigure}{0.45\textwidth}
          \includegraphics[width=\textwidth]{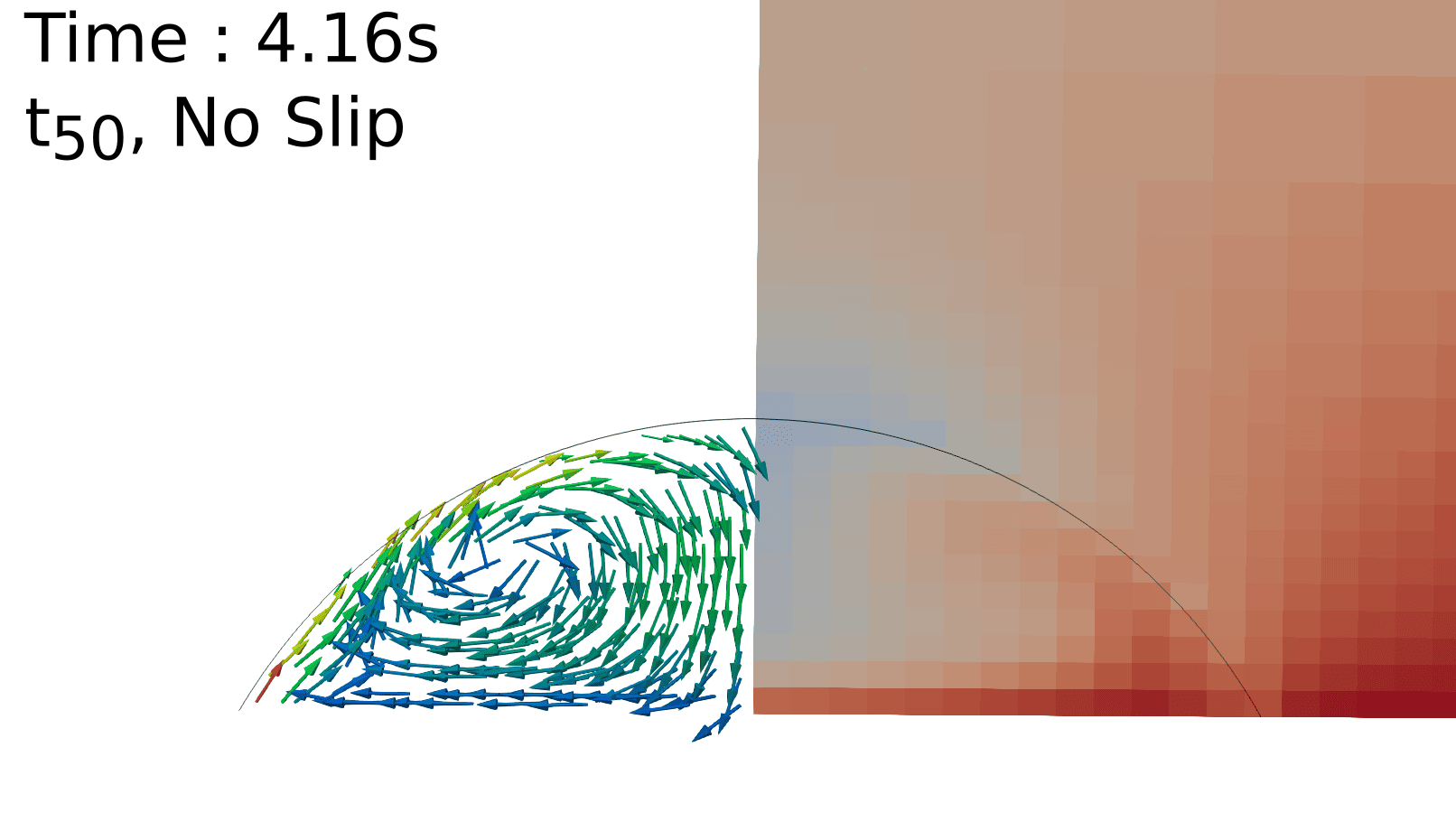}
                  \caption{}
         \label{fig:T2NoSlip}
         \end{subfigure}
             \begin{subfigure}{0.45\textwidth}
          \includegraphics[width=\textwidth]{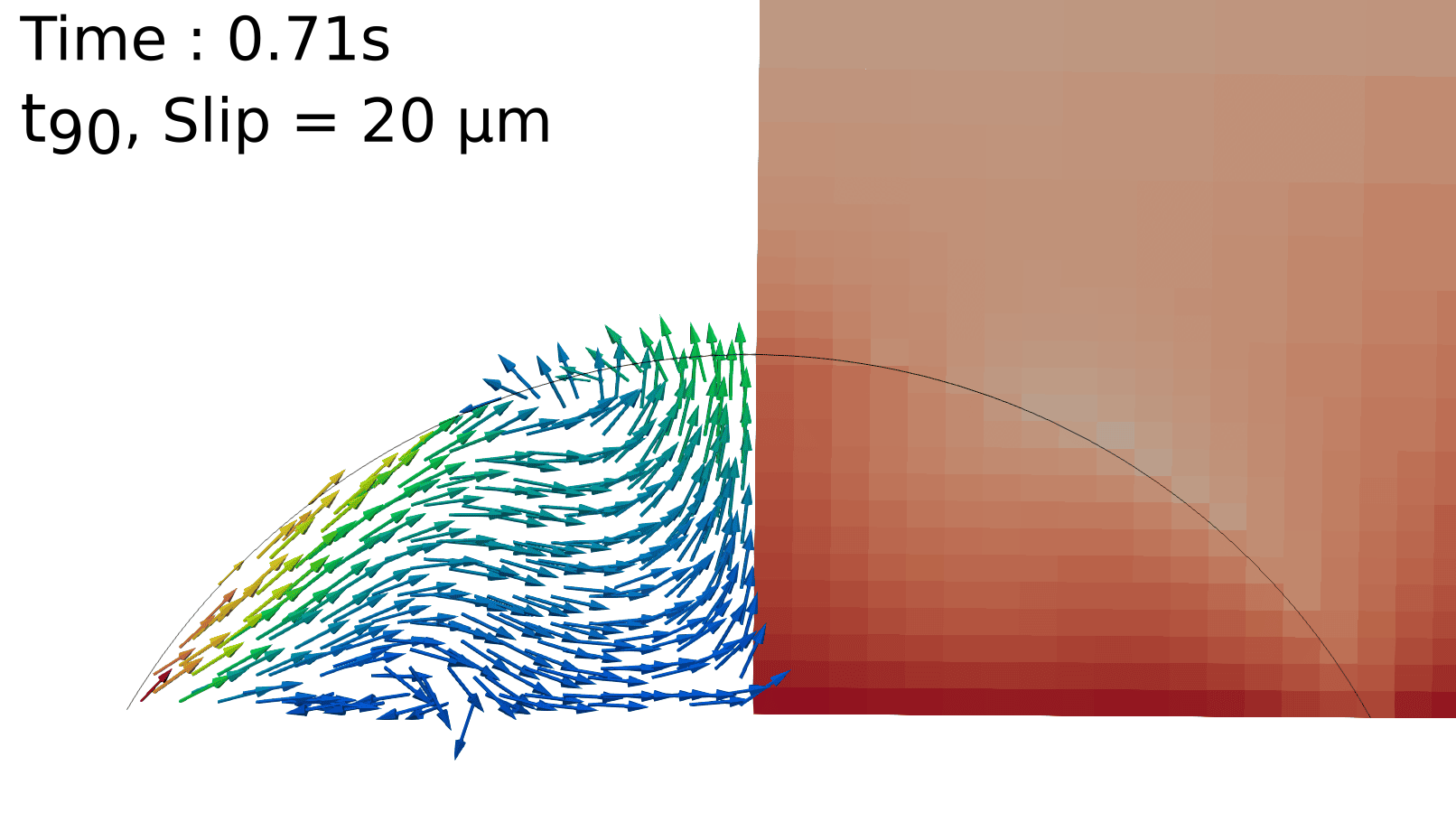}
                  \caption{}
         \label{fig:T1Slip2}
         \end{subfigure}
             \begin{subfigure}{0.45\textwidth}
          \includegraphics[width=\textwidth]{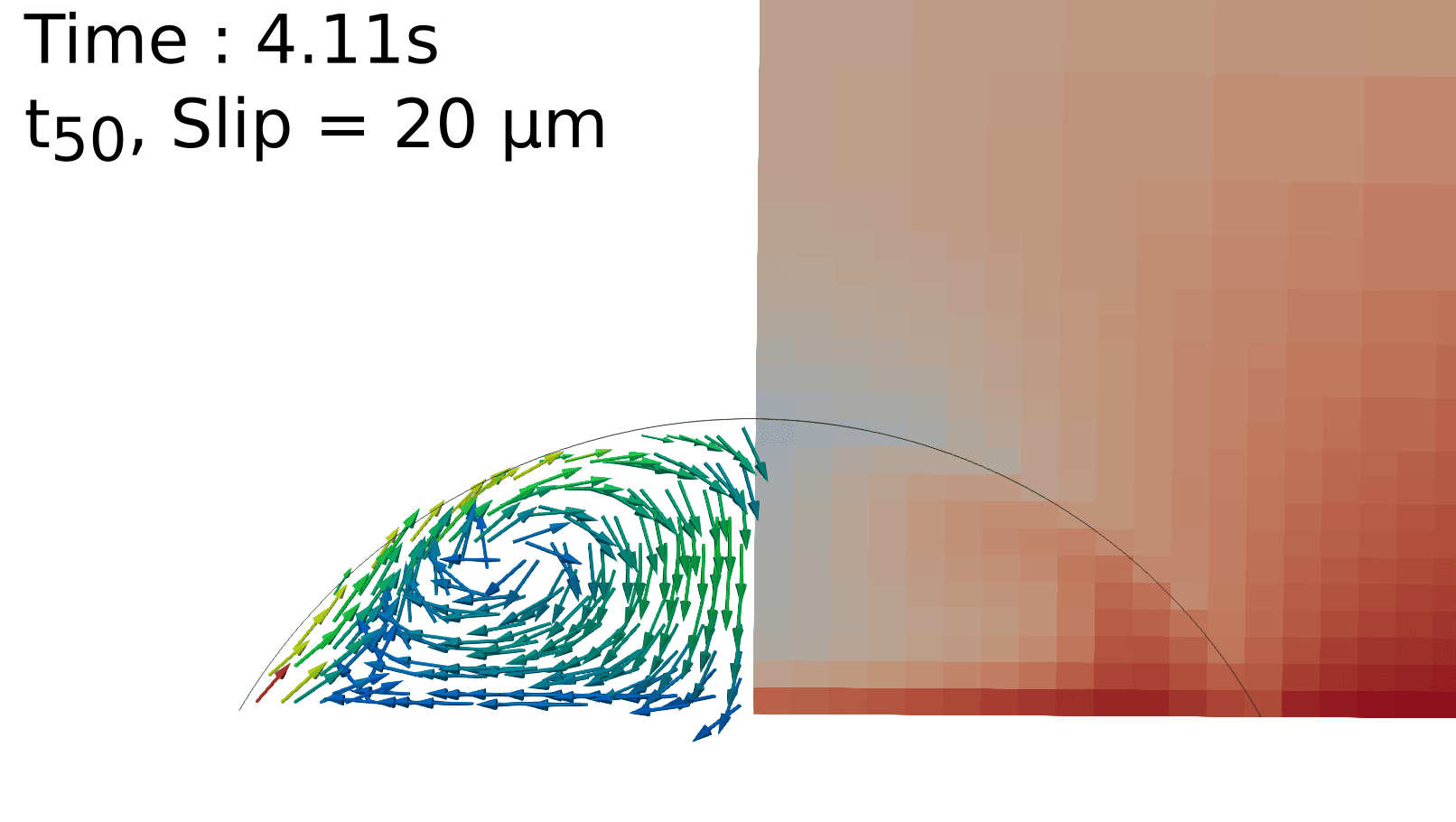}
                  \caption{}
         \label{fig:T2Slip2}
         \end{subfigure}
             \begin{subfigure}{0.45\textwidth}
          \includegraphics[width=\textwidth]{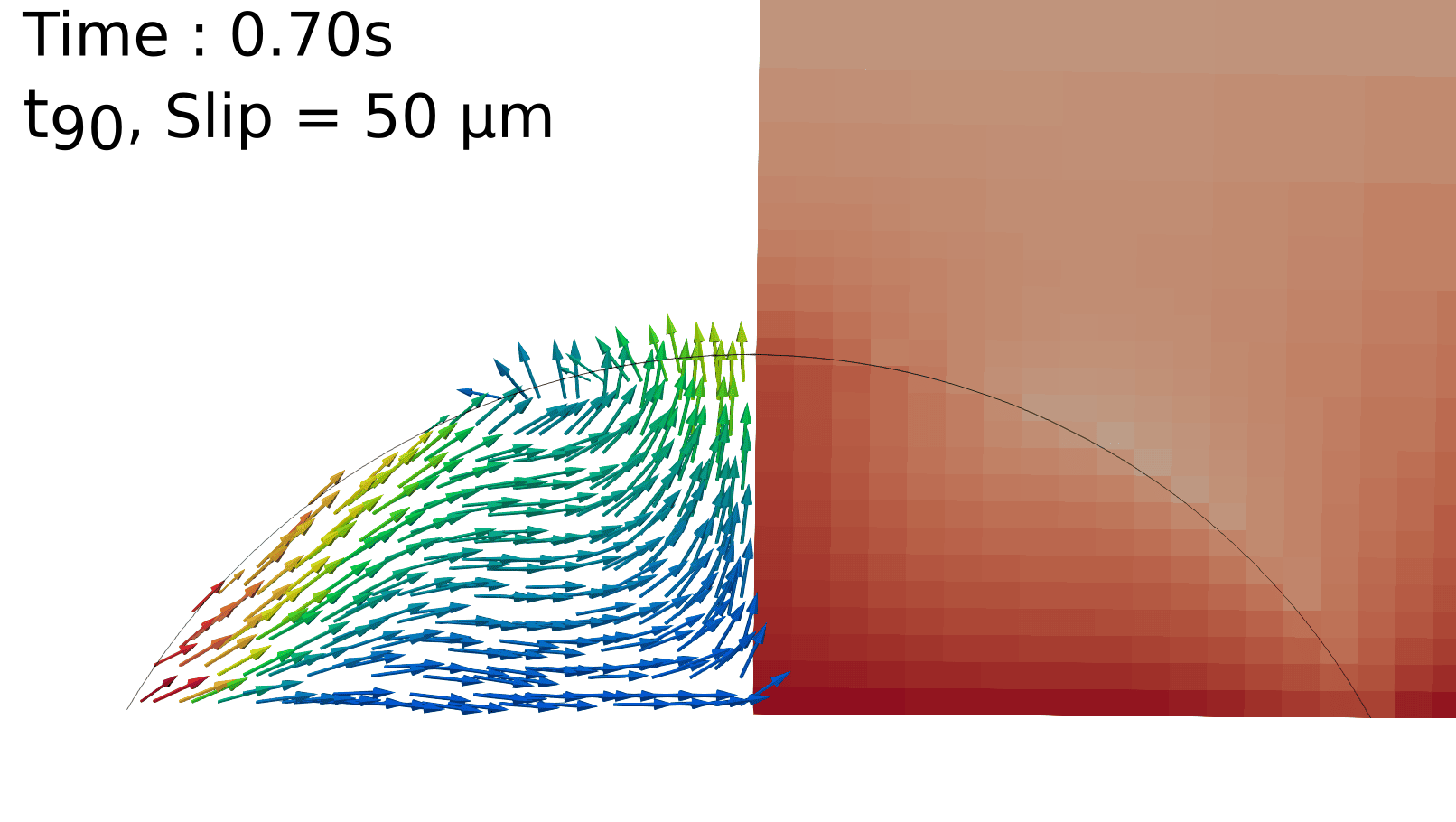}
                  \caption{}
         \label{fig:T1Slip5}
         \end{subfigure}
            \begin{subfigure}{0.45\textwidth}
          \includegraphics[width=\textwidth]{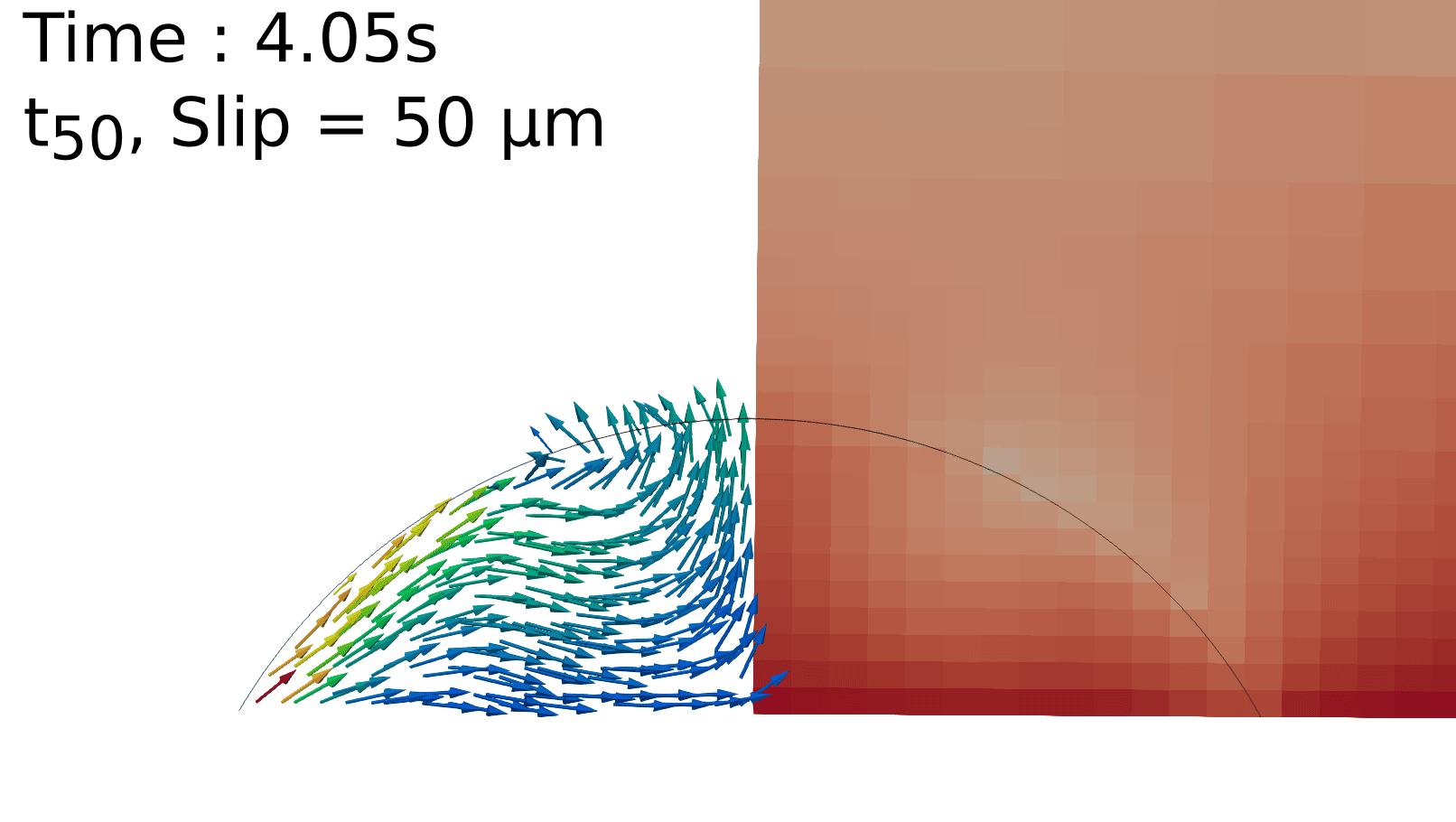}
                  \caption{}
         \label{fig:T2Slip5}
         \end{subfigure}
        \\
        \begin{subfigure}{0.5\textwidth}
         \includegraphics[clip,trim =10cm 5cm 10cm 15cm,width=\textwidth]{C2ColorMap.png}
         \end{subfigure}
    
                  \caption{Velocity and temperature plots for the cases including the Marangoni stresses, C2, where $\theta_\mathrm{c}=60\, ^\circ$, with different slip length. Velocity vector plots on left, where the color indicates the velocity magnitude. Contour plots of the temperature on the right. \textbf{(a)} C2, no-slip, $t_{90}$,\textbf{(b)} C2, no-slip, $t_{50}$,\textbf{(c)} C2, slip length = $20 \mu$m, $t_{90}$,\textbf{(d)} C2, slip length = $20 \mu$m, $t_{50}$,\textbf{(e)} C2, slip length = $50 \mu$m, $t_{90}$,\textbf{(d)} C2, slip length = $50 \mu$m, $t_{50}$}
                  \label{fig:figLast}
    \end{figure}
    
To study the slip-length model used to simulate the moving contact line for an evaporating sessile droplet, three different cases are considered for case C2 with a contact angle $60^\circ$ and accounting for Marangoni stresses: (i) a vanishing slip length (i.e.~no slip), see Figures \ref{fig:T1NoSlip} and \ref{fig:T2NoSlip}, (ii) a slip length of 20~$\mu$m, see Figures \ref{fig:T1Slip2} and \ref{fig:T2Slip2}, and (iii) a slip length of 50~$\mu$m, see Figures \ref{fig:T1Slip5} and \ref{fig:T2Slip5}.
In general, we can observe in Figure \ref{fig:figLast} that an increasing slip length suppresses the formation of the vortex ring in the droplet and increases the temperature at the apex of the droplet. This stands in contrast to available experimental measurements~\citep{Zhu2021,Zhu2023} and other numerical studies~\citep{Paul2023}, which show that for a volatile sessile droplet evaporating with a freely moving contact line, the coldest region appears at the apex of the droplet. We, therefore, conclude that a no-slip boundary condition is the most appropriate choice for our study.
% It is important to note that micro-liter ethanol droplets have an evaporation time of the order of $\mathcal{O} (10)$ s. Consequently, the mean speed of the contact line is $\mathcal{O}(10^{-5})$ m/s, whereas the flow velocity inside the droplet during evaporation is approximately $\mathcal{O}(10^{-3})$ m/s. As a result, in the constant contact angle (CCA) mode, the contact-line boundary condition does not interfere with the no-slip condition for the internal fluid domain. This has also been reported in \citet{Paul2023}.

\bibliographystyle{elsarticle-harv}

% \bibliography{SessileDroplet}
% \bibliography{/mnt/SHARE/BvWResearch.bib}

\end{document}